\documentclass[pdflatex,sn-basic]{sn-jnl}

\usepackage{graphicx}
\usepackage{multirow}
\usepackage{amsmath,amssymb,amsfonts}
\usepackage{amsthm}
\usepackage{mathrsfs}
\usepackage[title]{appendix}
\usepackage{xcolor}
\usepackage{textcomp}
\usepackage{manyfoot}
\usepackage{booktabs}
\usepackage{algorithm}
\usepackage{algorithmicx}
\usepackage{algpseudocode}
\usepackage{listings}
\usepackage{xspace}

\usepackage{upgreek}
\makeatletter
\input{aas_macros.sty}
\def\ref@jnl#1{{\jnl@style#1\ }}
\makeatother

\newcommand{\eqn}[1]{\text{Eq.~(\ref{#1})}}
\newcommand{\sect}[1]{\text{Sect.~\ref{#1}}}
\newcommand{\fig}[1]{\text{Fig.~\ref{#1}}}
\newcommand{\tab}[1]{\text{Table~\ref{#1}}}

\newcommand{\mtd}{\textlangle3D\textrangle}
\newcommand{\multitd}{\texttt{Multi3D}}
\newcommand{\balder}{\texttt{Balder}}
\newcommand{\nltetd}{\texttt{NLTE3D}}
\newcommand{\cobold}{\texttt{CO$^{5}$BOLD}}

\newcommand{\marcs}{\texttt{MARCS}}
\newcommand{\atlas}{\texttt{ATLAS}}
\newcommand{\stagger}{\texttt{Stagger}}

\newcommand{\phoenix}{\texttt{PHOENIX}}
\newcommand{\bifrost}{\texttt{Bifrost}}
\newcommand{\muram}{\texttt{MURaM}}

\newcommand{\medis}{\texttt{M3DIS}}

\newcommand{\lgeps}[1]{A(\mathrm{#1})}
\newcommand{\conc}[1]{C(\mathrm{#1})}
\newcommand{\awgt}[1]{\mu(\mathrm{#1})}
\newcommand{\lggf}{\log{gf}}
\newcommand{\nm}{\mathrm{nm}}
\newcommand{\eV}{\mathrm{eV}}
\newcommand{\dex}{\mathrm{dex}}
\newcommand{\kms}{\mathrm{km\,s^{-1}}}

\newcommand{\vmic}{\xi_{\mathrm{mic}}}

\newcommand\ion[2]{\text{#1\,\textsc{\lowercase{#2}}}}

\begin{document}

\title[Solar chemical composition]{Solar chemical composition}

\author[1]{\fnm{A. M.} \sur{Amarsi}}\email{anish.amarsi@physics.uu.se}
\equalcont{These authors contributed equally to this work.}

\author[2,3]{\fnm{N.} \sur{Grevesse}}\email{nicolas.grevesse@uliege.be}
\equalcont{These authors contributed equally to this work.}

\affil[1]{\orgdiv{Theoretical Astrophysics, Department of Physics and Astronomy}, 
\orgname{Uppsala University}, \orgaddress{\street{Box 524}, \city{Uppsala}, \postcode{75120},
\country{Sweden}}}

\affil[2]{\orgdiv{Centre Spatial de Li\`ege}, \orgname{Universit\'e de Li\'ege}, 
\orgaddress{\street{avenue Pr\'e Aily}, \postcode{B-4031 Angleur-Li\`ege},
\country{Belgium}}}

\affil[3]{\orgdiv{Space sciences, Technologies and Astrophysics Research (STAR)
Institute}, 
\orgname{Universit\'e de Li\`ege}, 
\orgaddress{\street{All\'ee du 6 ao\^ut, 17, B5C}, 
\postcode{B-4000 Li\`ege}, \country{Belgium}}}

\abstract{The Sun is one of the fundamental benchmarks in astronomy, and
there is
ever growing interest in precise and accurate determinations of its
chemical composition.
We present the current state-of-the-art in
spectroscopic determinations of the solar elemental abundances,
that now routinely employ
three-dimensional (3D) radiative-hydrodynamic simulations
of the solar photosphere and post-processing in
non-local thermodynamic equilibrium (non-LTE).
We critically review the recent literature
and present recommended present-day and protosolar
abundances of the 83 long-lived elements,
and discuss their implications
vis-\`{a}-vis primitive meteorites, helioseismology, neutrino fluxes,
and the Solar Modelling Problem.}


\maketitle

\newpage
\setcounter{tocdepth}{3}
\tableofcontents
\newpage

\section{Introduction}
\label{introduction}

The abundance pattern of 
the $83$ long-lived elements in the Sun is an important standard 
relevant to a broad range of astrophysical fields.
It is the yardstick in astronomy against which all other 
cosmic objects are compared 
\citep[e.g.][]{2025A&ARv..33....3G}; for instance,
it serves either as an input or a target for
models of cosmic chemical evolution
\citep[e.g.][]{2020ApJ...900..179K},
as a reference for comparing and interpreting 
the compositions of 
host stars of diverse exoplanetary systems
\citep[e.g.][]{2022A&A...664A.161B},
and as a precise baseline for studies of 
nucleosynthesis, for instance for the
rapid neutron capture process, the astrophysical sites of which
remain poorly understood 
\citep[e.g.][]{2021RvMP...93a5002C}.
The solar chemical composition
is implicit to stellar modelling:
the solar abundance pattern scaled by metallicity
serves as a standard input for calculating opacities
needed to model the photospheres
\citep[e.g.][]{2008A&A...486..951G} and interiors 
\citep[e.g.][]{2016ApJ...823..102C} of stars 
across the Hertzsprung-Russell diagram.
Needless to say, the detailed solar chemical composition also
plays an important role in modelling and understanding 
the various bodies in the solar system
\citep[e.g.][]{2023Sci...379n8671N},
the solar wind and upper solar atmosphere
\citep[e.g.][]{2019ApJ...879..124L};
and, not least, the Sun itself.

This last point is exemplified by the significant 
discrepancies between predictions for the stratification
of the solar interior compared with 
helioseismic inferences of the 
helium abundance in the convective envelope,
the sound speed profile,
and the depth of the convection zone
\citep[e.g.][]{2005ApJ...618.1049B,2021LRSP...18....2C}.
This problem was originally referred to as the
Convective Zone Problem
\citep{2004ApJ...614..464B,2004ApJ...615.1042M},
and later renamed the 
Solar Abundance Problem
\citep[e.g.][]{2008SoPh..251...53C,2009ApJ...704.1174P},
to highlight the sensitivity of the solar interior modelling
on the assumed chemical composition.
Here, we adopt a more agnostic label, namely the
Solar Modelling Problem, because
it seems unlikely that it can be resolved simply by 
tweaking the solar abundance pattern
\citep{2023A&A...669L...9B,2024A&A...686A.108B}.

Solar abundance analyses have a history
dating back one hundred years with the development 
of quantitative spectroscopy of the Sun's light
\citep{1925PhDT.........1P,1929ApJ....70...11R}
and mass spectrometry of a class of primitive meteorites
known as CI chondrites
\citep{1922NW.....10..918G,10007418258}.
A Standard Solar Composition
\citep{1989GeCoA..53..197A,1998SSRv...85..161G} emerged towards
the end of the 20th century, taking advantage of
exquisite observational data
\citep[e.g.][]{1995OptEn..34.2736D,1999SoPh..184..421N},
ever improving atomic data
\citep[e.g.][]{1993PhST...47..133G},
and helioseismic inversions for precise constraints on the helium
content \citep{1991Natur.349...49V}.
As precision and accuracy improved, the solar spectroscopic
abundances were found to approach those inferred
from CI chondrites, a notable accomplishment
following decades of discrepancies;
we refer the reader to the introductions of
\citet{2021A&A...653A.141A} and \citet{2024M&PS...59.3193J},
and references therein, for an overview of the history of these fields.

The turn of the century marked the beginning of a new era
of solar abundance determinations.  
The spectroscopic analyses of the 1990s were 
based on one dimensional (1D)
hydrostatic model photospheres, post-processed under the assumption
of local thermodynamic equilibrium (LTE).
In contrast, the new era was characterised by 
the use of 3D radiation-hydrodynamics
simulations, and the increasing use of 
non-LTE radiative transfer post-processing.
This post-processing was typically carried out in 1D,
although
consistent 3D non-LTE approaches were sometimes attempted for simple
species,
following the pioneering study of
\ion{O}{I} line formation by \citet{1995A&A...302..578K}.
With these developments,
the metal mass fraction of the solar photosphere turned out to be 
10\% to 30\%
lower than the Standard Solar Composition values of the 1990s 
\citep[e.g.][]{2005ASPC..336...25A,2009ARA&A..47..481A,2011SoPh..268..255C}.

\begin{figure}[ht]
\centering
    \includegraphics[width=1\textwidth]{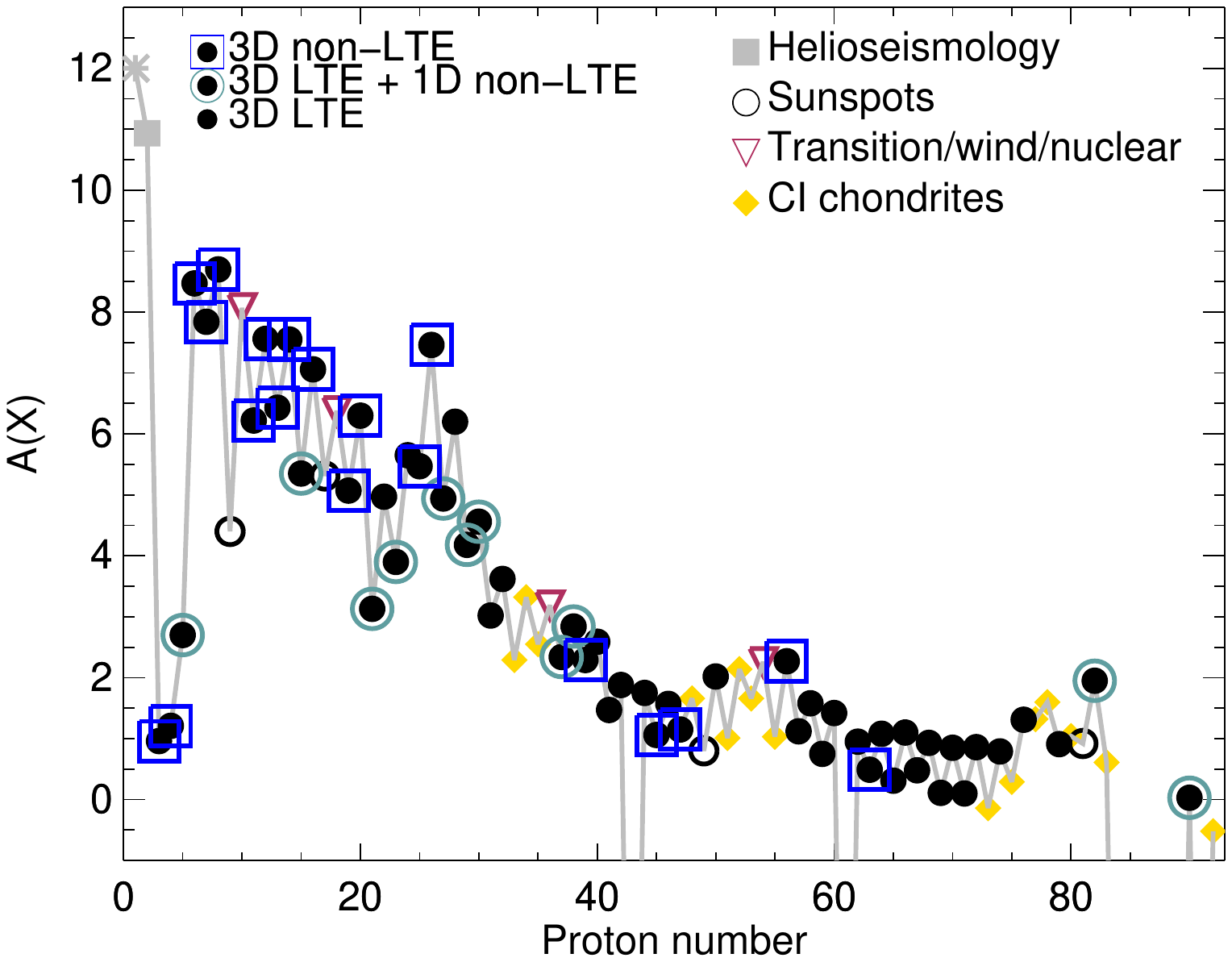}
    \caption{Abundance pattern of the $83$ long-lived
    elements (elements with at least one isotope of
    half-life longer than or comparable to the age of the solar system).
    Elemental abundances
    $\lgeps{X}\equiv\log N_{\mathrm{X}}/N_{\mathrm{H}}+12$,
    where $N_{\mathrm{X}}$ is the number of nuclei of element $\mathrm{X}$
    summed over all atoms, ions, and molecules.
    The reference element, hydrogen, is thus fixed at $\lgeps{H}=12$.
    For $58$ elements from lithium to thorium, the abundances
    are based on spectroscopy of the quiet solar photosphere
    (solid black circles): dark 
    blue squares highlight those results 
    based on consistent 3D non-LTE
    spectroscopic analyses of the solar photosphere, and
    light blue circles highlight those
    results based on 3D LTE spectroscopy with
    non-LTE versus LTE abundance corrections based on
    1D or \mtd{} model photospheres; and the remainder are based 
    on 3D LTE.
    For $24$ other long-lived elements from helium to uranium,
    the results are based on alternative methods (\sect{resultsother}).}
    \label{fig:abundancepattern}
\end{figure}

\begin{figure}[ht]
\centering
    \includegraphics[width=1\textwidth]{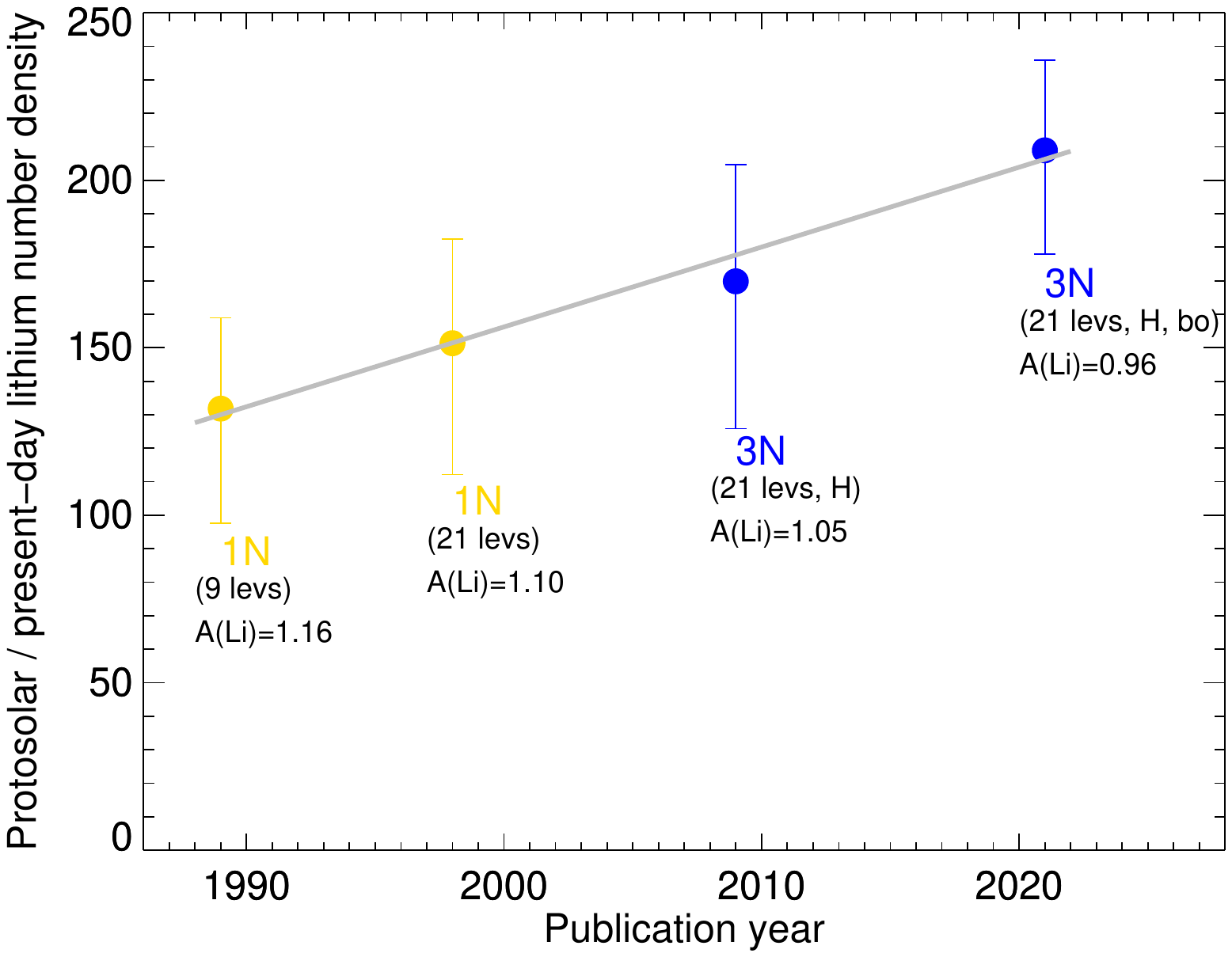}
    \caption{Estimates of the level of depletion of lithium 
    in the convective envelope in different spectroscopic compilations
    \citep{1989GeCoA..53..197A,1998SSRv...85..161G,
    2009ARA&A..47..481A,2021A&A...653A.141A},
    taking the protosolar nebula value to be $\lgeps{Li}=3.28$ 
    via CI chondrites (\sect{discussionproto}).
    The successive changes to the spectroscopic solar lithium abundance are
    related to a more complete non-LTE model atom
    (number of levels, ``levs''; \citealt{1994A&A...288..860C});
    a consistent 3D non-LTE treatment together with
    a realistic description of the inelastic collisions with
    neutral hydrogen (``3N'', ``H'';
    \citealt{2003A&A...409L...1B})
    and an improved treatment of
    the background opacities (``bo'';
    \citealt{2021MNRAS.500.2159W}).}
    \label{fig:lithium}
\end{figure}

Significant progress has been made
over the last ten to fifteen years.
An important milestone is that
results from 3D model photospheres are now folded into the
abundance analyses for all $58$ elements that can be reliably inferred
from the quiet solar spectrum \citep{2021A&A...653A.141A}.
Furthermore, this period is marked by the much broader application of
1D non-LTE abundance corrections,
together with the rapid development of consistent 3D non-LTE modelling
(\fig{fig:abundancepattern}).
It is also noteworthy that the non-LTE models developed
in recent years draw upon atomic data sets that are 
typically more complete and based on more realistic physics
\citep[e.g.][]{2016A&ARv..24....9B}.
These various improvements can impart significant 
shifts in inferred abundances (e.g.~\fig{fig:lithium}).
We reflect that many of these methodological
developments were propelled in part by
the so-called industrialisation of
stellar astrophysics \citep{2019ARA&A..57..571J},
specifically the need to model stars like the
Sun to greater levels of precision and accuracy 
so as to reach the full scientific potential of the
huge volumes of data collected via,
for example, LAMOST \citep{2015RAA....15.1095L},
SDSS \citep{2023ApJS..267...44A}, 
Gaia \citep{2023A&A...674A...1G}, GALAH \citep{2025PASA...42...51B},
and DESI \citep{2026OJAp....955260K};
and due to be collected in the near future
via, for example, 4MOST \citep{2019Msngr.175....3D}, 
and WEAVE \citep{2024MNRAS.530.2688J}.
This industrialisation helped
motivate
the calculation of key atomic data and the development of reliable
model atoms (see e.g.~references within
\citealt{2016ApJ...833..225Z},
\citealt{2020A&A...642A..62A}, and
\citealt{2023A&A...669A..43G})
and 3D non-LTE post-processing tools
\citep{2009ASPC..415...87L,
2015A&A...583A..57S,
2018A&A...615A.139A,
2026MNRAS.546f2085H}.
At the same time, a new generation
of grids are being developed 
for 3D model stellar photospheres
\citep[e.g.][]{2023A&A...669A.157W,
2024A&A...688A..27P,2024A&A...688A.212R,2025A&A...703A.199E,
2025A&A...699A.148A}, and for
stellar interiors in 1D
\citep[e.g.][]{2021ApJ...908..102P,
2022A&A...665A.126N,2024A&A...690A..91S,
2025A&A...704A..79M,2026ApJS..283...64D},
with rapid progress being made in 3D 
(e.g.~\citealt{2022A&A...659A.193A}, and references therein).
There is great interest in having a modern solar elemental abundance
pattern to feed into this next generation of stellar models: one that is
standard, so as to allow for ease of comparison between different studies; and
one that is accurate, so as to give reliable overall results.

We first present and discuss
most important data and methods
for spectroscopic determinations
of the photospheric abundances (\sect{method}).
We then compare commonly used 
solar abundance compilations,
elucidate from where the difference
arise in these different compilations, and present
our recommended values (\sect{results}).
We follow with a discussion of the abundances measured via 
CI chondrites,
the composition of the protosolar nebula,
and the Solar Modelling Problem (\sect{discussion}),
before presenting our closing reflections
(\sect{conclusion}).

\section{Methods}
\label{method}

We preface our comparison of different 
spectroscopic results for the solar chemical composition in \sect{results}
with a discussion of the relevant data and analysis methods.
We start with the observational data (\sect{methodspectrum}),
and subsequently the selection and measurement of 
diagnostic absorption lines (\sect{methodlines}).
We follow with a discussion of the key modelling considerations,
starting with the physical conditions in the solar layers where these
absorption lines are formed (\sect{methodatmosphere}).
Atomic data underpins solar and stellar spectroscopy:
the conclusions of LTE analyses are 
strongly sensitive to the accuracy of the
transition probabilities 
and broadening parameters of the individual
diagnostic absorption lines (\sect{methodlte}).
The demand for atomic data is even greater for non-LTE analyses,
because the complete set of radiative and collisional processes
that determine the statistical equilibrium 
has to be taken into account (\sect{methodnlte}).
This requires the construction of a
so-called model atom that is as complete as possible
including energy levels that reach
quite close to the ionisation limit,
possibly of more than one ionisation stage;
radiative bound-bound and bound-free transitions connecting these levels;
as well as cross-sections for
inelastic collisions with electrons and with
neutral hydrogen atoms.
We end with a brief discussion of how 3D non-LTE effects
typically affect solar elemental abundance
determinations (\sect{methodeffects}).

\subsection{Observational data sets}
\label{methodspectrum}

Analyses of the Sun are based on light
at near-UV to near-infrared wavelengths.
There are a variety of observational reasons for this;
from a modelling perspective, it is advantageous to 
carry out spectroscopic analyses on light that forms
in the deeper layers of the photosphere.
The primary reason is that
departures from LTE are less severe in such layers:
the decoupling of light from matter,
that push atomic, ionic, and molecular number populations away from
their Saha-Boltzmann distributions, is less extreme; and furthermore particle
densities are higher such that collision processes are more efficient at
restoring the populations to these distributions.  
Additionally, model photospheres that are lacking chromospheres
become less realistic 
in the upper layers close to the temperature minimum.
Continuum light from around $152\,\mathrm{nm}$ to around
$250\,000\,\mathrm{nm}$ forms below the chromosphere
\citep[e.g.][]{2019ARA&A..57..189C}.
The deepest-forming light has a
wavelength of around $1644\,\nm$, corresponding to the 
electron affinity of
the $\mathrm{H^{-}}$ anion, the major
opacity contributor in the solar photosphere.

Light observed at the centre of the disc (hereafter referred to as
the disc-centre intensity), rather than light integrated
across
the entire disc (the disc-integrated flux),
is often preferred for spectroscopic abundance analyses.
The disc-centre intensity is emitted at deeper radial layers
compared to off-centre intensities (as evidenced
by the limb darkening phenomenon),
and consequently the disc-integrated flux
samples higher layers of the photosphere on average.
Furthermore,
the disc-centre intensity better displays the minute 
asymmetries imparted on the absorption lines 
\citep[e.g.][]{1982ARA&A..20...61D}
by the up and down motions of the 
granules and inter-granular lanes
in the photosphere (\sect{methodatmosphere}).
In the disc-integrated flux, 
some of this information is lost because of sampling
the granules and inter-granular lanes
with a variety of different alignments relative to the observer.
Furthermore, the absorption lines in the disc-integrated flux 
suffer additional broadening
compared to the same absorption lines seen in the
disc-centre intensity,
due to a projected rotational velocity that reaches nearly $2\,\kms$.
This is important because 
slight distortions of the absorption line
shapes can help indicate the presence of unidentified blends. Thus,
analyses on disc-centre intensity are generally better placed to 
mitigate the impact of blends on abundance inferences.

Disc-resolved or centre-to-limb observations deserve special mention.
As mentioned above, light observed at the limb 
escapes from higher radial
layers than light formed at disc-centre.
This fact has been exploited to infer the atmospheric stratification
of the photosphere and construct semi-empirical 1D model photospheres
\citep[e.g.][]{1967ZA.....65..365H,1974SoPh...39...19H},
and to test the structures of theoretical models
\citep[e.g.][]{2013A&A...554A.118P}.
The centre-to-limb variation of absorption lines
has also been used to test, and sometimes calibrate, the inelastic 
collision cross-sections in non-LTE model atoms
(for example, for the 
\ion{O}{I} $777\,\nm$ triplet, see
\citealt{2009A&A...508.1403P},
\citealt{2015A&A...583A..57S}, and 
\citealt{2018A&A...616A..89A}).
This takes advantage of the strong depth dependence of the
ratio of the number densities of neutral hydrogen
to free electrons \citep{2004A&A...423.1109A},
the two dominant microscopic perturbers in the solar photosphere
\citep[e.g.][]{2016A&ARv..24....9B}.

\begin{table}[ht]
    \caption{Selected high resolution, high signal-to-noise
    ratio observations covering a wide spectral range
    that have recently been used for abundance analyses.}\label{tab:atlas}
\begin{tabular*}{\textwidth}{@{}lllrr@{}}
\toprule
    Reference & 
    Observatory &
    Type & 
    $\Delta\lambda/ \nm$ & 
    Approx.~$R=\lambda/\delta\lambda$\footnotemark[1] \\
\midrule
    {\citet{1973apds.book.....D}} &  
    Jungfraujoch &
    Disc-centre & 
    $300$--$1000$ &
    $\gtrsim500\,000$ \\
    (Li\`{e}ge atlas) & ($3580\,\mathrm{m}$) & & & \\ 
\noalign{\smallskip}
\hline
\noalign{\smallskip}
    {\citet{delbouille1981photometric}} &
    Kitt Peak & 
    Disc-centre & 
    $1000$--$5400$ & 
    $\sim300\,000$ \\
    & ($2096\,\mathrm{m}$) & & & \\
\noalign{\smallskip}
\hline
\noalign{\smallskip}
    {\citet{1984SoPh...90..205N}} & 
    Kitt Peak &
    Disc-centre \& &
    $329$--$1251$ &
    $\sim300\,000$ \\
    (Hamburg atlas) & ($2096\,\mathrm{m}$) & Disc-integrated & & \\
\noalign{\smallskip}
\hline
\noalign{\smallskip}
    {\citet{1989STIN...9013893F}} &
    ATMOS &
    Disc-centre &
    $2083$--$16510$ &
    $62\,500$--$480\,000$ \\
    & (Space) & & & \\ 
\noalign{\smallskip}
\hline
\noalign{\smallskip}
    {\citet{2010JQSRT.111..521H}} &
    ACE &
    Disc-centre &
    $2258$--$14290$ &
    $37\,500$--$220\,000$ \\
    & (Space) & & & \\
\noalign{\smallskip}
\hline
\noalign{\smallskip}
    {\citet{2005MSAIS...8..189K}} &
    Kitt Peak &
    Disc-integrated &
    $300$--$1000$ &
    $\sim300\,000$ \\
    & ($2096\,\mathrm{m}$) & & & \\
\noalign{\smallskip}
\hline
\noalign{\smallskip}
    {\citet{2015A&A...573A..74S}} &
    Kitt Peak &
    Centre-to-limb &
    $408$--$995$ &
    $\sim300\,000$ \\
    & ($2096\,\mathrm{m}$) & & & \\
\noalign{\smallskip}
\hline
\noalign{\smallskip}
    {\citet{2023A&A...673A..19E}} &
    G\"{o}ttingen &
    Centre-to-limb &
    $420$--$800$ &
    $521\,000$--$992\,000$ \\
    & ($160\,\mathrm{m}$) & & & \\
\botrule
\end{tabular*}
    \footnotetext[1]{Values stipulated in the reference;
    or, otherwise, nominal values drawn from other literature.
    See \citet{2016A&A...590A.118D} for a critical
    evaluation of the resolving powers of the
    Li\`{e}ge and Hamburg atlases.}
\end{table}

Observations of the solar spectrum at 
high resolution and high signal-to-noise ratio
that cover a wide spectral range are typically referred to as
atlases.  A selection of atlases that have been employed for 
solar abundance analyses are listed in \tab{tab:atlas}.
It is remarkable that the Li\`{e}ge and 
Hamburg atlases from around half a century ago 
remain the state-of-the-art for the disc-centre intensity at 
optical wavelengths;
we refer the reader to 
\citet{2016A&A...590A.118D} for an in-depth discussion of these
two important data sets.
\tab{tab:atlas} demonstrates that more recent progress 
includes extensions to the infrared
enabled by space-based observations, as well as more complete 
centre-to-limb data such as from the recent atlas of
\citet{2023A&A...673A..19E},
used for example in the 3D non-LTE analyses of
\citet{2024A&A...683A.242C} and \citet{2024A&A...683A.200S}
for \ion{Na}{I} lines 
and \ion{Y}{II} lines, respectively.
Omitted from the table but no less important are 
narrow-band observations targeting selected absorption lines.
For example, centre-to-limb observations of selected
\ion{O}{I} lines with the  Swedish Solar Telescope
by \citet{2009A&A...507..417P,2009A&A...508.1403P}
and more recently by 
\citet{2015A&A...583A..57S},
\citet{2018A&A...616A..89A},
\citet{2021MNRAS.508.2236B}, and
\citet{2023A&A...672L...6P}
have heavily influenced the debate about the 
solar oxygen abundance.

\subsection{Selection and measurement of absorption lines}
\label{methodlines}

Solar abundance analyses usually proceed on a line-by-line basis.
This means that separate abundances are inferred for selected
absorption lines of a given element.  
One advantage of this
approach is that it is possible to examine trends in the inferred 
abundances with line parameters.
Trends with wavelength, excitation energy (the energy of the lower level
relative to the ground state of the atom, ion, or molecule), and oscillator
strength, which are related to the formation depth, give clues on deficiencies
in the modelling of the solar photosphere.  Such trends may also reflect
deficiencies in the modelling of departures from LTE, although non-LTE effects
typically have a complex dependence on these and other line parameters that can
be difficult to disentangle.  Additionally, obvious outliers from the mean
trends may reflect the presence of a blend, or an unreliable oscillator strength
(\sect{methodlte}), and signal that
a closer inspection of that particular diagnostic absorption
line is needed.

\begin{figure}[ht]
\centering
    \includegraphics[width=0.51\textwidth]{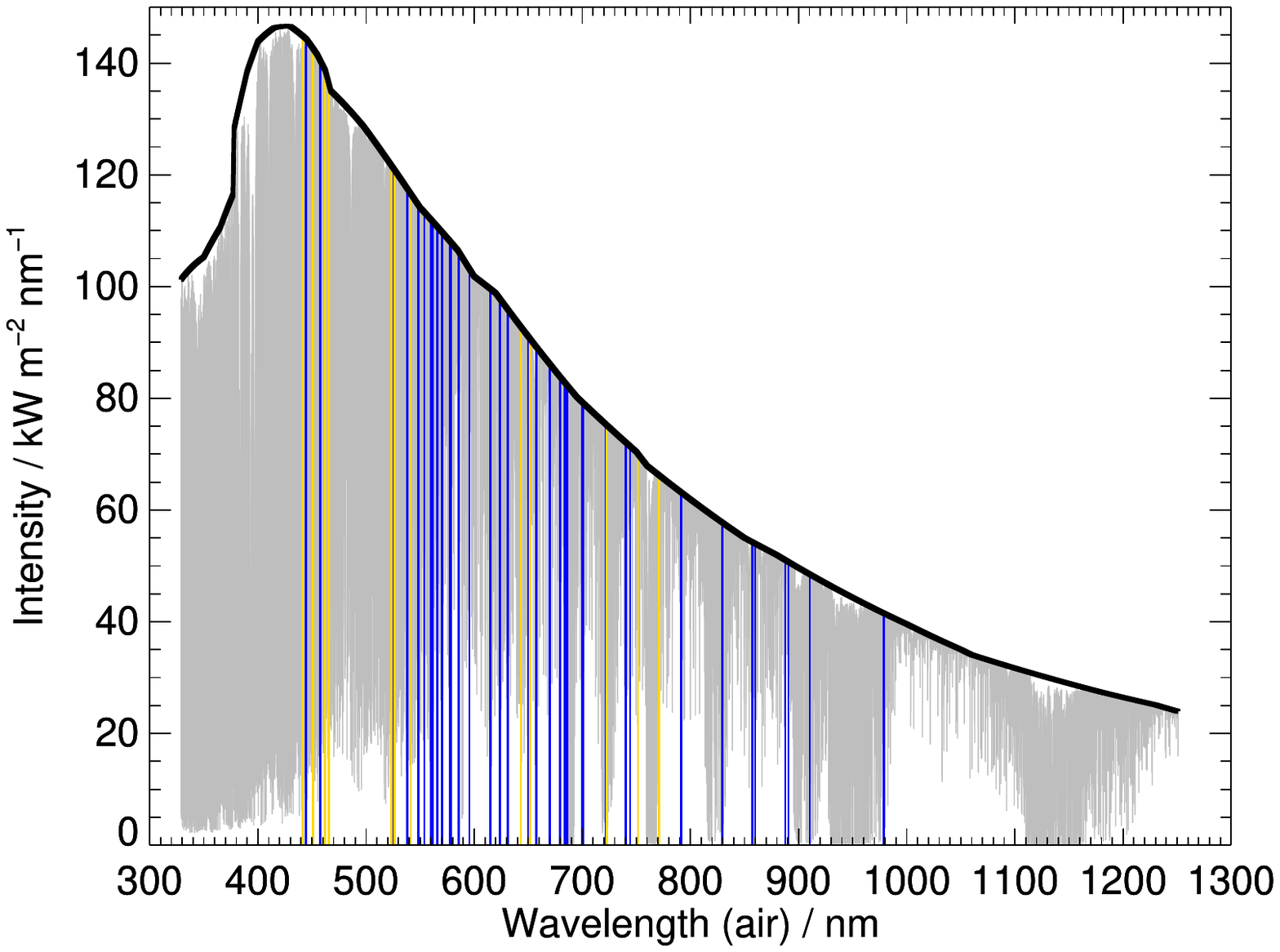}\includegraphics[width=0.51\textwidth]{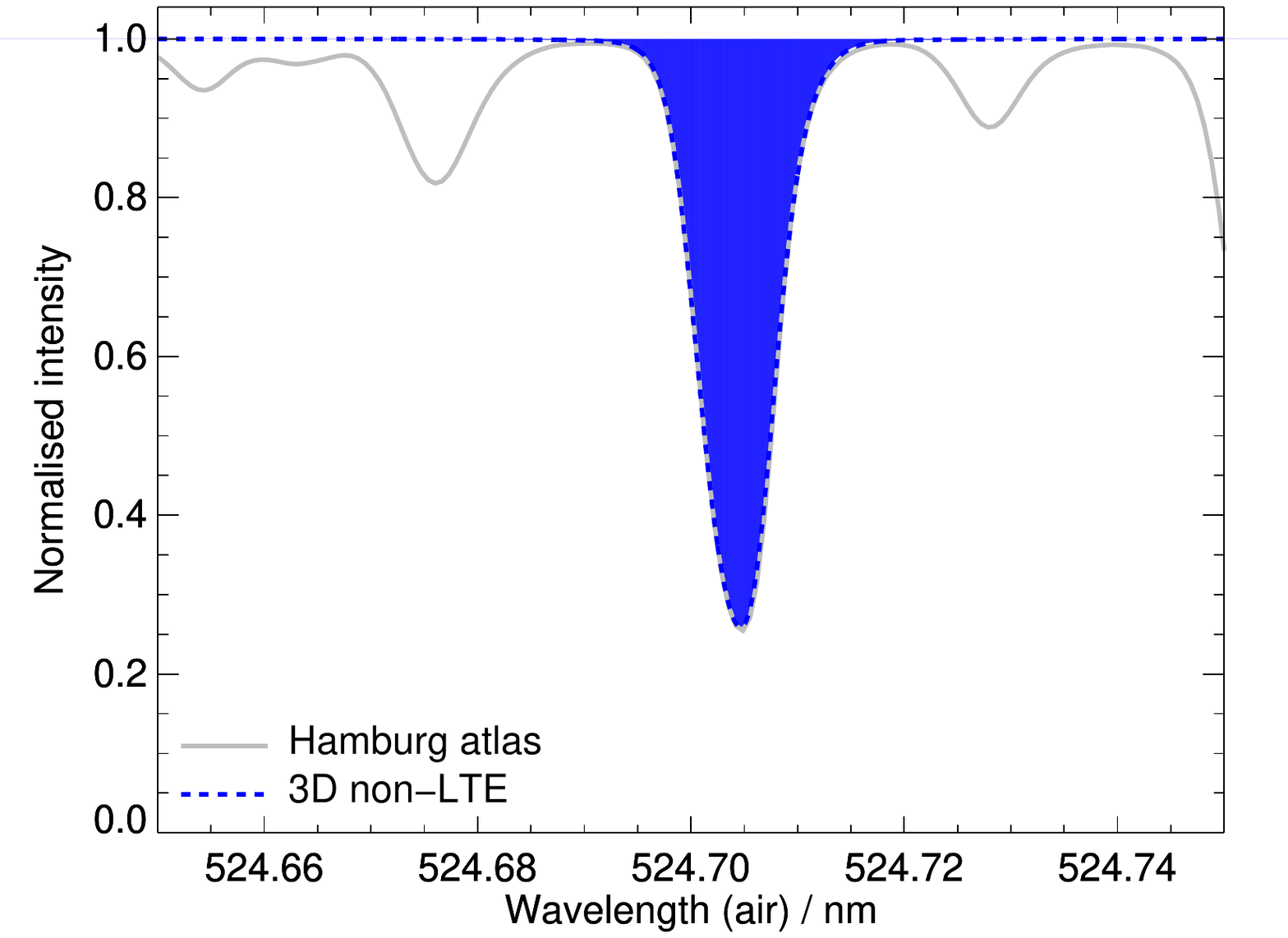}
    \caption{Absorption lines in the Hamburg  disc-centre intensity
    atlas.  \textbf{Left:} Absolute intensity from
    \citet{1984SoPh...90..205N}. Blue and yellow 
    vertical columns indicate 
    wavelength regions used in the analysis of \ion{Fe}{I}
    and \ion{Fe}{II} lines respectively by 
    \citet{2021A&A...653A.141A}.
    \textbf{Right:} The continuum-normalised
    \ion{Fe}{I} $524.70\,\nm$ line, with a
    3D non-LTE model spectrum based on the model of
    \citet{2021A&A...653A.141A} overplotted,
    assuming $\lgeps{Fe}=7.46$.
    The equivalent width of the line is the shaded area.}
    \label{fig:hamburg}
\end{figure}

The solar spectrum from the near UV to the near infrared
has of the order of several tens of thousands of
detectable absorption lines.
For example, $16\,000$ potentially detectable lines of \ion{Fe}{I} alone
were presented in \citet[][]{2022ApJS..260...28P}.
This can be contrasted against 
the iron abundance analysis presented in 
\citet{2021A&A...653A.141A}, for example,
which is based 
on just $40$ \ion{Fe}{I} lines 
(as well as $13$ \ion{Fe}{II} lines; see \fig{fig:hamburg}).
When there are multiple detectable 
absorption lines, usually a first selection for abundance determinations
is made based on the severity of blending
by other species in the solar photosphere as well as from absorbers in the
Earth’s atmosphere;
and a further selection based on the availability of
reliable oscillator strengths (\sect{methodlte}).
These two criteria tend to bias the selection towards 
absorption lines in optical wavelengths.
There are also constraints on the strength of
the absorption line: 
they must be deep enough such that
they can be measured
reliably; at the same time, they should ideally lie on the weak 
part of the curve-of-growth
(e.g.~Chapter 13 of \citealt{2022oasp.book.....G})
such that their equivalent widths
(the area enclosed by the continuum-normalised absorption line;
see \fig{fig:hamburg}) have
a linear dependence on the elemental abundance.
In the classical LTE picture,
the cores of absorption lines that are on the saturated
part of the curve-of-growth 
form close to 
the temperature minimum, and thus can no longer get deeper even
as the elemental abundance is increased.
Their apparent strengths are less sensitive to the elemental 
abundance and thus these absorption lines are less reliable abundance
diagnostics.

Analyses of a given absorption line in the solar spectrum
are either based on equivalent width measurements,
or a profile-fitting approach.
The profile-fitting approach can help
to validate the modelling: for example, reproducing
the line broadening and asymmetries helps verify that
the resolved velocity field is well described
in the 3D model photosphere
\citep[e.g.][]{2000A&A...359..729A}.
Furthermore, this approach is often necessary for those
elements with few detectable features in the solar spectrum
and for which it is necessary to analyse strongly blended features.
For example, for the \ion{Be}{II} $313\,\nm$
\citep[e.g.][]{2022A&A...657L..11K,2024A&A...690A.128A},
the shape of the absorption line 
helps constrain the contribution of blends.
One fitting strategy is to give more weight to the wings
of the absorption line, which form deeper in the photosphere
where the models are typically more reliable
\citep[e.g.][]{2000MNRAS.311..535B}.
There are also 
advantages for performing analyses based on equivalent widths. For instance, the
profile-fitting approach requires the model spectrum to be convolved with the
instrumental profile of the spectrograph, and (when using the disc-integrated
flux) rotational broadening must also be taken into account; whereas equivalent
widths are to first order conserved by these macroscopic broadening processes.
Furthermore, for profile-fitting of model spectra based
on 1D model photospheres, an additional macroturbulent parameter needs to be
included to account for broadening 
by large scale velocity field gradients.
This macroturbulent parameter conserves equivalent widths;
it follows that equivalent widths are less sensitive to the 
modelling of the large-scale velocity fields in the 3D model photosphere.
A corollary is that analyses employing 3D model photospheres 
may rely on coarser horizontal and temporal samplings when they
are based on equivalent widths, compared to when they are
based on profile fits \citep[][]{2024A&A...688A.212R}.

\subsection{Theoretical and semi-empirical
3D, 1D, and \mtd{} model photospheres} 
\label{methodatmosphere}

At optical wavelengths the solar surface appears 
granulated, with bright, hot areas of upflowing gas, or granules,
having mean
diameters of around $1\,\mathrm{Mm}$ 
and lifetimes of the order of several minutes,
surrounded by dark, cool inter-granular lanes of rapidly inflowing gas
\citep[e.g.][]{2009LRSP....6....2N}.
This is a consequence of convection visible also
at the optical surface.  Although the solar photosphere is
formally stable against convection, 
there is overshoot of mass, momentum, and energy into it.
This directly affects the emergent spectrum,
by introducing velocity fields
that causes Doppler shifts on small and large scales
that broaden absorption lines and introduce asymmetries.
There are also more indirect effects on the emergent
spectrum, via changes to the atmospheric structure and 
the introduction of temperature and density inhomogeneities.

\begin{figure}[ht]
\centering
    \includegraphics[width=0.87\textwidth]{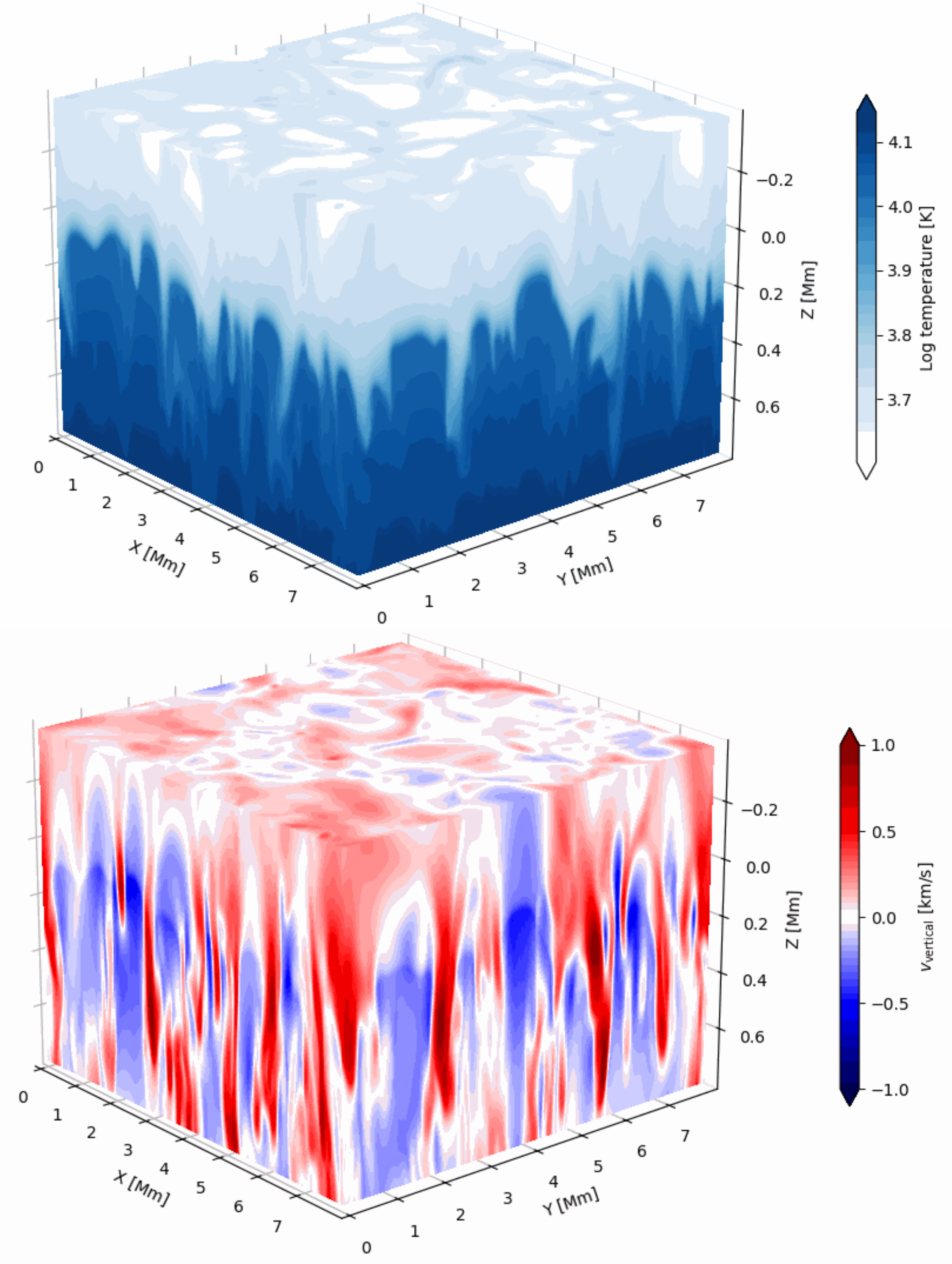}
    \caption{Snapshot of a 3D model photosphere.
    \textbf{Top:} Logarithm of the gas temperature.
    \textbf{Bottom:} Vertical component of the gas velocity.
    Based on the solar model from the \stagger{}-grid
    \citep{2013A&A...557A..26M,2024A&A...688A.212R}.}
    \label{fig:3dbox}
\end{figure}

To take such effects into account,
the state-of-the-art in solar abundance analyses 
has been based on theoretical 
3D radiation-hydrodynamics simulations of solar surface convection
(\fig{fig:3dbox}).  The literature results discussed in \sect{results}
are based on two families of 3D model photospheres,
and for clarity we refer to these via the names
of the codes with which they were computed:
the \cobold{} code \citep{2012JCoPh.231..919F} and the \stagger{} code 
\citep{2018MNRAS.475.3369C,2024ApJ...970...24S};
other 3D model photospheres 
have been discussed in the literature,
including those based on
the \bifrost{} code \citep{2011A&A...531A.154G,2025ApJ...990..195G},
the \medis{} code \citep{2024A&A...688A..52E},
and the \muram{} code \citep{2024A&A...681A..81W}.
Recent comparisons 
\citep[e.g.][]{2020A&A...636A.120A,2022A&A...668A..48D}
suggest that the absorption lines predicted by the 
\cobold{} and \stagger{} families of models are usually
in agreement to around $0.01\,\dex$
(in terms of logarithmic equivalent width ratio,
which, for weak lines, closely corresponds to the 
difference in the inferred abundance $\Delta\lgeps{X}$).
In both cases, solar abundance
work have been based on simulations of a small patch of the 
convection zone, which are sometimes referred 
to as local, or box-in-a-star, simulations
(in contrast to global, or star-in-a-box, simulations;
e.g.\ \citealt{2024A&A...692A.223F}).
For example, the model used by 
\citet{2018A&A...616A..89A} has a
vertical extent of $4\,\mathrm{Mm}$,
spanning around $12$ Rosseland-mean logarithmic optical depths 
or $14$ pressure scale heights
roughly centred at the optical surface;
this may be contrasted against, for example, the extent of
the full convective zone which is around $200\,\mathrm{Mm}$
\citep[e.g.][]{1991ApJ...378..413C}.
The same model spans an area of $6\times6\,\mathrm{Mm^{2}}$,
which is only around $0.001\%$ the hemispherical area of the solar surface,
but wide enough to encompass around ten granules.

The 3D simulations 
solve the hydrodynamic equations of mass, momentum, and energy
conservation.  Radiative heating and cooling is taken
into account by solving the radiative transfer equation
in 3D and at each time step. 
A key approximation is to use opacity
binning, where around a dozen opacity bins are first calculated by
considering the approximate formation depths
of light at different wavelengths
\citep[e.g.][]{2023A&A...675A.160P,2023A&A...677A..98Z}.
The monochromatic radiative transfer equation is then solved,
under the LTE approximation, for these few bins, in contrast to the
$10^{5}$ wavelength points typically used in the construction of
1D model photospheres \citep[e.g.][]{2008A&A...486..951G}.
The physical resolution of the simulations is such that 
only the largest flow structures are resolved
(large eddy simulations); 
this approach has been demonstrated
to properly reproduce the physics of solar surface convection
that are most relevant to spectral line formation
\citep[e.g.][]{1997A&A...328..229N}.
The current state-of-the-art for solar abundances
does not account for magnetic fields, but pilot studies
report that their effects on the equivalent widths
of absorption lines, when integrated over 
an area corresponding to at least several granules,
are only of the order of $0.01\,\dex$ 
\citep[e.g.][]{2015ApJ...799..150M,2015A&A...579A.112S,2016A&A...586A.145S}.

Spectroscopic analyses of stars not too different from
the Sun are usually
based on theoretical 1D model photospheres computed using 
for example the \marcs{} code
\citep{2008A&A...486..951G}, \atlas{} code
\citep{2003IAUS..210P.A20C,2005MSAIS...8...14K},
or \phoenix{} code \citep{2025A&A...698A..47H}.
These models typically resort to a parameterised treatment of 
the convective flux such as the mixing length theory 
\citep[e.g.][]{1958ZA.....46..108B,1965ApJ...142..841H}.
Spectral line formation on such models employ additional
broadening parameters labelled microturbulence 
and macroturbulence, that primarily reflect
velocity gradients on scales much smaller than or
much larger than an optical depth respectively,
but may also compensate for other deficiencies
in the 1D model photosphere
\citep[e.g.][]{2016AN....337..844L}.
Increasingly, horizontally- and temporally-averaged
3D model photospheres, hereafter \mtd{} model photospheres,
usually based on the \stagger{}-grid
\citep{2013A&A...560A...8M}, are employed for solar, and stellar
abundance analyses.
While an advantage of \mtd{} model photospheres
are a more realistic average structure,
\mtd{} model photospheres nevertheless suffer from many of the
same deficiencies as 1D models such as the lack of 
a realistic 3D velocity field and the lack of
inhomogeneities, both of which strongly impact spectral line formation
\citep[e.g.][]{2011ApJ...736...69U,2024ARA&A..62..475L};
furthermore, the averaging procedure is not unique,
and in particular \mtd{} models
are typically constructed without constraining the
effective temperature.
Spectroscopic analyses based on both 1D and \mtd{} models
are consistently outperformed
by those of 3D models \citep[e.g.][]{2009LRSP....6....2N,2013A&A...554A.118P}.
Differential star versus Sun comparisons may
benefit from employing the same 1D paradigm for the star as for the Sun;
but, the usefulness of 1D model photospheres
for determining the
absolute solar chemical composition is much more limited.

As mentioned in \sect{introduction}, solar abundance
work leading up to the turn of the century 
was largely based on semi-empirical 1D model photospheres,
in particular that of \citet{1974SoPh...39...19H}.
Even today this model is sometimes used,
primarily to facilitate comparisons between different groups
and older literature.
The model of \citet{1974SoPh...39...19H}, an updated
version of that presented in \citet{1967ZA.....65..365H},
was constructed to reproduce the centre-to-limb 
variations of continuum light, the wings of strong lines,
and the central intensities of around $900$ weak
atomic and ionic lines.
As noted above, the model is slightly too warm
in the line-forming layers, with potential reasons
for this discussed in \citet{2021A&A...656A.113A}.
In more recent times there have been efforts to construct 
semi-empirical 3D model photospheres
\citep[e.g.][]{2011A&A...529A..37S}
and to subsequently use them for abundance analyses
\citep[e.g.][]{2015A&A...577A..25S,2017A&A...600A..45C,2020A&A...643A.142C}.
This is a promising avenue of research,
but, as discussed in \citet{2021A&A...653A.141A}
these model photospheres have yet to face the same thorough testing
against independent solar benchmarks
that the 3D \stagger{} and \cobold{} models have been through.

\subsection{Spectral line synthesis in 3D LTE and key atomic data
for diagnostic absorption lines}
\label{methodlte}

In the calculation of spectra for comparison with observations, the
radiative transfer equation is solved through the model
photosphere to determine the emergent intensity $I_{\lambda}$:
\begin{equation}
    \frac{\mathrm{d}I_{\lambda}}{\mathrm{d}{\tau_{\lambda}}} = 
    S_{\lambda}-I_{\lambda}\,,
    \label{eq:rt}
\end{equation}
where $S_{\lambda}$ is the source function,
namely the ratio of emissivity $j_{\lambda}$ to extinction
$\alpha_{\lambda}$;
and $\mathrm{d}\tau_{\lambda}$ is the
incremental optical 
path along a given ray at wavelength $\lambda$ that is
related to the geometric path as $\alpha_{\lambda}\mathrm{d}s$.
The equation may be solved on inclined rays in order to study
the centre-to-limb variation, and in order to 
determine the disc-integrated flux.
The equivalent width of the emergent intensity for some viewing angle is:
\begin{equation}
    W_{\mathrm{line}}=\int_{\mathrm{line}}\left(1-I_{\lambda}/I^{\mathrm{c}}_{\lambda}\right)\,\mathrm{d}\lambda\,,
    \label{eq:eqw}
\end{equation}
where $I^{\mathrm{c}}_{\lambda}$ is the continuum intensity, that
can be calculated by solving the radiative transfer
equation without the inclusion of any bound-bound radiative
emissivity or extinction.
In what follows,  we briefly discuss the most relevant ingredients
for LTE abundance analyses, and refer to 
Chapter 2 of \citet{2003rtsa.book.....R}
for further details.

For a bound-bound radiative transition between
a lower level $l$ and an upper level $u$,
the extinction scales with
the number of absorbers $n_{l}$
while the emissivity scales with
the number of emitters $n_{u}$;
the source function thus goes as $n_{u}/n_{l}$.
In LTE, these numbers, or populations,
are almost completely described by the Boltzmann, Saha,
and molecular Saha equations for excitation,
ionisation, and molecular equilibrium respectively;
the source function reduces to the Planck function in this case.
Completing the equations is a normalisation factor,
namely the total elemental abundance;
this is an input into the models, and what is ultimately 
varied so as to fit the measured equivalent width 
or absorption line shape (\sect{methodlines}).

The extinction (and emissivity) are modulated by 
the line profile $\phi$, through which
the various broadening physics are included in the model.
Doppler broadening is characterised by a Gaussian profile,
\begin{equation}
    \phi \sim
    \exp\left(-(\lambda(1-v/c)-\lambda_{0})^2/\Delta\lambda_{\mathrm{D}}^2\right)\,,
\end{equation}
where $\lambda_{0}$ is the wavelength at the centre
of the line in the rest frame,
and the standard deviation is given by the Doppler width
$\Delta\lambda_{\mathrm{D}}$ divided by $\sqrt{2}$:
\begin{equation}
    \Delta\lambda_{\mathrm{D}}
    =\lambda_{0}\sqrt{\frac{2k_{\mathrm{B}}T}{mc^{2}}
    +\frac{\vmic^2}{c^{2}}}\,.
    \label{eq:doppler}
\end{equation}
In 3D model photospheres, the Doppler broadening is fully accounted for 
via the resolved velocity field along the direction of the ray $v$,
and the microturbulence parameter $\vmic$ is set to zero.
Analyses based on 1D and \mtd{} model photospheres, which lack
a description of the resolved velocity field,
instead have non-zero microturbulence parameter $\vmic$.
\citet[][]{1978A&A....70..537H} recommend
the depth-independent value of $1.0\,\kms$ for the disc-centre intensity.
They also recommend to use a larger
value of $1.6\,\kms$ at the limb;
this does not reflect the depth-dependence of microturbulence,
which actually decreases at higher radial layers
\citep[e.g.][]{1974SoPh...39...19H};
but rather reflects stronger velocity fluctuations
as rays emergent from the limb penetrate granules
that are seen edge-on \citep[e.g.][]{2022SoPh..297....4T}.
Similarly, 1D and \mtd{} analyses adopt additional
macroturbulence broadening, for example 
by convolution of the emergent spectrum with a Gaussian
corresponding to a velocity of $1.6\,\kms$ for disc-centre intensity,
or $2.1\,\kms$ at the limb \citep{1978A&A....70..537H}.
In contrast, and as with microturbulence broadening,
the macroturbulence broadening is naturally taken into account in analyses 
based on 3D model photospheres without resorting to 
tunable parameters \citep[e.g.][]{2000A&A...359..729A}.

Natural and pressure broadening are characterised by Lorentzian profiles.
For metal lines in the solar photosphere, pressure broadening due to
hydrogen collisions typically dominates, and these are well modelled by the
theory of Anstee, Barklem, and O’Mara (ABO; 
\citealt{1995MNRAS.276..859A,1997MNRAS.290..102B,1998MNRAS.296.1057B}).
Highly-excited lines may fall beyond the regime of validity of ABO theory,
and then the classical Lindholm--Foley--Uns\"{o}ld (LFU) theory is
sometimes employed; this systematically underestimates the line broadening
and it is usual to see empirical enhancement factors of $1.5$ or $2.0$
(Sect.~4.1.2 of \citealt{2016A&ARv..24....9B}).
In the absence of reliable broadening data, it is advisable to exclude
absorption lines from the abundance analysis if their equivalent widths are
sensitive to the pressure broadening.
The convolution of the Gaussian profile for Doppler broadening
and the Lorentzian profile for natural and pressure broadening
gives rise to a Voigt line profile overall.

The extinction (and emissivity) scale with the dimensionless 
absorption oscillator strength $f_{lu}$, which is closely related
to the Einstein coefficients $B_{lu}$, $B_{ul}$, and $A_{ul}$, 
as well as to the line strength $S$
\citep[e.g.][]{martin1999atomic}.
For example, for the Einstein coefficient of spontaneous emission,
or more commonly the transition rate, $A_{ul}$:
\begin{equation}
    g_{l}f_{lu}=\frac{2\pi e^2}{m_{\mathrm{e}}c\epsilon_{0}\lambda^{2}}
    g_{u}A_{ul}\,,
    \label{eq:gf}
\end{equation}
with statistical weights for the lower and upper levels
$g_{l}$ and $g_{u}$
(hereafter replacing $g_{l}f_{lu}$ with $gf$
for brevity);
the other symbols having their usual meanings.
Formally the oscillator strength is only defined for permitted (E1)
transitions; nevertheless, values 
may still be presented for forbidden transitions
via \eqn{eq:gf}.

\begin{figure}[ht]
\centering
    \includegraphics[width=0.5\textwidth]{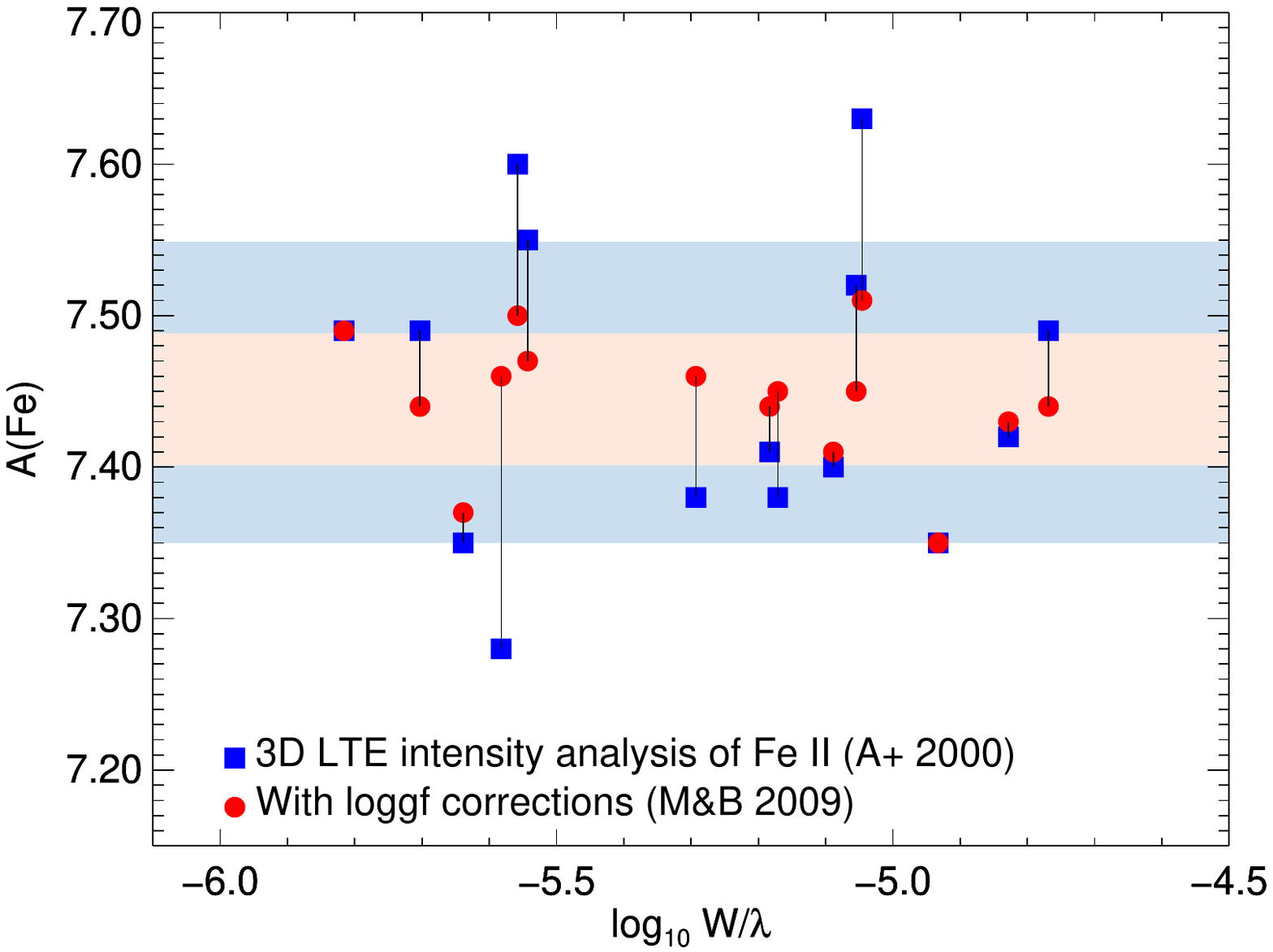}\includegraphics[width=0.5\textwidth]{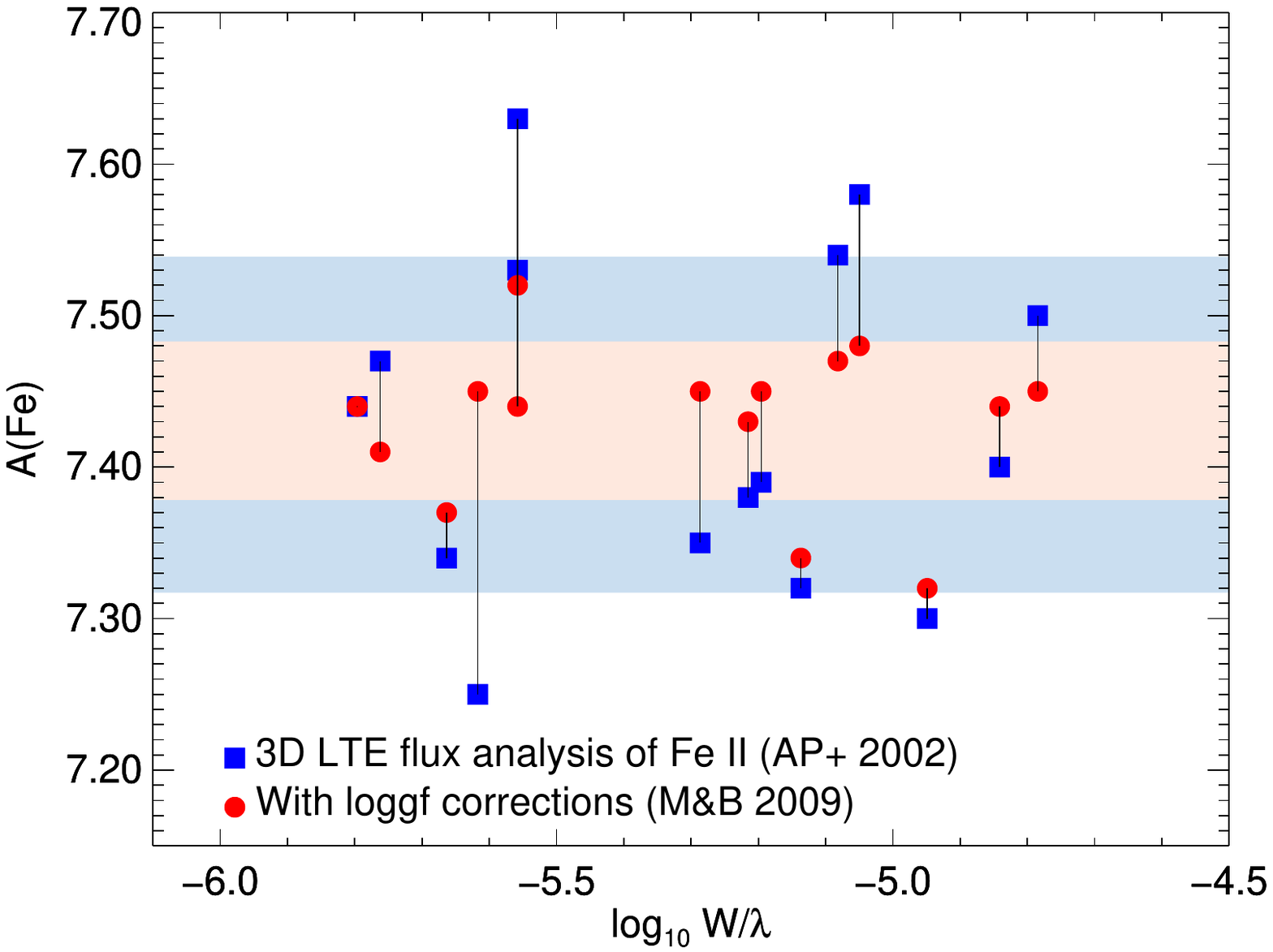}
    \caption{Reduction of line scatter through the use
    of improved oscillator strengths (blue versus red shaded areas).
    \textbf{Left:} Iron abundances as a function of
    logarithmic reduced equivalent width inferred
    by \citet{2000A&A...359..743A}
    from individual \ion{Fe}{II} lines
    in the Hamburg disc-centre intensity atlas.
    \textbf{Right:} Iron abundances inferred 
    by \citet{2002ApJ...567..544A} for the original Kurucz disc-integrated
    flux atlas \citep{1984sfat.book.....K}.
    Both analyses are based on the same 3D model photosphere,
    computed with an early version of the \stagger{} code.
    Red circles show the original results obtained
    using oscillator strengths from laboratory lifetimes 
    and laboratory or astrophysical branching fractions
    \citep{1987A&AS...67..225K,1990A&A...230..244H,
    1990A&A...231..536P,1992A&A...259..301H}.
    Blue squares show the same data, corrected for improved
    oscillator strengths from \citet{2009A&A...497..611M}
    that are based on laboratory lifetimes and
    theoretical branching fractions.
    Shaded blue and red areas show the respective sample standard
    deviations.}
    \label{fig:fe2}
\end{figure}

To first order the logarithmic elemental abundance inferred from an
absorption line
goes as the negative of the
logarithmic transition rate of that line.
It is often advisable to adopt
transition rates that were measured in the laboratory
\citep[e.g.~for \ion{Fe}{I} and 
\ion{Fe}{II} lines:][]{2014MNRAS.441.3127R,
2015A&A...584A..24H,2017ApJ...848..125B,2019ApJS..243...33D}.
Nevertheless, there exist transitions of astrophysical 
interest that are challenging to measure accurately, 
and for which theoretical data are routinely employed
\citep[as is the case for e.g.~some diagnostic
\ion{S}{I}
lines:][]{2006JPhB...39.2861Z,2008ADNDT..94..561D,2024ApJS..274...32C,2026A&A...707A.141L}.
Hybrid approaches based on, for example, measured branching
fractions together with theoretical radiative 
lifetimes, are also a promising way
for reliable oscillator strengths \citep[e.g.][]{2024A&A...682A.184P}.
Advances in both experimental techniques
\citep[e.g.][]{2023EPJD...77..104C} and computational methods 
\citep[e.g.][]{2025CoPhC.31209604S} have improved and continue
to significantly increase the precision of solar abundance analyses.
Line-by-line abundance results can thus be corrected:
\begin{equation}
    \Delta\lgeps{X}_{\mathrm{line}}=-\Delta\lggf_{\mathrm{line}}\,.
\end{equation}
This is demonstrated
in \fig{fig:fe2}, which shows how improved
$\lggf$ data significantly reduce the scatter
in line-by-line iron abundances from the older literature.
Such improvements are especially important
when the elemental abundance can only be inferred from
a small statistical sample of diagnostic absorption lines.

There are a host of other atomic parameters
that may significantly impact 3D LTE solar abundance
analyses, including wavelengths, partition functions,
and hyperfine and isotopic structure.
Curated data may be found within various atomic databases.
First and foremost may be the NIST Atomic Spectra Database
\citep{2020Atoms...8...56R}
for experimental energies and critically-compiled
oscillator strengths.
The VALD database \citep{1995A&AS..112..525P} contains 
a large collection of different atomic data with the flexibility
to choose between different sources, as well 
as key broadening parameters for many lines of astrophysical interests,
and, recently, parameters for hyperfine and isotopic structure
\citep{2019ARep...63.1010P}.
We refer the reader to Section 2 of \citet{2016A&ARv..24....9B}
for further discussion and references.

\subsection{Spectral line synthesis in 3D non-LTE and the need for complete
atomic data sets}
\label{methodnlte}

Relaxing the assumption of LTE
means that the populations of absorbers and emitters
are no longer assumed to be given by the Boltzmann and Saha
equations, and sometimes also the molecular Saha equation
(\sect{methodlte}).
Instead, these populations are calculated consistently with
the radiation field, by solving the equations
of statistical equilibrium at every point in the 3D model photosphere:
\begin{equation}
    n_{i}\sum_{j}\left(R_{ij}+C_{ij}\right)=
    \sum_{j}n_{j}\left(R_{ji}+C_{ji}\right)\,.
    \label{eq:nlte}
\end{equation}
This states that the population of level $i$ is
set by the rate of radiative ($R$) and collisional
($C$) transitions out of that level (left hand side),
balanced by radiative and collision transitions
with that level as the final state (right hand side).
As in LTE (\sect{methodlte}), the equations describing
the populations, \eqn{eq:nlte},
are closed by specifying a total elemental abundance.
The number of ionisation stages and number of levels
to include in the model atom depends
sensitively on which lines are being modelled.
For example, for modelling \ion{Fe}{I} lines
of low or intermediate excitation energy,
it is important to contain
energy levels of the minority neutral
species ($E_{\mathrm{ion}}=7.90\,\eV$) 
up to sufficiently close to the 
first ionisation limit to collisionally-couple
to the lowest levels of the majority singly-ionised state
(within $\sim0.5\,\eV$; \citealt{2011A&A...528A..87M})
that serve as a population reservoir. 
Whereas for modelling the \ion{Be}{II} resonance lines, 
it is necessary to include 
a complete description of not just the
singly-ionised species, but also
of the neutral species 
($E_{\mathrm{ion}}=9.32\,\eV$) owing to their comparable
populations in the solar photosphere,
as well as the lowest levels of the doubly-ionised
species that serve as a population sink
\citep[e.g.][]{2024A&A...690A.128A}.
For non-LTE studies of atomic or ionic lines, it is usual
to subtract the minority molecular contribution
from the total
elemental abundance; this molecular contribution can be estimated in LTE
at each point in the 3D model photosphere. In what follows, we briefly point out
the most important physics that will ultimately help with understanding
the diversity of solar abundance results in 
\sect{results}, and refer to Chapter
3 of \citet{2003rtsa.book.....R} for further details.

As well as on the oscillator strengths
or photoionisation cross-sections, discussed below, 
the radiative rates $R$ depend on the angle-averaged radiation field,
\begin{equation}
    J=\frac{1}{4\pi}\int I\,\mathrm{d}{\Omega}\,.
    \label{eq:angleaverage}
\end{equation}
The radiation field is not known ahead of time; as it depends on
the populations of absorbers and emitters 
(\sect{methodlte}). The radiation field also connects different
parts of the photosphere with each other, within the photon
mean free path (or the effective photon mean free path,
when there is significant scattering).
Thus, by relaxing the assumption of LTE, the problem becomes
more non-linear, and non-local.
The equations of statistical equilibrium,
\eqn{eq:nlte}, have to be solved iteratively together with 
the radiative equation \eqn{eq:rt} in three dimensions.
Codes such as \multitd{} 
\citep{2009ASPC..415...87L,2026MNRAS.546f2085H} 
and its offshoot \balder{} \citep{2018A&A...615A.139A},
as well as \nltetd{} \citep{2015A&A...583A..57S}
employ the MALI algorithm of \citet{1992A&A...262..209R}, with a local
operator as this is simple and efficient to run on 
supercomputers with a distributed memory approach (e.g.\ with the
Message Passing Interface, MPI).
For similar reasons, the angle-averaged radiation field
is calculated on short characteristics; once the populations
have converged, the emergent
spectrum is calculated with a less-diffusive long characteristics
solver (for a discussion of such solvers
see e.g.\ \citealt{2021ApJ...912...63D}).

The radiative rates $R$ appearing in the statistical equilibrium
equations take into account both bound-bound and
bound-free transitions.  In principle, atomic data, in particular the
oscillator strengths for bound-bound transitions
and the photoionisation cross-sections for bound-free transitions,
are needed for all pairs of levels included in the model atom.
This is in stark contrast to the LTE case (\sect{methodlte})
where only the information about the diagnostic absorption
line itself is important.
Atomic databases are invaluable here.
The Kurucz Smithsonian Atomic and Molecular Database
\citep{1995ASPC...78..205K} complements data from NIST 
(described in \sect{methodlte}) with
predicted energy levels and semi-empirical
oscillator strengths, which are important for completeness
\citep[e.g.][]{2017MNRAS.468.4311L}.
TIPTOPbase \citep{2020Atoms...8...30M},
which is
the database of the Opacity Project \citep[e.g.][]{1992RMxAA..23...19S}
and the Iron Project \citep[e.g.][]{1993A&A...279..298H},
and its extension NORAD \citep{2020Atoms...8...68N},
are also useful in this regard, as well as
for providing photoionisation cross-sections
based on R-matrix calculations 
that may be up to several orders of magnitude more reliable than
the usual hydrogenic approximation.
Tailored atomic structure calculations may also be necessary.
For example, in their study of lead,
\citet{2012A&A...540A..98M} calculated energy levels
via relativistic Hartree-Fock (HFR) calculations with
the Cowan code \citep{1981tass.book.....C};
and in their study of silver,
\citet{2026A&A...711A.155C} employed oscillator 
strengths from \citet{2026A&A...709A..31J}
based on solving the 
Multiconfiguration Dirac--Hartree--Fock (MCDHF)
equations.
We refer the reader to Sects.~2.2.1 and 2.2.2 of 
\citet{2024ARA&A..62..475L} for further discussion and references.

The radiative transitions of species other than those in the model
atom also influence the statistical equilibrium solution.
Their contribution to the extinction modifies
the radiative rates via the
angle-averaged radiation field. 
In recent years there have been improvements
to the background contribution to the extinction
so as to better take into account
the blending of bound-bound radiative transitions
in the model atom by atomic, ionic, and molecular lines.
\citep[e.g.][]{2015MNRAS.454L..11A,2016MNRAS.455.3735A}.
For example, taking this contribution into account
led to the recent downwards revision
of the solar lithium abundance (\fig{fig:abundancepattern}):
extra extinction from background lines
reduces the angle-averaged radiation field
in \ion{Li}{I} transitions in the UV,
and this reduces the photon pumping effect
that would otherwise act to deplete the low-lying levels
and weaken the \ion{Li}{I} $670\,\nm$ resonance line
\citep{2021MNRAS.500.2159W}.
The background contribution is typically taken
into account in LTE.
The models of \citet{2020A&A...637A..80O}
suggest that departures from LTE in \ion{Mg}{I} transitions
may impact the statistical equilibrium for calcium;
but this was not found to be significant in independent
calculations presented in \citet{2021A&A...653A.141A},
at least for the \ion{Ca}{I} lines considered in that analysis.

\begin{table}[ht]
    \caption{Inelastic collision reactions
    often included in statistical equilibrium calculations.}\label{tab:collisions}
\begin{tabular*}{\textwidth}{@{}lll@{}}
\toprule
    Perturber & Reactions & Approaches and example references \\
\midrule
    Electrons & 
    $\mathrm{X}_{l}+\mathrm{e^{-}}\leftrightarrow
    \mathrm{X}_{u}+\mathrm{e^{-}}$, & 
    Semi-classical
    {\citep[e.g.][]{1962ApJ...136..906V,1962PPS....79.1105S}},\\
    & 
    $\mathrm{X}_{l}+\mathrm{e^{-}}\leftrightarrow
    \mathrm{X}_{u}^{+}+2\mathrm{e^{-}}$. &
    $R$-matrix 
    {\citep[e.g.][]{2011rmta.book.....B,2020A&A...634A...7M}},\\
    & &
    BSR
    {\citep[e.g.][]{2006CoPhC.174..273Z,
    2022ApJS..259...52T}},\\
    & &
    CCC 
    {\citep[e.g.][]{1996PhRvL..76.2674B,2024ADNDT.15601634D}}.\\
\noalign{\smallskip}
\hline
\noalign{\smallskip}
    Hydrogen &
    $\mathrm{X}_{l}+\mathrm{H}\leftrightarrow \mathrm{X}_{u}^{*}+\mathrm{H}$, & 
    Drawin {\citep{1984A&A...130..319S,1993PhST...47..186L}},\\
    &
    $\mathrm{X}_{l}+\mathrm{H}\leftrightarrow
    \mathrm{X}_{u}^{+}+\mathrm{H}+\mathrm{e^{-}}$, & 
    Free Electron
    \citep{1991JPhB...24L.127K},\\
    &
    $\mathrm{X}_{l}+\mathrm{H}\leftrightarrow
    \mathrm{X}_{u}^{+}+\mathrm{H}^{-}$. & 
    Asymptotic+LZ
    {\citep[e.g.][]{2013PhRvA..88e2704B,2016PhRvA..93d2705B}},\\
    & &
    Full Quantum
    {\citep[e.g.][]{2010CPL...488..145G,2012PhRvA..85c2704B}}.\\
\botrule
\end{tabular*}
\end{table}

Larger inelastic collision rates $C$ 
push the overall statistical equilibrium towards LTE,
because the Maxwellian velocity distributions
of the perturbing species ensures detailed
balance (Chapter 4 of \citealt{2014tsa..book.....H}).
The main perturbers in the solar photosphere are 
the free electrons, by virtue of their speed,
and neutral hydrogen atoms, by virtue of their abundance
\citep{2016A&ARv..24....9B}.
Some of the key reactions commonly included in model atoms
are shown in \tab{tab:collisions}.
For a given process, the
rates of the forwards and reverse
reactions should be calculated internally
in the non-LTE spectrum synthesis code,
such that detailed balance is strictly enforced
because the time scales for the atomic processes are 
much smaller than those of macroscopic changes:
\begin{equation}
    n^{\mathrm{LTE}}_{i}C^{\mathrm{reaction}}_{ij}=
    n^{\mathrm{LTE}}_{j}C^{\mathrm{reaction}}_{ji}\,.
    \label{eq:detailedbalance}
\end{equation}
An infinitely large collisional rate coefficient
then guarantees
relative LTE between levels $i$ and $j$,
where they share the same departure coefficients
$b=n^{\mathrm{non-LTE}}/n^{\mathrm{LTE}}$.
Relative LTE is often enforced in non-LTE modelling for
fine structure levels having small energy separations, as per
the Massey criterion \citep{1949RPPh...12..248M}.
Moreover, \eqn{eq:detailedbalance} 
ensures that the statistical equilibrium
correctly converges on the global LTE solution
if all of the collisional rate coefficients
become infinitely large.
Conversely, not enforcing detailed balance
may lead to spurious non-LTE effects.
\tab{tab:collisions} also lists several common approaches
for the numerical calculation of these transition rates;
we refer the reader to Sect.~2.2.3 
of \citet{2024ARA&A..62..475L} for more discussion and
additional references.
Here, we just stress that it is often the case that 
the inelastic collisions with neutral hydrogen
are the dominant uncertainty in the non-LTE models.
The state-of-the-art has 
improved tremendously in the last fifteen years:
Asymptotic+LZ models, pioneered by 
\citet{2013PhRvA..88e2704B}
and \citet{2016PhRvA..93d2705B},
are now routinely employed.
These use asymptotically-valid descriptions of 
the potential energy surfaces of the 
$\mathrm{X+H}$ quasimolecule, combined with
the semi-classical
Landau--Zener approach for the collision dynamics;
they may be expected to be reliable to within an
order of magnitude for transitions involving
low-lying levels \citep[e.g.][]{2016A&ARv..24....9B},
and the propagated uncertainties in the inferred abundances
are less than $0.01\,\dex$ for the alkali elements lithium and
sodium \citep{2021ApJ...908..245B}.
However, this approach does not describe interactions
on short inter-nuclear distances \citep[e.g.][]{2018A&A...616A..89A}.
The approach of including rate coefficients
calculated with the Free Electron model of \citet{1991JPhB...24L.127K}
has so far led to better consistency with
astrophysical observables
\citep[e.g.][]{2018A&A...616A..89A,2019A&A...624A.111A,
2019A&A...631A..80B}.
This is not completely satisfactory when the Free Electron
model is extrapolated out of the Rydberg regime, and
further work, both theoretical 
\citep[e.g.][]{2026ChPhL..43b0303W}
and experimental 
\citep[e.g.][]{2024PhRvA.109e2820S},
are still needed to better constrain the cross-sections for 
inelastic collisions with neutral hydrogen.

\subsection{The impact of 3D non-LTE modelling on the solar abundances}
\label{methodeffects}

\begin{figure}[ht]
\centering
    \includegraphics[width=1\textwidth]{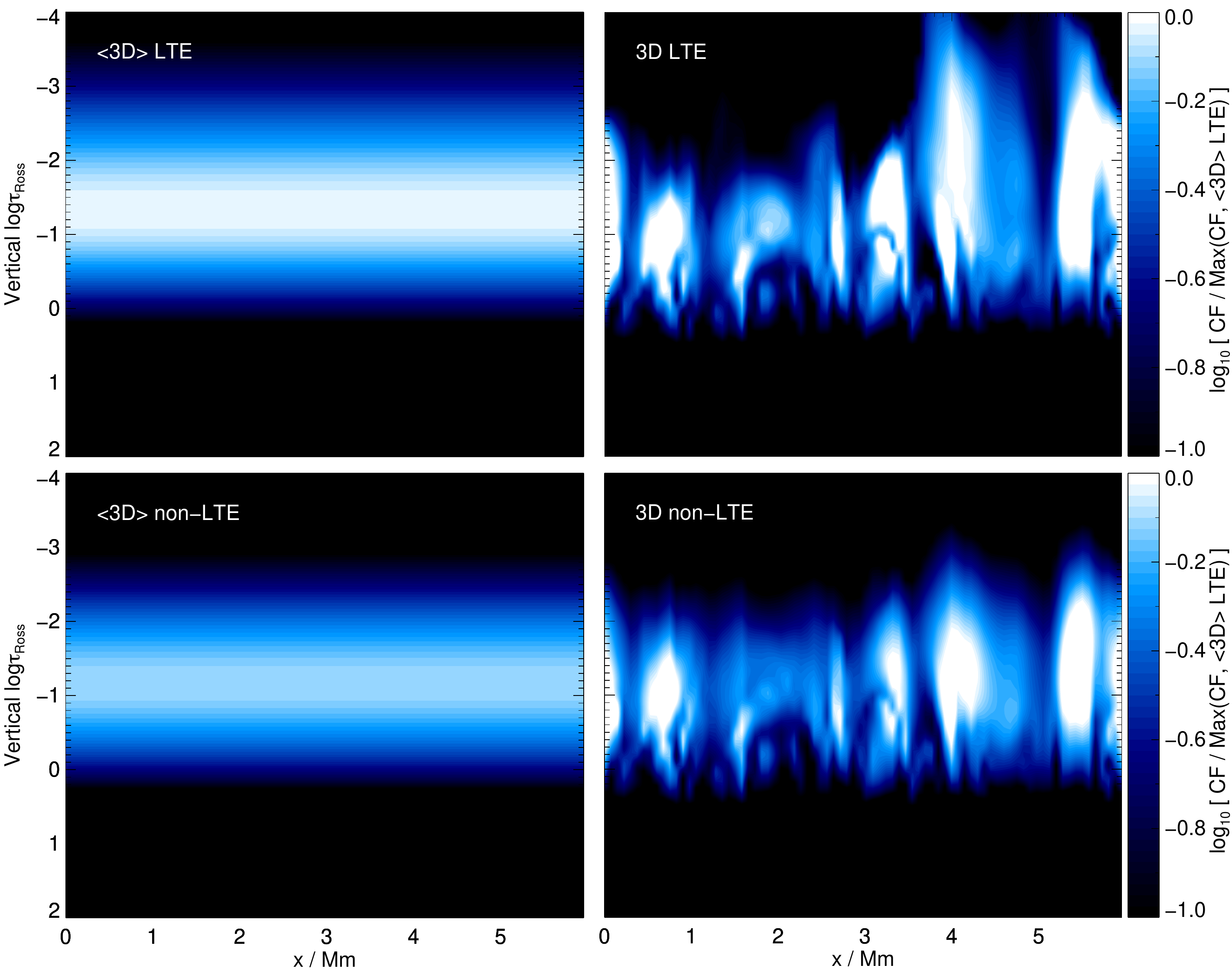}
    \caption{Formation of the 
        \ion{Ag}{I} $328\,\nm$ resonance line
        within \mtd{} and 3D model photospheres.
    \textbf{Top left:} Contribution function to the 
        depression in vertical intensity 
        \citep[e.g.][]{1986A&A...163..135M,
1996MNRAS.278..337A,2015MNRAS.452.1612A}
        in the \mtd{} model photosphere,
        in LTE.
    \textbf{Top right:} Contribution function to the 
        depression in vertical intensity in a vertical 
        slice of a snapshot of a 3D model,
        in LTE.
    \textbf{Bottom left:} Same as top left, but in
        non-LTE.
    \textbf{Bottom right:} Same as top right,
        but in non-LTE.
        Same normalisation applied in all four panels.
        Based on calculations presented in 
        \citet{2026A&A...711A.155C}.}
    \label{fig:cf}
\end{figure}

It is well known that the introduction of 3D and non-LTE modelling
in the early 2000s coincided with 
the lowering of the metal mass fraction as inferred by
spectroscopy of the solar photosphere (\sect{introduction}).
However, as we discuss below, the 3D and non-LTE effects on elemental abundances
are not systematically negative.
Thus the reasons for the lowering of the metal mass fraction
are nuanced, and we attempt to clarify some of these reasons below.

Before doing so, it is helpful to decompose the 3D effects into those due to difference in
the mean stratification (the indirect 3D effect), and those due to 
inhomogeneities (the direct 3D effect). Figure 6 of
\citet{2024ARA&A..62..475L}
illustrates
this for lines of different excitation potential and reduced equivalent width,
for majority and minority atomic and ionic species, based on calculations
on 3D, \mtd{}, and theoretical 1D model photospheres. A key finding is that
the indirect 3D effect is such that the 3D versus 1D abundance corrections
are almost always positive. This is because the mean stratification of 3D
model solar photospheres is shallower (the temperature-depth relation is
less steep) than those of corresponding theoretical 1D model solar 
photospheres
\citep[e.g.][]{2009LRSP....6....2N,
2013A&A...554A.118P}, giving rise to 
shallower absorption lines in the model spectrum (e.g.~Chapter 13 of
\citealt{2022oasp.book.....G}).  The indirect 3D effect becomes more positive
for lines of minority species (that are more temperature sensitive) and for
lines of larger reduced equivalent width (that form in the higher layers of the
photosphere). In contrast, the direct 3D effect may have either sign as
discussed in \citet{2024ARA&A..62..475L}, 
but there is again a tendency for lines of larger
reduced equivalent width to show positive 3D versus 1D abundance 
corrections. Inhomogeneities can have a strong impact on molecular lines
via their influence on the molecular equilibrium. Carbon monoxide (CO)
presents a striking example: its formation is highly temperature sensitive,
and it is able to form copiously in cool pockets of gas in inhomogeneous
3D model photospheres, such that its absorption lines get deeper in the
model spectrum, and carbon
abundances inferred from weak CO lines are around $0.15\,\dex$ lower in 3D
\citep[e.g.][]{2006A&A...456..675S,2021A&A...656A.113A}.

Turning to departures from LTE, the effects
are fundamentally driven by photon pumping and
photon losses in different radiative transitions
(e.g.~Section 2.5.2 of \citealt{2024ARA&A..62..475L}).
For example, neutral atoms with small ionisation energies 
$E_{\mathrm{ion}}\lesssim8.5\,\eV$
are minority species
in the main line-forming regions of the solar photosphere,
per the Saha equation.
For such species, pumping
by hot non-local photons escaping from the deeper layers
tends to cause the absorbing levels
to become less populated
(driving overexcitation or overionisation), relative to LTE
(as is the case for neutral iron;
e.g. \citealt{2001ApJ...550..970S}).
Taking this effect into account leads to 
shallower absorption lines in the model spectrum,
and hence higher inferred abundances.
On the other hand, strong resonance lines,
deep lines in the optical or near infrared,
and subordinate lines from metastable lower levels
such as the \ion{O}{I} $777\,\nm$ triplet
\citep[e.g.][]{1993A&A...275..269K},
often suffer from the non-LTE effect known as
photon losses.
This is where the lack of radiation in the absorption line means
that photoexcitation cannot keep up with spontaneous
emission.
Taking this effect into account leads to
deeper absorption lines in the model spectrum, and hence lower inferred
abundances. In practice, competing non-LTE effects in
multi-level atoms make it difficult to make more general statements about
how departures from LTE may systematically alter elemental
abundance determinations.

Finally, 
seeing that the 3D and non-LTE effects may be both positive and negative, 
we turn back to the question of
why the use of 3D and non-LTE models led to a general lowering of the solar
elemental abundances.
A key reason is that
semi-empirical 1D model photospheres, rather than theoretical 1D model
photospheres, were being used for solar spectroscopy in the 1990s. The commonly
used semi-empirical 1D model photosphere of
\citet{1974SoPh...39...19H}
is too hot in the upper layers, and thus a
temperature stratification that is too shallow, partly as a consequence of
its construction on lower resolution observational data and its assumption
of LTE
\citep{1999A&A...347..348G,2021A&A...656A.113A}.
This means it predicts absorption lines that are too shallow, 
and so it leads to overestimated solar abundances.
A second important reason
for why 3D and non-LTE
modelling led to a general lowering of the solar elemental abundances is
that they improved our understanding of the shapes of absorption lines
in the solar spectrum.
Such models enabled more precise measurements
of absorption lines with a better accounting for blends
(typically reducing the equivalent width values).
The auroral
[\ion{O}{I}] $630\,\nm$ line presents a good example,
for which \citet{2001ApJ...556L..63A} demonstrated
the significance of a \ion{Ni}{I} blend via the use
of a 3D model photosphere.
Their advocated abundance was $0.14\,\dex$
lower than the standard of \citet{1998SSRv...85..161G},
which by itself amounted to a 12.6\% reduction
in the metal mass fraction.

We end this section by discussing the need for consistent 3D non-LTE 
modelling. 
For computational and practical reasons, it is common to
adopt an inconsistent 3D plus non-LTE approach to solar abundance 
analyses, where abundances derived with a 3D LTE analysis are post-corrected
using differential non-LTE and LTE abundance corrections calculated on
1D or \mtd{} model photospheres.
However, treating the 3D effects and the non-LTE effects
separately like this is an approximation
to a holistic line-formation problem.
The inconsistent approach can give reasonable answers when the 
non-LTE effects are less sensitive to the atmospheric structure
(when the 3D and non-LTE coupling is weak), for
example for \ion{N}{I} lines as discussed
in \citet{2020A&A...636A.120A}.
But this approximation breaks down in other cases.
Taking the \ion{Ag}{I} $328\,\nm$ resonance line as an example,
\fig{fig:cf} illustrates how the complexities of 
spectral line formation are naturally washed away in 
1D and \mtd{} model photospheres.
As a consequence, an inconsistent 3D
plus non-LTE approach is unable to take into account the steep 
gradients
associated with the inhomogeneities that are apparent in the figure.
These gradients enhance the UV radiation field, driving photon pumping
which depopulates the ground state of neutral silver and weakens the
\ion{Ag}{I} $328\,\nm$ resonance line.
In this case, inconsistent approaches underestimate
the silver abundance by around $0.06\,\dex$
relative to consistent 3D non-LTE modelling
\citep{2026A&A...711A.155C}.

\section{Recommended abundances}
\label{results}

To focus this discussion we consider three solar abundance compilations
that are commonly used or discussed in recent literature:
the A21 compilation \citep{2021A&A...653A.141A},
the C11 compilation \citep{2011SoPh..268..255C},
and the L25 compilation 
\citep{2025SSRv..221...23L}.
The first of these, the A21 compilation, 
is a significant update of 
\citet{2009ARA&A..47..481A} that remains frequently used today,
which in turn can be viewed as an evolution of the 
Standard Solar Composition
of \citet{1989GeCoA..53..197A} and \citet{1998SSRv...85..161G},
following the development and application of
3D and non-LTE models. 
In A21 there are recommendations for the 
abundances for the 83 long-lived elements;
this number includes the reference element hydrogen
with $\lgeps{H}\equiv12$.
For 58 elements from lithium to thorium, these are based
on spectroscopy of the quiet solar photosphere (excluding sunspot
analyses),
where abundances, or abundance corrections,
were calculated based on 3D \stagger{} model photospheres,
with many of them being presented in
a series of articles in 2015
\citep{2015A&A...573A..27G,2015A&A...573A..25S,2015A&A...573A..26S}.

The C11 compilation
includes spectroscopic abundances for $12$ elements
based on 3D \cobold{} model photospheres
\citep{2007A&A...467L..11C,2007A&A...470..699C,
2007A&A...473L...9C,2008A&A...483..591C,
2008A&A...488.1031C,2009A&A...498..877C,
2010A&A...514A..92C,2008A&A...484..841M}.
In this section 
we also consider the studies of
zirconium \citep{2011AN....332..128C},
oxygen \citep[][]{2015A&A...579A..88C,2015A&A...583A..57S},
and silicon \citep{2022A&A...668A..48D},
that are by the same C11 group and are also
based on 3D \cobold{} model photospheres.

The L25 compilation
includes spectroscopic and meteoritic abundances
for the $83$ long-lived elements,
the latest in a series of compilations
\citep[e.g.][]{2003ApJ...591.1220L,
2009LanB...4B..712L,2021SSRv..217...44L}.
On the spectroscopic side, the abundances in L25 
are more inhomogeneous than in A21 and C11,
with many elemental abundances being based on averages of 
data presented by different groups
employing different model photospheres,
in particular with weight given to the
study of \citet{2022A&A...661A.140M}
who presented abundances for eight elements based on
a \mtd{} model photosphere.
The L25 spectroscopic abundances were adopted
in the recent compilation of \citet{2026enap....2..387B}.

As already mentioned above, out of the $83$ long-lived elements, 
there are around 58 elements from lithium to thorium 
that can be reliably derived from 
spectroscopy of the quiet solar photosphere.
We discuss 34 of these 
in \sect{resultslight} to \sect{resultsheavy}.
The other $58-34=24$ are not considered in C11,
and in L25 their abundances are
drawn directly from A21; thus for brevity we simply
recommend the results from A21.
In \sect{resultsother}
we also briefly discuss recommendations for
the $82-58=24$ elements from helium to uranium 
where abundances relative
to hydrogen must be derived from alternative methods:
namely, helioseismology, sunspots, the upper solar atmosphere,
nuclear physics, or CI chondrites.

The recommended abundances $\lgeps{X}$ 
are gathered in \tab{tab:abundances}.
These quantities have both values and uncertainties.
In the following discussion we mainly focus on discrepancies between
the values advocated by the three compilations A21, C11, and L25.
The estimation of the $1\sigma$ uncertainty is non-trivial
because it is dominated by systematics.
Except where otherwise indicated, we 
adopt the uncertainties given in A21,
which are based on propagating rough estimates of the modelling 
uncertainties onto the inferred abundances
(see the discussion in Section 4.1 of 
\citealt{2015A&A...573A..25S}).

\input{{table/abundances}.tex}

\subsection{Light elements: Lithium, beryllium, and boron}
\label{resultslight}

These three fragile elements of low absolute abundance display
only one useful absorption line each in the solar spectrum.
This makes their inferred abundances susceptible to large changes
as model spectra improve over time.
In the last five years the solar
abundances of both lithium and beryllium were both revised significantly;
the solar abundance of 
boron has unfortunately not been revisited in over twenty years,
and this is
reflected in its large stipulated $1\sigma$ uncertainty.

\subsubsection*{Lithium ($Z=3$).}
The solar abundance
of lithium can be inferred from the 
\ion{Li}{I} $670.8\,\nm$ resonance doublet;
profile-fitting is required so as to take into account 
various blends, particularly those
due to CN and \ion{Fe}{I} lines.
The values in L25 ($1.04\,\dex$) and C11 ($1.03\,\dex$)
are in close agreement with the
value of $1.05\,\dex$ given in
\citet{2009ARA&A..47..481A}.
More recent 3D non-LTE modelling presented in
\citet{2021MNRAS.500.2159W}, and adopted in A21, gave rise to
a significant downwards revision 
to $0.96\,\dex$, with the primary reason being 
an improved treatment of background line opacities 
in their statistical equilibrium calculations
(\sect{methodnlte}).
This analysis was based on the disc-integrated flux,
and a 3D non-LTE analysis of the centre-to-limb variation
could help better constrain the blending
contributions \citep[e.g.][]{2017MmSAI..88...45C}.
For now, our recommended result is that of A21, which is
based on the value from \citet{2021MNRAS.500.2159W} 
with a larger uncertainty: $\lgeps{Li}=0.96\pm0.06$.

\subsubsection*{Beryllium ($Z=4$).}
The most reliable diagnostic for the solar beryllium abundance is
the \ion{Be}{II} $313.1\,\nm$
resonance line, with the neighbouring
\ion{Be}{II} $313.0\,\nm$ resonance line from the 
same doublet being more saturated and more severely blended.
In A21 the 3D LTE value of \citet{2004A&A...417..769A} is adopted
($1.38\,\dex$), although the authors
noted that accounting for blends would probably 
lead to a significant downwards revision.
This was later confirmed in \citet{2024A&A...690A.128A},
who found a value of $1.21\,\dex$,
based on consistent 3D non-LTE profile-fitting,
calibrating the blends on the centre-to-limb variation,
and taking into account the impact of missing UV continuous opacity.
Tests in 1D indicate excellent agreement
between their non-LTE model and that 
presented in \citet{2022A&A...657L..11K},
and the \citet{2024A&A...690A.128A} abundance value is adopted in L25.
We thus recommend the result of
\citet{2024A&A...690A.128A}: $\lgeps{Be}=1.21\pm0.05$.

\subsubsection*{Boron ($Z=5$).}
The value of $2.70\,\dex$ is given in both A21 and L25.
This value originates from the 1D LTE analysis of the 
\ion{B}{I} $249.7\,\nm$ resonance line
by \citet{1999ApJ...512.1006C}, who used
observations at disc-centre and at the limb
to calibrate the UV continuous opacity. 
\citet{2005ASPC..336...25A} reported 3D LTE versus
1D LTE abundance corrections of $+0.04\,\dex$,
while \citet{1996A&A...311..680K} reported
1D non-LTE versus 1D LTE abundance corrections
of $-0.05\,\dex$ albeit for the disc-integrated flux.
It would be worthwhile to carry out a modern
3D non-LTE analysis similar to that 
of beryllium \citep{2024A&A...690A.128A},
taking advantage of recent Asymptotic+LZ model data
for the inelastic collisions with neutral hydrogen
\citep{2023MNRAS.520..107V}.
For now, our recommended result is the same as in A21:
$\lgeps{B}=2.70\pm0.20$.

\subsection{Low mass elements: Carbon, nitrogen, and oxygen}
\label{resultscno}

Although these three highly abundant elements are discussed in 
A21, C11, and L25, it is only in the first of these compilations that
the abundances draw on the full set of 
available diagnostics.
In A21, the abundances are based on
3D non-LTE analyses of a number of atomic lines;
these include highly excited permitted transitions,
as well as low-excitation
forbidden ones of [\ion{C}{I}] and [\ion{O}{I}].
These values are combined with those from molecular
lines, based on simultaneous 3D LTE modelling
of $408$ electronic or rovibrational lines of
C$_{2}$, CH, CN, CO, NH, and OH,
corresponding to $12$ diagnostic groups
\citep{2021A&A...656A.113A}.
These various atomic and molecular diagnostics 
have different formation depths and model sensitivities,
and taking them all into account therefore helps mitigate
systematic modelling errors.
In contrast, the values presented 
in C11 and L25 are based purely on atomic diagnostics.
In the case of C11 this is for reasons of homogeneity:
analyses of the solar carbon, nitrogen, or oxygen
abundances based on molecular lines have not been 
carried out with a 3D \cobold{} model photosphere.
In the case of L25, the authors state that 
molecular analyses are unreliable unless departures from LTE are 
taken into account.
But, this statement seems
too pessimistic: as discussed in Section 2.5.2 of 
\citet{2024ARA&A..62..475L}, departures from LTE 
are predicted to be small, particularly for 
rovibrational transitions within the ground electronic
state.  This was recently corroborated
in the 3D non-LTE modelling of the CH molecule
\citep{2026MNRAS.546f2085H}, where the 3D non-LTE 
versus 3D LTE abundance corrections
were found to be only of the order of $0.01\,\dex$.
Below we focus on a comparison of
the atomic diagnostics,
and ultimately combine them with the values from the
comprehensive molecular analysis of \citet{2021A&A...656A.113A}
to present advocated abundances for
carbon, nitrogen, and oxygen.

\subsubsection*{Carbon ($Z=6$).}
There are numerous \ion{C}{I} lines in the visible and near infrared 
of the solar spectrum. Unfortunately, many of
these lines are blended, especially in the near infrared.
The atomic value presented in
A21 originates from \citet{2019A&A...624A.111A},
who presented consistent 3D non-LTE modelling of equivalent
widths in the disc-centre intensity
for $14$ permitted \ion{C}{I} lines and 
one forbidden [\ion{C}{I}] line.
The value in A21 for the permitted lines 
includes a $+0.03\,\dex$ upwards revision due to
new theoretical oscillator strengths for $10$ of the $14$ lines
as discussed in \citet{2021MNRAS.502.3780L}.
This leads to $8.47\,\dex$ for the \ion{C}{I} lines,
and $8.45\,\dex$ for the [\ion{C}{I}] line.

The carbon abundance in C11 originates from \citet{2010A&A...514A..92C},
who presented a 3D LTE analysis
with \mtd{} non-LTE versus \mtd{} LTE abundance corrections 
of equivalent widths in the disc-centre intensity
and disc-integrated flux 
for $44$ permitted \ion{C}{I} lines ($8.50\,\dex$)
plus the forbidden [\ion{C}{I}] line ($8.41\,\dex$).
\citet{2019A&A...624A.111A} demonstrated that,
after correcting for differences in assumed oscillator strengths,
the differences in their average values
are primarily related to the C11 equivalent widths being
systematically larger.  This likely reflects
the challenge in mitigating the effects of blends
for many of these diagnostic lines,
and this could also help explain
the large range of their abundance values from the permitted lines
that exceeds $0.5\,\dex$.

The carbon abundance in L25 is the mean value of
those in C11, A21, and that of
\citet{2022A&A...661A.140M} who presented
a \mtd{} LTE profile-fitting analysis 
in the disc-integrated flux for 
three permitted \ion{C}{I} lines.
Considering just these three lines, updating to the 
theoretical oscillator strengths of \citet{2021MNRAS.502.3780L},
it turns out that C11 and A21 are in excellent agreement,
with mean values of $8.49\,\dex$ and $8.48\,\dex$ respectively.
In contrast,
the \mtd{} LTE analysis of \citet{2022A&A...661A.140M} gives $8.56\,\dex$.
In fact, the 1D LTE analysis of \citet{2022A&A...661A.140M} gives
a much lower value of $8.48\,\dex$,
in good agreement with C11 and A21
(\citealt{2010A&A...514A..92C} and
\citealt{2019A&A...624A.111A} both agree that there are only slight
3D/non-LTE effects for these lines in the Sun).
Thus, the $0.08\,\dex$ discrepancy of 
\citet{2022A&A...661A.140M} points to issues with their
\mtd{} modelling, which is 
inherently limited as discussed in \sect{methodatmosphere}
and which has not been validated against solar constraints
in the same way that the 3D models have been
\citep[e.g.][]{2013A&A...554A.118P}.

For carbon, we therefore base the value
on the atomic ($8.47\,\dex$, permitted; 
$8.45\,\dex$, forbidden) 
and molecular ($8.47\,\dex$) values of A21.
We briefly note that, for the lines in common,
the values of \citet{2026MNRAS.546f2085H} for CH lines
agree with those given by
\citet{2021A&A...656A.113A} after correcting
for differences in equivalent widths.
In A21 the single forbidden line was given the same
weight as the full set of permitted lines, whereas here
we opt to group them together ($8.47\,\dex$).
Our new recommended result is thereby slightly higher than that
of A21: $\lgeps{C}=8.47\pm0.04$.

\subsubsection*{Nitrogen ($Z=7$).}
There are several 
\ion{N}{I} lines that may serve as abundance diagnostics.
They are all faint, of high excitation potential
($10.33\,\eV$ or higher), and most are significantly blended,
making them challenging from both 
the modelling and analysis perspectives.
The atomic value presented in A21 originates from
\citet{2020A&A...636A.120A}, who presented
consistent 3D non-LTE modelling
of equivalent widths in the disc-centre intensity
for five \ion{N}{I} lines.
Four are blended by CN lines, whose relative contributions 
are as much as $33\%$ at disc-centre, and grow towards the limb.
The blends were accounted for by noting that
the branching fractions in the theoretical CN transition data
are of higher precision than the absolute oscillator 
strengths; and also noting 
that the CN molecule forms in the upper layers of the photosphere
whereas the high-excitation \ion{N}{I} lines form in
the deep layers.
Thus, the pure equivalent widths of the blended \ion{N}{I} lines
were estimated by subtracting the CN contributions
that were estimated from the observed 
equivalent widths of neighbouring CN lines,
scaled by the branching fractions in \citet{2014ApJS..210...23B}.
This analysis led to a value of $7.77\,\dex$.
As for the \ion{C}{I} lines discussed above,
the value from \ion{N}{I} lines 
should now be shifted upwards in light of
improved theoretical oscillator strengths \citep{2023ApJS..265...26L},
in this case by $0.02\,\dex$ to $7.79\,\dex$.

The nitrogen abundance in C11 originates from
\citet{2009A&A...498..877C}, who presented a
3D LTE analysis with \mtd{} non-LTE versus \mtd{} LTE abundance corrections 
of equivalent widths in the disc-centre intensity
for $12$ \ion{N}{I} lines.
Their advocated value ($7.86\,\dex$) is significantly higher than
the atomic value presented in A21.
Similar to the case for \ion{C}{I} lines, 
\citet{2020A&A...636A.120A} demonstrated that,
after correcting for differences in assumed oscillator strengths,
this discrepancy is mainly related to the 
measurement of equivalent widths, and
the large range of $0.4\,\dex$ in their abundance values could reflect the
challenges imposed by blends.

The nitrogen abundance in L25 originates 
from \citet{2022A&A...661A.140M}, who presented
a \mtd{} LTE profile-fitting analysis 
in the disc-integrated flux for 
two \ion{N}{I} lines ($7.98\,\dex$).
In L25 a 3D LTE versus \mtd{} LTE abundance correction
of $-0.04\,\dex$ is applied to the disc-integrated flux, taken from
the disc-centre correction of \citet{2009A&A...498..877C}, 
and leads to their advocated value of 
$7.94\,\dex$.  Similar to carbon, the large discrepancy
with A21 and C11 may again be due to the \mtd{} model,
complicated further by the presence of the CN blends.
With the model presented in \citet{2020A&A...636A.120A},
the de-blended \ion{N}{I} lines in disc-integrated flux
requires a 3D non-LTE versus \mtd{} LTE abundance
correction of $-0.08\,\dex$.
Furthermore, Table~3 of \citet{2021A&A...656A.113A} shows
that, at disc-centre, the \mtd{} model underestimates
the strength of CN lines by around $0.10\,\dex$;
correcting for this lowers their average nitrogen abundance 
by around $0.024\,\dex$, and the correction
would be even more severe in the disc-integrated flux.
Combined, this suggests a value 
that is somewhat lower than
$7.87\,\dex$.  This still remains
around $0.08\,\dex$ higher than the values of A21
and C11 (after updating the equivalent widths
and oscillator strengths in C11),
just as was found for the \ion{C}{I} lines.
This residual difference probably again reflects
problems with the
\mtd{} modelling of \citet{2022A&A...661A.140M}
that may have a strong impact on the temperature-sensitive high-excitation
\ion{C}{I} and \ion{N}{I} lines.  
In fact, the 1D LTE values of 
\citet{2022A&A...661A.140M} agree
with those of A21 and C11 after applying 
the various corrections described above,
which suggests
that the \mtd{} model is the origin for the residual $0.08\,\dex$ 
discrepancy.

We thus argue that the atomic value in A21
is probably the more reliable one, which
is $7.79\,\dex$ after updating the \ion{N}{I} values
with new theoretical oscillator strengths from \citet{2023ApJS..265...26L}.
We combine this with the molecular value of $7.89\,\dex$.
Adoptng the same uncertainty estimate as in A21
leads to our new recommended result:
$\lgeps{N}=7.84\pm0.07$.

\subsubsection*{Oxygen ($Z=8$).}
\label{resultsoxygen}
Oxygen abundance diagnostics from atomic lines include
the high-excitation permitted \ion{O}{I} multiplets
at $616\,\nm$, $777\,\nm$, $845\,\nm$, and $927\,\nm$
(although several components of these various 
multiplets are strongly blended and should be given zero weight);
and the low-excitation forbidden [\ion{O}{I}] lines
at $557.7\,\nm$, $630.0\,\nm$, and $636.4\,\nm$.
The atomic value in A21 originates from
a consideration of all of these features
\citep[][]{2016PhDT.......279A}, but
most weight was given to the cleanest of these features: namely,
to the \ion{O}{I} $777\,\nm$ triplet
based on \citet{2018A&A...616A..89A},
who presented consistent 3D non-LTE modelling 
of the centre-to-limb variation;
and to the [\ion{O}{I}] $630.0\,\nm$ line, that is well-modelled
in 3D LTE \citep{2016MNRAS.455.3735A},
but is severely blended with 
a \ion{Ni}{I} line \citep[][]{2001ApJ...556L..63A,2003ApJ...584L.107J}.
The \ion{Ni}{I} blend can be treated in 3D LTE
if the adopted nickel abundance is derived from a 
comparable 3D LTE analysis of other \ion{Ni}{I} lines
suffering similar non-LTE effects; in any case,
\citet{2022A&A...661A.140M} report non-LTE effects
only of the order of $0.01\,\dex$.
In A21 the permitted \ion{O}{I} lines give 
$8.69\,\dex$, only
slightly lower than the forbidden
[\ion{O}{I}] lines ($8.70\,\dex$).
With the new theoretical oscillator strengths
from \citet{2023A&A...674A..54L},
the weighted 
permitted and forbidden diagnostics as well as the molecular diagnostics
\citep{2021A&A...656A.113A} now point to $8.70\,\dex$.

The C11 group have presented a long series of articles on the solar
oxygen abundance, based on 3D \cobold{} model photospheres. We focus on
the
two most recent articles from that group.
In the first article, \citet{2015A&A...583A..57S} presented
consistent 3D non-LTE modelling of
the centre-to-limb variation of the
\ion{O}{I} $777\,\nm$ triplet, and advocated a value of $8.76\,\dex$. 
The difference with the study
of \citet{2018A&A...616A..89A} that informs the value
presented in A21 is related to the
different treatment of the inelastic hydrogen collisions:
\citet{2015A&A...583A..57S} calibrated the simple Drawin
recipe on the centre-to-limb variation,
whereas \citet{2018A&A...616A..89A}
calibrated a more physically-motivated 
approach based on the Asymptotic+LZ and Free
Electron models (\sect{methodnlte}),
that gives rise to more negative non-LTE effects.
In the second article, \citet{2015A&A...579A..88C}
presented a 3D LTE profile-fitting analysis of 
the centre-to-limb variation of the [\ion{O}{I}] $630.0\,\nm$ line 
to arrive at $8.73\,\dex$,
using several observational data sets including
space-based data from the Solar Optical Telescope
on board the Hinode satellite \citep{2007SoPh..243....3K}.
The difference with A21 can be attributed to their 
small nickel abundance of $6.10\,\dex$, which they calibrated 
in a novel way
via the centre-to-limb variation
of the blended [\ion{O}{I}]+\ion{Ni}{I} feature itself.
This is $0.10\,\dex$ lower
than the 3D LTE value of $6.20\,\dex$ of A21 via \citet{2015A&A...573A..26S},
and may reflect the
limitations of this method to derive the \ion{Ni}{I} blending contribution,
given that the uncertainty on the laboratory oscillator
strength of the blend is just $0.06\,\dex$ \citep{2003ApJ...584L.107J}.
Correcting for the nickel abundance brings the value
presented in \citet{2015A&A...579A..88C} into agreement with that in A21.

The oxygen abundance in L25 is the mean value
of those in seven different studies or compilations: 
the value in A21 ($8.69\,\dex$); three
values from the C11 group (the two aforementioned studies,
giving
$8.76\,\dex$ and $8.73\,\dex$ respectively;
as well as the older study of \citealt{2013A&A...554A.126C}, $8.81\,\dex$);
the value of $8.80\,\dex$ from
\citet{2020A&A...643A.142C} who presented
a 3D LTE profile-fitting analysis of the 
disc-centre intensity for the
[\ion{O}{I}] $630.0\,\nm$ line, based on an empirical
3D model photosphere; 
the value of $8.77\,\dex$ from
\citet{2022A&A...661A.140M} who presented a
\mtd{} non-LTE profile-fitting analysis of 
the disc-integrated flux for
the [\ion{O}{I}] $630.0\,\nm$ line and the
\ion{O}{I} $777\,\nm$ triplet;
and the value of $8.75\,\dex$
from \citet{2021MNRAS.508.2236B} who presented
3D non-LTE modelling of the centre-to-limb variation of
the [\ion{O}{I}] $630.0\,\nm$ line and the
\ion{O}{I} $777\,\nm$ triplet.
The mean value is $8.76\,\dex$.
The most recent articles from the C11 group were discussed above,
and we briefly note that the 
study of \citet{2020A&A...643A.142C} is based on
an unvalidated model photosphere
(\sect{methodatmosphere})
and so should be treated with caution,
as should the study of \citet{2022A&A...661A.140M}
given the discrepancies in their 
\mtd{} modelling discussed above for
\ion{C}{I} and \ion{N}{I}.
Thus, for brevity, we focus on
the investigation of \citet{2021MNRAS.508.2236B},
that is based on a 3D \stagger{} model photosphere
in the column-by-column approach.
For the \ion{O}{I} $777\,\nm$ triplet,
their analysis has been superseded
by that of \citet{2023A&A...672L...6P},
who report $8.73\,\dex$. 
The residual difference
with A21 amounts to just around $0.03\,\dex$,
and we speculate that this could be due to differences in the 
non-LTE model atom and the treatment of background line opacities
(\sect{methodnlte}).
Thus, the [\ion{O}{I}] $630.0\,\nm$ line is what drives the discrepancies.
We argue that the value of $8.77\,\dex$ presented by
\citet{2021MNRAS.508.2236B} for this line should be lowered
by $0.10\,\dex$ for two reasons.  
First, the equivalent width for the 
blended [\ion{O}{I}]+\ion{Ni}{I} feature in
the disc-centre intensity is reported to be
$0.532\,\mathrm{pm}$; this is around $0.06\,\dex$
larger than the mean value
found by A21 and C11 (via \citealt{2015A&A...579A..88C}),
who report
$0.461\,\mathrm{pm}$ and $0.470\,\mathrm{pm}$ respectively.
Secondly, 
as discussed in Section 4.4 of \citet{2021A&A...656A.113A},
the contribution of the \ion{Ni}{I} blend to the feature
is underestimated because of assuming a solar nickel abundance
rather than deriving it in a self-consistent way:
because of this,
\citet{2021MNRAS.508.2236B} find
a contribution of $0.149\,\mathrm{pm}$,
compared to $0.173\,\mathrm{pm}$ in A21
and $0.18\,\mathrm{pm}$ in 
\citet{2008A&A...488.1031C}.
Combined, the $0.10\,\dex$ downwards revision brings
the value of \citet{2021MNRAS.508.2236B} to
$8.67\,\dex$ for the [\ion{O}{I}] $630.0\,\nm$ line,
and the solar oxygen abundance to
$8.70\,\dex$ when also including the \ion{O}{I} $777\,\nm$ triplet.

Overall, we therefore argue to base the solar
oxygen abundance on that presented in A21, after revising
the oscillator strengths of the 
permitted \ion{O}{I} oscillator strengths
based on new theoretical data from
\citet{2023A&A...674A..54L}.
The A21 value has the advantage of being based on
the full set of permitted and forbidden atomic lines
as well as molecular lines, and
(for atomic lines) the slight differences with
other studies, particularly those of C11, are
now largely understood.
The permitted and forbidden atomic diagnostics
as well as the molecular diagnostics 
\citep{2021A&A...656A.113A} point to $8.70\,\dex$.
Adopting the same uncertainty estimate as in A21
leads to our new recommended result: $\lgeps{O}=8.70\pm0.04$.

\subsection{Intermediate-mass elements: Sodium to calcium}
\label{resultsintermediate}

These elements of intermediate absolute abundances
can be studied from several weak absorption lines
of the neutral atomic species, and in some cases also from weak
absorption lines of the
singly-ionised species. 
A21 presented consistent 3D non-LTE modelling of
equivalent widths in the disc-centre intensity for
\ion{Na}{I} lines, \ion{Mg}{I} and \ion{Mg}{II} lines,
\ion{K}{I} lines, and 
\ion{Ca}{I} and \ion{Ca}{II} lines;
while for aluminium \citep{2017A&A...607A..75N},
silicon \citep{2017MNRAS.464..264A}, and
phosphorous and sulphur \citep{2015A&A...573A..25S},
they draw on earlier literature values; all are based on 
3D \stagger{} model photospheres.
Below, these are compared
with the values in the C11 compilation
for phosphorous, sulphur, and potassium,
and with a recent determination from
that group for silicon; all are based on
3D \cobold{} model photospheres;
as well as with the values in L25 that are generally
based on averages of different literature.
The abundances of the halogens fluorine and chlorine,
and the noble gases neon and argon,
are not based on spectroscopy of the quiet photosphere
and are left to \sect{resultsother}.

\subsubsection*{Sodium ($Z=11$).} 
In A21, the value of $6.22\,\dex$
is based on consistent 3D non-LTE modelling
of equivalent widths in the disc-centre intensity
for five \ion{Na}{I} lines.
The model atom is based on that presented in
\citet{2011A&A...528A.103L},
and the Full Quantum treatment of inelastic hydrogen collisions
therein compares well against
experimental data 
\citep{2021ApJ...908..245B}.
The sodium abundance in L25 ($6.29\,\dex$) originates
from \citet{2016ApJ...833..225Z}, who presented a
1D non-LTE profile-fitting analysis
of the disc-integrated flux for six \ion{Na}{I} lines.
As \citet{2016ApJ...833..225Z} had the goal of
deriving stellar abundances down to low metallicities,
four of their six lines are on the saturated part
of the curve of growth for the Sun and thus are less reliable 
solar abundance diagnostics (\sect{methodlines}).
Nevertheless, for the two weak \ion{Na}{I} $615.4\,\nm$ and $616.0\,\nm$
lines, their 1D non-LTE value is
$0.09\,\dex$ higher than those presented in A21.
Unfortunately, sodium is not included in the C11 compilation,
but other literature may be examined instead.
The 1D non-LTE value from \citet{2016ApJ...833..225Z}
for the two weak absorption lines are $0.04\,\dex$ 
higher than those of \citet{2014AstL...40..406A},
and $0.08\,\dex$ higher than those of \citet{2024A&A...683A.242C},
based on profile-fitting of the disc-integrated flux
and disc-integrated intensity, respectively.
In fact, based on a combination of strong and weak absorption lines,
these latter two studies advocated
of $6.20\,\dex$ and $6.23\,\dex$
overall, using 1D non-LTE and 3D non-LTE modelling respectively.
This is in good agreement with the A21 value based on 3D non-LTE
modelling of weak absorption lines.
We therefore recommend the A21 result:
$\lgeps{Na}=6.22\pm0.03$.

\subsubsection*{Magnesium ($Z=12$).}
In A21, the value of $7.55\,\dex$ is based on
a weighted mean from consistent 3D non-LTE modelling
of equivalent widths in the disc-centre intensity
for eight \ion{Mg}{I} lines ($7.56\,\dex$) and six \ion{Mg}{II} lines
($7.52\,\dex$). 
The \ion{Mg}{II} lines
show large departures from LTE that are extremely sensitive
to the inelastic hydrogen collisions, that are poorly constrained
for the ionised system (in contrast to the neutral system
for which Full Quantum model data are available).
The three \ion{Mg}{II} lines that have the least severe
departures from LTE (according to the model of A21)
give an average abundance of $7.56\,\dex$,
in perfect agreement with the \ion{Mg}{I} lines.
The magnesium abundance in
L25 ($7.58\,\dex$) is the mean
of the 3D non-LTE value from A21 ($7.55\,\dex$) as well as the values of
$7.66\,\dex$, $7.56\,\dex$, and $7.55\,\dex$
from
\citet{2015A&A...579A..53O}, \citet{2017ApJ...847...15B},
and \citet{2022A&A...661A.140M}, respectively,
all based on \mtd{} non-LTE modelling.
The first and oldest of these studies,
\citet{2015A&A...579A..53O}, has a value that is
$0.10\,\dex$ higher than the others and drives
the discrepancy between A21 and L25.
The goal of \citet{2015A&A...579A..53O} 
was to study departures from LTE
across stellar parameter space,
and thus naturally they
included absorption lines that 
are strong or blended in the Sun;
there are only two \ion{Mg}{I} lines
and two \ion{Mg}{II} lines in common with A21.
After updating their values (K.~Lind, priv.~comm.) with
more reliable \ion{Mg}{I} oscillator strengths
from \citet{2017A&A...598A.102P}
that are based on a combination of laboratory measurements
and theoretical calculations, their \mtd{} non-LTE value is reduced to
$7.60\,\dex$, but show a difference of $0.24\,\dex$ between
their largest and smallest abundance.
We thus recommend to give less weight to \citet{2015A&A...579A..53O}.
Doing so, the mean value from L25 closely matches that of A21.
Since A21 is the only one based on consistent 3D non-LTE modelling, we advocate
to combine their values for
the \ion{Mg}{I} lines together with their values for the three \ion{Mg}{II}
lines that have the least severe departures from LTE,
and to adopt their uncertainty estimate.
This leads to a new recommended result:
$\lgeps{Mg}=7.56\pm0.03$.

\subsubsection*{Aluminium ($Z=13$).} 
In A21 and L25, the value of $6.43\,\dex$
originates from
\citet{2017A&A...607A..75N}, who
presented consistent 3D non-LTE 
modelling of the disc-centre intensity
for six \ion{Al}{I} lines.
This profile-fitting analysis is based on a 3D \stagger{} model photosphere
(and the same 3D non-LTE code, \balder{}) as used in A21.
We also recommend this result here, taking
the original estimate for the uncertainty presented
in \citet{2017A&A...607A..75N}:
$\lgeps{Al}=6.43\pm0.03$.

\subsubsection*{Silicon ($Z=14$).} 
In A21, the value of $7.51\,\dex$ draws on
\citet{2015A&A...573A..25S} and 
\citet{2017MNRAS.464..264A}.
The first study presented a 3D LTE analysis of 
equivalent widths in the disc-centre
intensity for the nine \ion{Si}{I} lines
as well as the \ion{Si}{II} $637.1\,\nm$ line;
while the second study presented 
3D non-LTE versus 3D LTE abundance
corrections for these \ion{Si}{I} lines.
We here propose two updates to the value in A21.
First, we give zero weight to the \ion{Si}{II} line
on the basis that
the 3D LTE value ranges from $7.47\,\dex$ at disc-centre
to $7.68\,\dex$ in the disc-integrated flux
\citep{2022A&A...668A..48D}; this could indicate
that it suffers from non-LTE effects.
Secondly, we update the oscillator strengths
for seven of the nine \ion{Si}{I} lines 
using new data from \citet{2024A&A...682A.184P}
that are based on a combination of laboratory measurements
and theoretical calculations,
and that significantly reduce the range
in the values from different \ion{Si}{I} lines
from $0.14\,\dex$ down to $0.09\,\dex$.
The mean value from A21 is thus raised to $7.53\,\dex$.

While silicon is not discussed in the original C11
compilation, \citet{2022A&A...668A..48D}
recently presented a study 
based on 3D \cobold{} model photospheres.
They carried out a 3D LTE profile-fitting analysis 
of the disc-centre intensity and disc-integrated flux
for $10$ \ion{Si}{I} lines and the \ion{Si}{II} line,
and added the line-averaged 3D non-LTE versus 3D LTE abundance
correction of $-0.01\,\dex$ from 
\citet{2017MNRAS.464..264A} to arrive at a value of
$7.57\,\dex$.
For the seven \ion{Si}{I} lines in common,
there is a $0.02\,\dex$ of the difference
with the here-updated A21 value due to systematic difference
in the equivalent widths.
The remaining difference is due to the three additional lines
in the infrared that give higher
values (mean of $7.60\,\dex$); we caution that such lines 
were not included in the non-LTE study of 
\citet{2017MNRAS.464..264A}, and may be
susceptible to more severe negative non-LTE abundance corrections
via photon losses.

The silicon abundance in L25 ($7.56\,\dex$)
is the mean of that in A21 ($7.51\,\dex$)
the value of $7.57\,\dex$ in \citet{2022A&A...668A..48D},
and the value of $7.59\,\dex$ in \citet{2022A&A...661A.140M}.
The latter is based on \mtd{} non-LTE 
profile-fitting of the disc-integrated flux
for nine \ion{Si}{I} lines.
The \ion{Si}{I} lines used here are 
of moderate excitation potential ($4.93\,\eV$ or higher)
and somewhat temperature sensitive,
meaning that the errors associated with the \mtd{} model
discussed for the highly excited \ion{C}{I} lines and \ion{N}{I} lines
(\sect{resultscno}) may be contributing
to an upwards bias here as well.

Overall, we advocate to take the mean of 
the here-updated A21 value and the
value of \citet{2022A&A...668A..48D}
because they are based on full 3D modelling.
We fold the slight discrepancy
between them into the uncertainty estimate from A21.
This leads to a new recommended result:
$\lgeps{Si}=7.55\pm0.04$.

\subsubsection*{Phosphorous ($Z=15$).}
In A21 and C11, the values of $5.41\,\dex$ and $5.46\,\dex$
originate from
\citet{2015A&A...573A..25S}
and \citet{2007A&A...473L...9C} respectively,
who performed 3D LTE analyses
of equivalent widths for 
faint \ion{P}{I} lines in the infrared.
The first study considered eight lines in the disc-centre
intensity; while the second considered
only five lines, but also looked at the disc-integrated flux.
The phosphorous abundance in L25 is the mean
of these two values.
Restricting to the five lines in common, and
to the values inferred from the disc-centre intensity,
there is good agreement between A21 and C11:
$5.426\,\dex$ and $5.418\,\dex$, respectively,
with a mean value of $5.422\,\dex$.
We take the mean of these two to be our 3D LTE value.
Recently, \citet{2026A&A...708A.296A} presented
a 1D non-LTE model for \ion{P}{I},  with rather careful consideration
of the input atomic data.  
From their data we estimate a line-averaged 1D non-LTE
versus 1D LTE abundance correction of $-0.07\,\dex$ for
the five \ion{P}{I} lines in the disc-centre intensity.
Adopting the same uncertainty as estimated in
\citet{2026A&A...708A.296A}
leads to our new recommended result:
$\lgeps{P}=5.35\pm0.04$.

\subsubsection*{Sulphur ($Z=16$).}
In A21, the value of $7.12\,\dex$ 
originates from \citet{2015A&A...573A..25S},
who presented a 3D LTE analysis,
together with 1D non-LTE versus 1D LTE abundance corrections
(calculated using the model of \citealt{2005PASJ...57..751T}),
of equivalent widths in the disc-centre intensity
of the \ion{S}{I} $1045\,\nm$ triplet that suffers
from strong non-LTE effects 
\citep{2024ARep...68.1159K}
but for which laboratory
oscillator strengths exist \citep{1997PhyS...56..459Z};
as well as five other \ion{S}{I} lines that are less
sensitive to departures from LTE 
but for which there are only theoretical
oscillator strengths which show
large discrepancies \citep{2026A&A...707A.141L}.
The sulphur abundance in C11 
($7.16\,\dex$) originates from \citet{2007A&A...467L..11C} and 
\citet{2007A&A...470..699C}, in fairly
good agreement with A21; in L25 the value is drawn from C11.
More recently, \citet{2025A&A...703A..35A} 
advocated a lower value ($7.06\,\dex$),
based on consistent 3D non-LTE modelling, using a modern model
atom that was found to agree well 
with that independently constructed by \citet{2024ARep...68.1159K}.
The reason for the downwards revision compared to 
A21 and C11 was discussed in
Section 4.2 of \citet{2025A&A...703A..35A}, and is related to 
improved modelling of 
the \ion{S}{I} $1045\,\nm$ triplet
(consistent 3D non-LTE, drawing on Asymptotic+LZ model data for the inelastic
hydrogen collisions),
together with a better handle on the
theoretical oscillator strengths of the five other 
diagnostic \ion{S}{I} lines.
We thus recommend the \citet{2025A&A...703A..35A}
result here, with the uncertainty being
dominated by systematics: $\lgeps{S}=7.06\pm0.04$.

\subsubsection*{Potassium ($Z=19$).} 
In A21, the value of $5.07\,\dex$ is based on 
consistent 3D non-LTE modelling of equivalent widths in the
disc-centre intensity for 
three weak \ion{K}{I} lines.
The potassium abundance in C11 ($5.11\,\dex$) is based on
a 3D LTE analysis,
adopting 1D non-LTE versus 1D LTE abundance
corrections from \citet{2006A&A...453..723Z},
of equivalent widths in the
disc-integrated flux for
six strong and weak \ion{K}{I} lines.
L25 simply adopt the mean of the values in A21 and C11.
Limiting to the three weak lines, the C11 value
becomes $5.06\,\dex$, in good agreement with A21.
Given the large non-LTE effects 
(C11 reported a line-averaged
1D non-LTE versus 1D LTE abundance
correction of $-0.16\,\dex$), especially in the disc-integrated flux 
(\sect{methodspectrum}), we recommend the result
of A21 because it draws on modern atomic data
\citep{2019A&A...627A.177R}, and because it
employs consistent 3D non-LTE modelling of weak lines in 
the disc-centre intensity: $\lgeps{K}=5.07\pm0.03$.

\subsubsection*{Calcium ($Z=20$).}
In A21, the value of $6.30\,\dex$ is based on
a weighted mean from 
consistent 3D non-LTE modelling of equivalent widths in the
disc-centre intensity for 
nine \ion{Ca}{I} lines ($6.30\,\dex$) and
five \ion{Ca}{II} lines ($6.30\,\dex$).
The calcium abundance in L25 ($6.35\,\dex$) is 
the mean of the 3D non-LTE value from A21 ($6.30\,\dex$),
the value of $6.37\,\dex$ presented in the 
\mtd{} non-LTE study of
\citet{2022A&A...661A.140M},
and the values of $6.33\,\dex$ and $6.40\,\dex$
for \ion{Ca}{I} lines and \ion{Ca}{II} lines
presented in the 1D non-LTE study of
\citet{2017A&A...605A..53M}.
Both \citet{2022A&A...661A.140M}
and \citet{2017A&A...605A..53M} are based on the disc-integrated
flux for which there are significant departures from LTE
($-0.07\,\dex$ and $-0.13\,\dex$ for \ion{Ca}{I} lines
and \ion{Ca}{II} lines, respectively,
according to \citealt{2017A&A...605A..53M}).
In contrast, A21 is based on the disc-centre intensity 
and report 3D non-LTE versus 3D LTE differences
of $+0.01\,\dex$ for \ion{Ca}{I}
and $-0.02\,\dex$ for \ion{Ca}{II} after line-averaging.
\citet{2022A&A...661A.140M} reported a large range in abundance
values of $0.12\,\dex$ (compared to $0.07\,\dex$ in A21),
which may reflect limitations in the non-LTE modelling:
the Grotrian diagram in their 
Figure 3 suggests that only around $50$ levels of \ion{Ca}{I} were
included (compared to $141$ in A21).
\citet{2017A&A...605A..53M} reported a large ionisation imbalance
of $0.07\,\dex$ (no imbalance was found in A21),
and this again could be due to the non-LTE modelling:
they relied on the simple Drawin
recipe for hydrogen collisions with ionised calcium
(in contrast to A21 who adopt Asymptotic+LZ model data from 
\citealt{2018ApJ...867...87B}).
We thus recommend the 
A21 result, that is based on consistent 3D non-LTE 
modelling of the disc-centre intensity with a modern model
atom: $\lgeps{Ca}=6.30\pm0.03$.

\subsection{Iron-group elements: Scandium to nickel}
\label{resultsirongroup}

In A21, the abundances of seven of these eight elements
originate from \citet{2015A&A...573A..26S}, who presented
3D LTE analyses of equivalent widths
at disc-centre intensity; for many species, non-LTE versus LTE 
abundance corrections based on 1D or \mtd{} model photospheres
were applied.
Iron is instead based on consistent 3D non-LTE
modelling presented within A21.
All eight elements are based on 3D \stagger{} model photospheres.
Of these elements, only
iron is included in C11, based on a 3D LTE analysis of 
\ion{Fe}{II} lines using a 3D \cobold{} model photosphere.
In general, L25 present averages from across the literature.
Between A21 and L25, the abundances of the
first three of these elements agree to $0.01\,\dex$
We discuss these elements together,
before inspecting the more discrepant 
elements chromium to nickel more closely,
for which the differences reach up to $0.12\,\dex$.

\subsubsection*{Scandium, titanium, and vanadium ($Z=21$ to $23$).}
In A21, the values of $3.14\,\dex$, $4.97\,\dex$, and $3.90\,\dex$
are updated from \citet{2015A&A...573A..26S},
who presented a 3D LTE analysis of 
equivalent widths in the disc-centre
intensity for a large selection of
lines of the neutral and singly-ionised species
of these elements.
Rather uncertain 1D non-LTE versus 1D LTE abundance corrections
were applied: for scandium these were drawn from 
the 1D non-LTE calculations of \citet{2008A&A...481..489Z},
for titanium these were calculated with the model of
\citet{2011MNRAS.413.2184B} on an \mtd{} model photosphere;
and for \ion{V}{I} lines these were assumed
to be $+0.1\,\dex$ by considering the typical
non-LTE effects predicted for 
\ion{Sc}{I}, \ion{Ti}{I}, and \ion{Cr}{I} lines.
For scandium and vanadium, A21 modified the 
values of \citet{2015A&A...573A..26S} to account for
new laboratory oscillator strengths
\citep{2014ApJS..215...20L,2019ApJS..241...21L},
amounting to corrections of 
$-0.02\,\dex$ and $+0.01\,\dex$, respectively.
For titanium, A21 assigned full weight to the
\ion{Ti}{II} lines which have only a small sensitivity to departures from LTE
\citep{2022A&A...668A.103M,2020AstL...46..120S},
amounting to an upwards shift of $0.07\,\dex$
relative to  \citet{2015A&A...573A..26S}.
The abundances in L25 appear to be based on a similar approach,
differing from those in A21 by at most $0.01\,\dex$.
We follow L25 in assigning zero weight to the \ion{Sc}{I} lines,
owing to their sensitivity to departures from LTE
and to the reliability of the 
\ion{Sc}{II} lines as suggested by their low standard deviation,
and otherwise advocate the A21 values and uncertainty 
estimates.  This leads to our recommended results: $\lgeps{Sc}=3.13\pm0.04$, 
$\lgeps{Ti}=4.97\pm0.05$, and $\lgeps{V}=3.90\pm0.08$.

\subsubsection*{Chromium ($Z=24$).} 
In A21, the value of $5.62\,\dex$ originates from
\citet{2015A&A...573A..26S}, who presented
a 3D LTE analysis,
together with \mtd{} non-LTE versus \mtd{} LTE abundance corrections
of $+0.03\,\dex$ for the neutral species 
(line-averaged; calculated using the model of \citealt{2010A&A...522A...9B}),
of equivalent widths in the disc-centre intensity for
$29$ \ion{Cr}{I} lines ($5.60\,\dex$) and $10$ \ion{Cr}{II} lines
($5.65\,\dex$).
The chromium abundance in L25 ($5.74\,\dex$) originates from
\citet{2010A&A...522A...9B} directly, based on
a 1D non-LTE profile-fitting analysis
of the disc-integrated flux for 
\ion{Cr}{I} lines, including some stronger lines 
that are also visible in metal-poor stars.
Part of the discrepancy is related to the
treatment of inelastic hydrogen collisions:
in \citet{2015A&A...573A..26S}, the simple Drawin
approximation was adopted, whereas the advocated
value of \citet{2010A&A...522A...9B} is based on setting
these collision cross-sections to zero,
which gives more positive 1D non-LTE versus
1D LTE abundance corrections ($+0.08\,\dex$, in the disc-integrated
flux).
The latter model was shown to slightly increase the scatter
in the study of \citet{2025A&A...700A.127C}, which
may be a further reflection of 
the importance of realistic treatments
of inelastic collisions in non-LTE models
(\sect{methodnlte}).
Owing to the lack of a reliable non-LTE model for
\ion{Cr}{I} lines, and the lower sensitivity
to departures from LTE in the \ion{Cr}{II} lines,
we propose to base the chromium abundance
on the 3D LTE value from \ion{Cr}{II} lines
presented by \citet{2015A&A...573A..26S}.
We also propose to remove the saturated \ion{Cr}{II} $458.82\,\nm$ line,
and to correct the abundances for the
improved laboratory oscillator strengths from
\citet{2017ApJS..228...10L} that 
reduces the standard deviation from $0.07\,\dex$ to $0.03\,\dex$.
This leads to our new recommended result:
$\lgeps{Cr}=5.65\pm0.04$.

\subsubsection*{Manganese ($Z=25$).} 
In A21, the value of $5.42\,\dex$ originates from
\citet{2015A&A...573A..26S}, who presented
a 3D LTE analysis,
together with \mtd{} non-LTE versus \mtd{} LTE abundance corrections
of $+0.06\,\dex$ (line-averaged; 
calculated using the model of \citealt{2007A&A...473..291B}),
of equivalent widths in the disc-centre intensity 
for $14$ \ion{Mn}{I} lines.
The manganese abundance in L25 ($5.52\,\dex$) originates
from \citet{2019A&A...631A..80B}, 
who presented a 3D non-LTE profile-fitting analysis of 
the disc-integrated flux for
$13$ \ion{Mn}{I} lines. The line-averaged 1D LTE values
between the two studies are in fair agreement
($5.32\,\dex$ and $5.34\,\dex$, respectively).
Despite significant differences in the model atoms,
the 1D non-LTE versus 1D LTE abundance corrections
are also fairly close
($+0.06\,\dex$ for the disc-centre intensity in
\citealt{2015A&A...573A..26S},
compared to $+0.07\,\dex$ for the disc-integrated flux
in \citealt{2019A&A...631A..80B}).
Thus, the difference of $0.1\,\dex$ may be related
to their non-LTE abundance corrections being strongly enhanced in the 3D model
photosphere, and there are arguments
presented within A21 that this is overestimated.
An independent 3D non-LTE study of the disc-centre intensity
would be welcome.
For now, our recommended result is based on the mean of the two studies,
folding the discrepancy into the uncertainty estimate from 
A21: $\lgeps{Mn}=5.47\pm0.08$.

\subsubsection*{Iron ($Z=26$).} 
In A21, the value of $7.46\,\dex$
is based on a weighted mean from
consistent 3D non-LTE modelling
of equivalent widths in the disc-centre intensity
for $40$ \ion{Fe}{I} lines ($7.46\,\dex$),
and $13$ \ion{Fe}{II} lines ($7.47\,\dex$).
Each species shows small
standard deviations of around $0.03\,\dex$.
The A21 value is in good agreement with the 3D non-LTE 
value of $7.48\,\dex$ reported in \citet{2017MNRAS.468.4311L},
who used the same 3D \stagger{} model photosphere
and radiative transfer code (\balder{}) employed in A21,
but with an older version of the model atom 
as well as a different selection of lines.

The iron abundance in C11 ($7.52\,\dex$) is based on
a 3D LTE analysis of
equivalent widths in the disc-centre intensity 
($7.53\,\dex$)
and disc-integrated flux ($7.52\,\dex$)
for $15$ \ion{Fe}{II} lines.
The discrepancy with A21 originates from the
equivalent widths, that are larger in C11.
In fact, when adopting these same
equivalent widths in the disc-centre intensity for the 
$11$ \ion{Fe}{II} lines in common,
the model of A21 gives a value of $7.51\,\dex$, 
in good agreement with C11.
Adopting the same oscillator strengths 
\citep{2009A&A...497..611M},
the standard deviation in the C11 values is
$0.06\,\dex$, much larger than the $0.03\,\dex$ in the A21 values.
This could indicate that the equivalent width measurements are
less affected by blends in A21.

The iron abundance in L25 ($7.51\,\dex$) is the mean
of the three aforementioned values ($7.46\,\dex$ from A21,
$7.48\,\dex$ from \citealt{2017MNRAS.468.4311L},
and $7.52\,\dex$ from C11),
as well as the values of
$7.56\,\dex$ and $7.50\,\dex$
from \citet{2011A&A...528A..87M}
and \citet{2022A&A...661A.140M}, respectively.
\citet{2011A&A...528A..87M} presented
a 1D non-LTE profile-fitting analysis of the disc-integrated flux,
and the large standard deviation in their \ion{Fe}{I} values
($0.09\,\dex$) 
is likely driven by strong, potentially blended absorption lines 
that were included so as to allow for differential comparisons
with metal-poor stars.
\citet{2022A&A...661A.140M} presented a
\mtd{} non-LTE profile-fitting analysis of 
the disc-integrated flux for
$20$ \ion{Fe}{I} lines and five \ion{Fe}{II} lines.
Their Table A1 (correcting for 
the fact that the $554.3936\,\nm$ line is of \ion{Fe}{I}
rather than \ion{Fe}{II})
indicates $7.53\,\dex$ for the neutral species
and $7.48\,\dex$ for the ionised species.
Their \ion{Fe}{I} values have an extremely large standard
deviation ($0.13\,\dex$)
and the difference between the largest and smallest
values is $0.62\,\dex$, or more than a factor of four.
This could again be related to problems with the \mtd{}
model that are more severe for
temperature-sensitive diagnostics (\sect{resultscno}).
We therefore focus on their values from the ionised species,
which appear to be more robust.
The values from
two of the five \ion{Fe}{II} lines are based on oscillator strengths from
\citet{1998A&A...340..300R}, which 
\citet{1999A&A...347..348G} and
\citet{2009A&A...497..611M} showed to be too small by around $0.11\,\dex$.
Correcting the oscillator strengths of
these two lines, the \ion{Fe}{II} abundance
becomes $7.44\,\dex$, and the standard deviation is reduced
from $0.08\,\dex$ to $0.03\,\dex$.
We ran the model
of A21 for the disc-integrated flux and found the 
line-averaged 3D non-LTE versus \mtd{} non-LTE 
abundance correction 
to be $+0.03\,\dex$;
applying this correction brings the \ion{Fe}{II} value of 
\citet{2022A&A...661A.140M} to $7.47\,\dex$, in perfect agreement with A21.

In summary, the value in A21 appears to be the most reliable, benefiting
from precise
equivalent width measurements and consistent
3D non-LTE modelling.
We therefore recommend their result: $\lgeps{Fe}=7.46\pm0.04$.

\subsubsection*{Cobalt ($Z=27$).} 
In A21, the value of $4.94\,\dex$ originates from
\citet{2015A&A...573A..26S}, who presented
a 3D LTE analysis,
together with \mtd{} non-LTE versus \mtd{} LTE abundance corrections
of $+0.09\,\dex$
(line-averaged; calculated using the model of \citealt{2010MNRAS.401.1334B}),
of equivalent widths in the disc-centre intensity 
for $13$ \ion{Co}{I} lines. 
The analysis of \citet{2015A&A...573A..26S} was updated
in A21 for the laboratory oscillator strengths of 
\citet{2015ApJS..220...13L}, which led to
a small abundance change ($0.01\,\dex$ upwards), 
but a significant reduction in the standard deviation.
The cobalt abundance in L25 ($4.95\,\dex$) appears to 
originate from \citet{2010MNRAS.401.1334B} directly,
who presented a 1D non-LTE profile-fitting analysis
of the disc-integrated flux for 
$11$ \ion{Co}{I} lines
with a line-averaged 1D non-LTE versus 1D LTE 
abundance correction of $0.14\,\dex$.
Here, we advocate the A21 value, but following L25 we
increase the systematic uncertainty due to the sensitivity
to departures from LTE combined with their
rather uncertain modelling (in particular, adopting 
inelastic hydrogen collisions with the simple Drawin recipe).
This leads to our new recommended result:
$\lgeps{Co}=4.94\pm0.07$.

\subsubsection*{Nickel ($Z=28$).} 
In A21, the value of $6.20\,\dex$ originates from
\citet{2015A&A...573A..26S}, who presented
a 3D LTE analysis
of equivalent widths in the disc-centre intensity 
for $17$ \ion{Ni}{I} lines.
The nickel abundance in L25 ($6.24\,\dex$) originates from
\citet{2022A&A...661A.140M}, 
who presented a \mtd{} non-LTE 
profile-fitting analysis
of the disc-integrated flux for $10$ \ion{Ni}{I} lines
and one \ion{Ni}{II} line.
The values are not strongly sensitive
to 3D and non-LTE effects:
\citet{2015A&A...573A..26S} reported a 3D versus \mtd{}
difference of at most $0.03\,\dex$ and a mean difference
close to zero, while 
\citet{2022A&A...661A.140M} reported
non-LTE effects of the order of $0.01\,\dex$ for the disc-integrated flux,  and these are presumably
even less severe for the disc-centre intensity.
The standard deviations in the two studies
for the \ion{Ni}{I} lines is $0.02\,\dex$ in
\citet{2015A&A...573A..26S} and
$0.07\,\dex$ in \citet{2022A&A...661A.140M}.
Assuming that the standard deviation
reflects the accuracy with which the 
solar spectrum was measured (for example as regards
the continuum placement and the
mitigation of perturbing blends), this motivates us to
recommend the result of A21, albeit with a slightly
increased uncertainty to account for these
differences: $\lgeps{Ni}=6.20\pm0.05$.

\subsection{Heavy elements: copper to thorium}
\label{resultsheavy}

In this group of $61$ elements,
around $36$ can be measured reliably in the spectrum
of the quiet photosphere.
Many of the values in A21 originate from
\citet{2015A&A...573A..27G}, who, using a 3D \stagger{} model photosphere,
presented
a 3D LTE analysis of equivalent widths
in the disc-centre intensity, or applied
3D LTE versus 1D LTE abundance
corrections to 1D LTE literature abundances from the Wisconsin
group \citep[e.g.][]{2009ApJS..182...80S};
1D non-LTE abundance corrections were applied for some species
where available.
In A21, 
revisions were presented for some elements
to account for improved equivalent width measurements or new abundance
corrections.
In C11, or subsequent studies from that group
employing 3D \cobold{} model photospheres,
five of these $36$ elements have been considered,
namely zirconium, europium, hafnium, osmium, and thorium.
In L25, the majority of the
abundances in this group are drawn from A21, and we do not explicitly
discuss them for brevity; the recommended results from A21 are included
in \tab{tab:abundances}.
We briefly note here that in L25 the abundances of cadmium and
iridium originate from
\citet{2015A&A...573A..27G}, whereas
no values were presented in A21 after judging
the lines to be too badly affected by blends.
Also, while it is not discussed here,
it is worth noting the study of 
\citet{2020A&A...634A..55G} who presented 
consistent 3D non-LTE modelling for barium, and their
value is adopted in both A21 and L25 as well as in the
present work.
Below, we discuss the $13$ elements 
for which there are differences between A21, C11, and L25,
or significant updates to present.

\subsubsection*{Copper ($Z=29$).}
In A21, the value of $4.18\,\dex$ 
originates from \citet{2015A&A...573A..27G},
who presented a 3D LTE analysis,
together with 1D non-LTE versus 1D LTE abundance corrections 
of $-0.01\,\dex$ (drawn from \citealt{2014ApJ...782...80S}),
of equivalent widths in the disc-centre intensity
for five \ion{Cu}{I} lines.
The copper abundance in L25 ($4.24\,\dex$)
originates from \citet{2014ApJ...782...80S} directly,
who presented a 1D non-LTE profile-fitting 
analysis of the disc-integrated flux
for nine \ion{Cu}{I} lines.  Much of the
discrepancy is related to the oscillator strengths:
\citet{2015A&A...573A..27G} adopted laboratory
values from
\citet{1968ZA.....69..180K} and \citet{1975JQSRT..15..463B},
whereas \citet{2014ApJ...782...80S} presented and
used theoretical oscillator strengths based on R-matrix calculations.
Adopting the theoretical oscillator strengths of  
\citet{2014ApJ...782...80S}, 
the value in \citet{2015A&A...573A..27G} would be raised
to $4.22\,\dex$, but the standard deviation
would be greatly increased from $0.05\,\dex$ to $0.13\,\dex$;
this suggests that the laboratory oscillator strengths are more reliable
overall.
Recent 1D non-LTE modelling from \citet{2025A&A...696A.210C}
with a more realistic treatment of inelastic hydrogen collisions
(using Asymptotic+LZ model data, compared to the simple Drawin recipe
in \citealt{2014ApJ...782...80S})
confirms that the abundance corrections have different
sign for the two near-infrared lines in \citet{2015A&A...573A..27G}.
This helps to mitigate the non-LTE effects on the line-averaged value.
We thus recommend the A21 result: $\lgeps{Cu}=4.18\pm0.05$.

\subsubsection*{Zinc ($Z=30$).}
In A21, the value of $4.56\,\dex$ originates from \citet{2015A&A...573A..27G},
who presented a 3D LTE analysis,
together with 1D non-LTE versus 1D LTE abundance corrections 
of $-0.02\,\dex$
(line-averaged; calculated using the model of \citealt{2005PASJ...57..751T}),
of equivalent widths in the disc-centre intensity
for five \ion{Zn}{I} lines.
The zinc abundance in L25 ($4.55\,\dex$) originates
from \citet{2022MNRAS.515.1510S}, who presented a
1D non-LTE profile-fitting analysis of three 
\ion{Zn}{I} lines in the disc-integrated flux.
The input atomic data in the non-LTE modelling is far improved in 
\citet{2022MNRAS.515.1510S}, who
find a mean non-LTE abundance correction
of $-0.05\,\dex$ for the disc-integrated flux;
for the same three lines,
\citet{2015A&A...573A..27G} report $-0.03\,\dex$;
this is nominally consistent with the expectation of
less severe corrections for the disc-centre intensity.
Consequently, we recommend the A21 result without updating the 1D
non-LTE versus 1D LTE abundance correction:
$\lgeps{Zn}=4.56\pm0.05$.

\subsubsection*{Rubidium ($Z=37$).}
In A21, the value of $2.32\,\dex$ is based on
the two infrared \ion{Rb}{I} resonance lines:
\citet{2015A&A...573A..27G} derived an abundance
of $2.47\,\dex$ via a 3D LTE analysis
of equivalent widths in the disc-centre intensity, which 
was revised downwards by $0.15\,\dex$ by A21 on
reinspecting the equivalent widths
($-0.03\,\dex$) and on applying 
a 1D non-LTE versus 1D LTE abundance correction
($-0.12\,\dex$) that \citet{2020AstL...46..541K}
reported for the disc-integrated flux.
One potential issue is that 
the abundance correction for the disc-centre intensity
may be somewhat less severe than that
for the disc-integrated flux.
The rubidium abundance in L25 ($2.35\,\dex$)
originates from \citet{2020AstL...46..541K} directly,
on their 1D non-LTE profile-fitting analysis of the 
disc-integrated flux.
Taking the mean of these two values and the uncertainty
estimate from A21 leads to our 
new recommended result:
$\lgeps{Rb}=2.34\pm0.08$.

\subsubsection*{Strontium ($Z=38$).}
In A21, the value of $2.83\,\dex$ originates from
\citet{2015A&A...573A..27G}, 
who presented a 3D LTE analysis
of equivalent widths in the disc-centre intensity
for two \ion{Sr}{I} lines ($2.80\,\dex$)
and three \ion{Sr}{II} lines
($2.85\,\dex$). These values include
\mtd{} non-LTE versus \mtd{} LTE abundance corrections 
calculated using the model of \citet{2012A&A...546A..90B}
amounting to $+0.11\,\dex$ and
$-0.17\,\dex$ for the two species, respectively.
These non-LTE abundance corrections are severe, and the model
draws on the simple Drawin recipe for the inelastic 
hydrogen collisions; nevertheless, this uncertainty
is to some extent mitigated by the two ionisation stages 
suffering non-LTE effects that go
in opposite directions.
The new laboratory oscillator strength
for the \ion{Sr}{II} $1091.5\,\nm$ line
from  \citet{2016PhRvA..93e2507L} amounts to a 
$+0.096\,\dex$ correction, and brings that line
to $2.84\,\dex$; the weighted mean
for the \ion{Sr}{II} lines becomes $2.87\,\dex$,
and the A21 abundance value is updated to $2.84\,\dex$.
The strontium abundance in L25 ($2.93\,\dex$) originates
from \citet{2012A&A...546A..90B} directly, who presented
a 1D non-LTE profile fitting analysis
of the disc-integrated flux.
Their Table 1 and Figure 6 suggests that they used
the same five lines as \citet{2015A&A...573A..27G}
as well as one additional line,
the \ion{Sr}{II} $416.18\,\nm$ (given
as $416.78\,\nm$ in their Table 1).
The 1D LTE values for the \ion{Sr}{I} lines 
in Figure 6 of \citet{2012A&A...546A..90B} 
appear to be offset from the 1D LTE values listed in 
\citet{2015A&A...573A..27G} by around $0.15\,\dex$.
This may be related to
the measurement of the spectrum, and we note that the equivalent widths
reported by
\citet{2015A&A...573A..27G} agree with
those measured by \citet{2000MNRAS.311..535B} to $0.03\,\dex$.
This agreement, together with
the principle that values inferred from the disc-centre
intensity tend to be more precise and accurate 
than those inferred from the disc-integrated flux (\fig{fig:fe2}
and \sect{methodspectrum}), leads us to advocate the 
here-updated A21 value discussed above; we fold
the discrepancy with L25 into the A21 uncertainty estimate.
This leads to our new recommended result:
$\lgeps{Sr}=2.84\pm0.07$.

\subsubsection*{Yttrium ($Z=39$).}
In A21, the value of $2.21\,\dex$ 
originates from \citet{2015A&A...573A..27G},
who presented a 3D LTE analysis
of equivalent widths in the disc-centre intensity
for nine \ion{Y}{II} lines.
The yttrium abundance in L25 ($2.30\,\dex$) originates
from \citet{2024A&A...683A.200S},
who presented consistent 3D non-LTE modelling of
the centre-to-limb variation of 
three \ion{Y}{II} lines (after giving zero weight to their
fourth line, the \ion{Y}{II}
$520.0\,\nm$ line, on account of its poor fit).
We propose to use the values presented by
\citet{2024A&A...683A.200S} at disc-centre:
this is primarily to mitigate the impact of non-LTE
effects (\sect{methodspectrum}), but also because
their Figure 3 displays a jump in inferred abundance
away from the limb that reflects interpolation problems
in their spectrum synthesis code
for inclined rays in the 3D geometry
(Hoppe et al., submitted).
In that case, the standard deviation is small ($0.02\,\dex$)
and their 3D LTE value ($2.19\,\dex$)
is in fair agreement with that of \citet{2015A&A...573A..27G}.
We thus propose to take the 3D non-LTE disc-centre value from 
\citet{2024A&A...683A.200S}, together with their estimated
uncertainty.  This leads to our new recommended result:
$\lgeps{Y}=2.29\pm0.07$.

\subsubsection*{Zirconium ($Z=40$).}
In A21, the value of $2.59\,\dex$ originates from
\citet{2015A&A...573A..27G}, 
who presented a 3D LTE analysis
of equivalent widths in the disc-centre intensity
for ten \ion{Zr}{II} lines.
Therein are also abundances derived
from up to eight \ion{Zr}{I} lines, although these
are given zero weight;
the four weakest and possibly least-blended of these
\ion{Zr}{I} lines lead to an abundance of $2.32\,\dex$,
that is consistent with the expectation of 
large non-LTE effects for the neutral species
\citep{2010AstL...36..664V,2011AstL...37..440V}.

While zirconium is not discussed in the original C11
compilation, \citet{2011AN....332..128C}
later presented a study based on a 3D \cobold{} model 
photosphere.
They advocated a value of $2.62\,\dex$,
based on a 3D LTE profile-fitting
analysis of the disc-centre intensity
for $15$ \ion{Zr}{II} lines.
They adopted laboratory oscillator strengths from
\citet{2006A&A...456.1181L},
whereas \citet{2015A&A...573A..27G} took the mean
of those data with the laboratory oscillator strengths from
\citet{1981ApJ...248..867B}.
Focusing on the nine lines in common,
adopting the equivalent widths derived
via the fits of \citet{2011AN....332..128C}
slightly reduces the standard deviation
in the \citet{2015A&A...573A..27G} abundances.
On the other hand, adopting 
the oscillator strengths used by 
\citet{2015A&A...573A..27G}
slightly reduces the standard deviation
in the \citet{2011AN....332..128C} abundances.
With these two changes, the two studies
are in good agreement:
$2.59\,\dex$ and $2.60\,\dex$, respectively.

The zirconium abundance in L25 ($2.68\,\dex$) is the mean of 
the two aforementioned values without modification
($2.59\,\dex$ and $2.62\,\dex$), together with a
1D non-LTE versus 1D LTE abundance correction of $+0.08\,\dex$
for \ion{Zr}{II} lines 
from \citet{2010AstL...36..664V,2011AstL...37..440V}.
This correction is uncertain, and may well be overestimated.
This is because \citet{2010AstL...36..664V} found that 
the \ion{Zr}{II} lines are 
strongly sensitive to the inelastic hydrogen
collisions. The authors relied on the Drawin recipe downscaled by a 
factor of ten
based on arguments that no longer hold following updates to the
model given in \citet{2011AstL...37..440V}.
Without any scaling  of the inelastic
hydrogen collisions, the corrections are just of the order
of $0.01\,\dex$.
In fact, one might expect ionised zirconium
(ground state configuration $\mathrm{4d^{2}\,5s}$)
to behave somewhat similarly as ionised titanium
(ground state configuration $\mathrm{3d^{2}\,4s}$),
and recent studies with modern atomic data indicate departures from
LTE that are only of the order of $\pm0.01\,\dex$
for ionised titanium \citep{2020AstL...46..120S,2022A&A...668A.103M}.

Thus, we propose to take the mean of the 3D LTE 
value for \ion{Zr}{II} lines from
\citet{2011AN....332..128C} and 
\citet{2015A&A...573A..27G},
with equivalent width and oscillator strength
corrections as described above.
We fold an additional $0.03\,\dex$ into the
$0.04\,\dex$ uncertainty stated
in A21 to account for the possible influence of neglected non-LTE effects.
This leads to our new recommended result:
$\lgeps{Zr}=2.59\pm0.05$.

\subsubsection*{Rhodium and silver ($Z=45,47$).}
\label{resultsrhodiumsilver}
These two elements were flagged in A21 as likely
suffering from large systematic errors (see their Figure 5).
The values in A21 ($0.78\,\dex$ and $0.96\,\dex$, respectively)
both originate from \citet{2015A&A...573A..27G},
who presented a 3D LTE analysis of equivalent widths
in the disc-centre intensity 
for the $5s\rightarrow 5p$ resonance lines of the neutral species
in the UV:
the \ion{Rh}{I} $343.5\,\nm$ and $369.2\,\nm$ lines
and
the \ion{Ag}{I} $328.1\,\nm$ and $338.3\,\nm$ lines.
In A21, the rhodium abundance was revised downwards
by $0.11\,\dex$ relative to \citet{2015A&A...573A..27G}
after adopting new laboratory oscillator
strengths from \citet{2015MNRAS.450..223M},
and giving full weight to the less-blended \ion{Rh}{I} $369.2\,\nm$ line.
The rhodium and
silver abundances in L25 are taken from A21.
Recently, 
\citet{2026A&A...711A.155C} presented
consistent 3D non-LTE modelling for silver and found
that the 3D non-LTE versus 3D LTE abundance correction
is $0.28\,\dex$. 
After also reconsidering the equivalent widths,
\citet{2026A&A...711A.155C} proposed a silver
abundance of $1.15\,\dex$, and an uncertainty
of $0.08\,\dex$ that takes into account both the measurements
and the models.
Owing to the similarity of their diagnostic lines,
we propose to apply the 3D non-LTE
versus 3D LTE abundance correction of $0.28\,\dex$ from silver 
also to rhodium. We make a coarse estimate that 
applying this 
correction to rhodium might either be underestimated or overestimated
by a factor of two and thereby 
fold $0.14\,\dex$ 
in quadrature with the uncertainty of $0.11\,\dex$ from A21.
This leads to our new recommended
results: 
$\lgeps{Rh}=1.06\pm0.18$ and
$\lgeps{Ag}=1.15\pm0.08$.

\subsubsection*{Lanthanum ($Z=57$).}
In A21, the value of $1.11\,\dex$ originates 
from \citet{2001ApJ...556..452L} and
\citet{2015A&A...573A..27G}.
The first study presented a value of $1.14\,\dex$ based on
a 1D LTE analysis of equivalent widths in the 
disc-integrated flux
for $14$ \ion{La}{II} lines, using 
the semi-empirical 1D model photosphere of \citet{1974SoPh...39...19H}.
\citet{2015A&A...573A..27G} presented 3D LTE versus 
1D LTE abundance corrections, specific to that semi-empirical
1D model photosphere, that amount to $-0.03\,\dex$.
The lanthanum abundance in L25 ($1.10\,\dex$)
is based on a similar approach, but
starting with $1.13\,\dex$ instead of $1.14\,\dex$
from \citet{2001ApJ...556..452L},
which is the mean 
value from different semi-empirical and theoretical 1D model photospheres.
The value in A21 is thus more differential.
Our new recommended result is based 
on the same approach, but starting with the value inferred from the
disc-centre intensity value by \citet{2001ApJ...556..452L} 
which is $0.01\,\dex$ higher than that inferred from the disc-integrated
flux: $\lgeps{La}=1.12\pm0.04$.

\subsubsection*{Europium ($Z=63$).}
In A21, the value of $0.52\,\dex$ originates from
\citet{2001ApJ...563.1075L} and \citet{2015A&A...573A..27G}.
The first study presented a value of $0.52\,\dex$ based on
a 1D LTE profile-fitting analysis of the 
disc-centre intensity
for $14$ \ion{Eu}{II} lines,
using the semi-empirical 1D model photosphere of \citet{1974SoPh...39...19H}.
\citet{2015A&A...573A..27G} presented 3D LTE versus 
1D LTE abundance corrections, specific to that semi-empirical
1D model photosphere, that amount to $-0.03\,\dex$;
this cancels with a 1D non-LTE versus 1D LTE abundance correction
of $+0.03\,\dex$ from \citep{2000A&A...364..249M}.

The europium abundance in C11 ($0.52\,\dex$)
originates from \citet{2008A&A...484..841M}, who presented
a 3D LTE analysis of 
equivalent widths in the disc-centre intensity and disc-integrated
flux for five \ion{Eu}{II} lines.
Their value on the semi-empirical
1D model photosphere 
is $0.51\,\dex$, in good agreement with the value of
$0.52\,\dex$ from \citet{2001ApJ...563.1075L} for these five lines.
This lends confidence that the measurements are precise in both studies.

The europium abundance in L25 ($0.57\,\dex$) originates from
\citet{2024A&A...683A.200S}.
This is based on consistent 3D non-LTE modelling of a single
line, the \ion{Eu}{II} $664.5\,\nm$ line, in
disc-centre intensity and disc-integrated flux.
They report a negative 3D non-LTE versus 3D LTE 
abundance correction for this line,
which suggests that the 
value in A21 is corrected for non-LTE effects
in the wrong direction.
One puzzling observation is that
the 3D LTE disc-centre value for the \ion{Eu}{II} $664.5\,\nm$ line
in \citet{2024A&A...683A.200S}, $0.59\,\dex$,
is much higher than in the compilations
of A21 ($0.51\,\dex$) and C11 ($0.52\,\dex$).
A 3D non-LTE analysis based on more lines
would be worthwhile.

For now, we propose to apply the 3D non-LTE versus 3D LTE abundance
correction from \citet{2024A&A...683A.200S}, $-0.02\,\dex$,
to the 3D LTE abundances of A21 and C11 based at disc-centre.
We fold the discrepancy in the 3D LTE abundances 
of A21/C11 versus \citet{2024A&A...683A.200S} into
the uncertainty estimate in A21.
This leads to our new recommended result:
$\lgeps{Eu}=0.49\pm0.05$.

\subsubsection*{Hafnium ($Z=72$).}
In A21, the value of $0.85\,\dex$ originates from 
\citet{2007ApJS..169..120L} and
\citet{2015A&A...573A..27G}.
The first study presented a value of $0.88\,\dex$ based on
a 1D LTE profile-fitting analysis of the 
disc-centre intensity
for four \ion{Hf}{II} lines,
using the semi-empirical 1D model photosphere of \citet{1974SoPh...39...19H}.
\citet{2015A&A...573A..27G} presented 3D LTE versus 
1D LTE abundance corrections, specific to that semi-empirical
1D model photosphere, that amount to $-0.03\,\dex$.
The hafnium abundance in C11 ($0.87\,\dex$)
originates from \citet{2008A&A...483..591C},
who presented a 3D LTE analysis of 
equivalent widths in the disc-centre intensity for the
same four \ion{Hf}{II} lines;
As with osmium below, L25 appears to take the mean of A21 and C11.
C11 also presents a value from the 
semi-empirical 1D model photosphere of \citet{1974SoPh...39...19H},
$0.88\,\dex$, which 
agrees exactly with that of \citet{2007ApJS..169..120L}.
Thus, the slight difference between the 3D values of A21 and C11
reflects a difference
in the predicted 3D effects: $-0.03\,\dex$ versus $-0.01\,\dex$.
Our new recommended result is based on the mean of A21 and C11:
$\lgeps{Hf}=0.86\pm0.05$.

\subsubsection*{Osmium ($Z=76$).}
In A21, the value of $1.35\,\dex$ originates from 
\citet{2015A&A...573A..27G}, who presented
a 3D LTE analysis of the equivalent width in the disc-centre intensity
for the \ion{Os}{I} $330.16\,\nm$ line;
the equivalent width was revised in A21,
corresponding to a $0.05\,\dex$ reduction in the abundance.
The osmium abundance in C11 ($1.36\,\dex$)
is based on a 3D LTE analysis of 
equivalent widths in the  disc-centre intensity from 
\citet{1984A&A...135...59K}
for nine \ion{Os}{I} lines.
As with hafnium above, L25 appears to take the mean of A21 and C11.
The large standard deviation of
$0.22\,\dex$ reflects the challenges imposed by
blends \citep[][]{2006A&A...448.1207Q}.
Just considering the \ion{Os}{I} $330.16\,\nm$ line,
and with the same equivalent width as
in A21, the value in C11 becomes $1.23\,\dex$.
This value is 
$0.12\,\dex$ lower than that of A21;
but, as discussed in \citet{2015A&A...573A..27G},
lowering the ionisation potential used in C11 from $8.7\,\eV$ to 
the value of
$8.4382\,\eV$ measured by \citet{1997PhRvA..55.1526C}
would raise their inferred abundance by around $0.15\,\dex$ and thus
bring the two studies into excellent agreement.
From the same line, \citet[][]{2006A&A...448.1207Q}
arrive at $1.25\,\dex$ from the 1D \marcs{} model photosphere,
which should be raised to around $1.27\,\dex$
according to the 3D versus 1D corrections 
in \citet{2015A&A...573A..27G} and C11.
Our new recommended result is based on the mean
of the values from A21 and 
\citet[][]{2006A&A...448.1207Q}: $\lgeps{Os}=1.31\pm0.12$.

\subsubsection*{Thorium ($Z=90$).}
In A21, the value of $0.03\,\dex$ originates from
\citet{2015A&A...573A..27G}, who presented a 3D LTE analysis,
together with a 1D non-LTE versus 1D LTE abundance correction 
of $+0.01\,\dex$ (drawn
from \citealt{2012A&A...540A..98M}),
of the equivalent width in the disc-centre
intensity for the \ion{Th}{II} $401.9\,\nm$ line.
The thorium abundance in C11 ($0.08\,\dex$) originates
from \citet{2008A&A...483..591C}, 
who presented a 3D LTE profile-fitting analysis of 
the disc-centre intensity for the
same \ion{Th}{II} line.
L25 does not provide a spectroscopic value for thorium.
In the disc-centre intensity, the de-blended equivalent width
of \citet{2008A&A...483..591C} differs by $0.05\,\dex$ with those of
\citet{2015A&A...573A..27G}, which explains most of the difference
between C11 and A21.  \citet{2015A&A...573A..27G} attribute
this difference to an improved estimation 
of the contribution of \ion{Co}{I} and \ion{V}{I} blends.
We thus recommend the result of A21: $\lgeps{Th}=0.03\pm0.10$.

\subsection{Elements not measured in the quiet photosphere spectrum}
\label{resultsother}

We have discussed the abundances of 
the $58$ elements from lithium to thorium
that can be inferred from spectroscopy of the quiet photosphere.
For completeness, in this section we briefly
discuss our recommendations for
$24$ other long-lived elements from helium to uranium,
for which alternative methods
for abundance determinations
must be used.

\subsubsection*{Helium ($Z=2$); helioseismology.}
\label{resultshelium}

Helium transitions
between singly- and doubly-ionised form
in the second helium ionisation zone:
\begin{equation}
    \mathrm{He^{+}\leftrightarrow He^{2+}+e^{-}}\,.
\end{equation}
This happens
at around $0.98$ solar radii where it causes 
the first adiabatic exponent $\Gamma_{1}$ to dip away from the
value of $5/3$ expected for a monatomic ideal gas
(right panel of Figure~3 of \citealt{2024A&A...681A..57B});
this in turn imparts a
signature in the $p$-mode oscillation frequencies
\citep{1991Natur.349...49V}.
Assuming that the equation of state is well-constrained
\citep[e.g.][]{1994ApJ...426..801A}, helioseismology
can be used to decode this signature so as to 
infer the present-day helium mass fraction in the solar convective envelope
which in turn gives the most precise estimate of
helium in the solar photosphere.
\citet{2026A&A...708A.147B}
derived a present-day helium mass fraction
in the solar convective envelope as $\mathcal{Y}_{\mathrm{present}}=0.2575\pm0.0025$,
based on the metal-to-hydrogen mass
fraction ratio $(\mathcal{Z}/\mathcal{X})_{\mathrm{present}}=0.0187$ from
A21, in the usual scale where $\mathcal{X}+\mathcal{Y}+\mathcal{Z}=1$.
The latter ratio agrees within $1\sigma$ with the value of
$(\mathcal{Z}/\mathcal{X})_{\mathrm{present}}=0.0191$ that 
we recommend in \sect{discussionproto}.
The systematic uncertainty is dominated
by that of the equation of state.
Their Table 3 compiles results from
different groups employing different,
yet fairly modern, equations of state
(OPAL, SAHA-S, FreeEOS, and MHD), as well as different techniques
and data sets
\citep{1998A&A...338..756R,
2002A&A...384..666D,
2004ApJ...606L..85B,
2013MNRAS.430.1636V,2014MNRAS.441.3296V}.
The mean value is
$\mathcal{Y}_{\mathrm{present}}=0.2506$, with a standard deviation of $0.0042$.
With the metal abundances found in this work, this leads to 
our recommended result for helium: $\lgeps{He}=10.934\pm0.010$.

\subsubsection*{Fluorine, 
chlorine,
indium, and thallium ($Z=9$, $17$, $49$, and $81$): sunspot analyses.}

Umbra atlases \citep[e.g.][]{2001sus..book.....W} can be used
for spectroscopic abundance analyses.
Such analyses are typically performed with LTE spectrum synthesis and
theoretical 1D hydrostatic model photospheres; it should be noted that,
like for the quiet photosphere, these assumptions
may impart some systematic error 
\citep[e.g.][]{2021A&A...647A..46S,2023A&A...669A.144S}.
Fluorine and chlorine do not present atomic features
in the spectrum of the quiet photosphere,
and abundances are therefore recommended based on features of
of HF \citep{2014ApJ...788..149M}
and HCl \citep{2016AJ....152..196M}
in the umbra spectrum.
For indium, \citet{2008MNRAS.384..370V} presented an abundance from
the \ion{In}{I} $451.13\,\nm$ line 
arguing that it is less blended in the umbra spectrum
than in the spectrum of the quiet photosphere.
For thallium, \citet{1969MNRAS.142...71L}
first presented an upper limit based on the 
\ion{Tl}{I} $535.05\,\nm$ line in the 
spectrum of the quiet photosphere.
\citet{1972SoPh...26..250L} revisited
this line in the umbra spectrum, and A21 revised
their value upwards by $0.02\,\dex$ to account for
the precise laboratory oscillator strength of
$\lggf=-0.218$ from \citet{1964PhRv..136...87G}.
Our recommended results are unchanged from A21:
$\lgeps{F}=4.40\pm0.25$, $\lgeps{Cl}=5.31\pm0.20$,
$\lgeps{In}=0.80\pm0.20$, and $\lgeps{Tl}=0.92\pm0.17$.

\subsubsection*{Neon,
argon,
krypton,
and xenon ($Z=10$, $18$, $36$, and $54$): upper solar atmosphere and
neutron-capture cross-sections.}

These noble gases do not present atomic features in the
usual spectrum of the quiet photosphere.
Neon can be inferred from the upper solar atmosphere
by several different methods 
(Section 2.8 of A21),
and here we take the mean value from two of
these methods. The first approach is
via the bulk solar wind,
for which there exist in-situ measurements by
the Genesis sample return mission
\citep{2019M&PS...54.1092B}.
These measurements are strongly correlated with the measured
hydrogen to helium ratio, which may be interpreted
as being due to fractionation effects correlated with the
first ionisation potential
\citep[FIP; e.g.][]{2015LRSP...12....2L}.
Extrapolating 
to the helioseismic hydrogen to helium ratio (\sect{resultshelium})
and accounting also for the minor isotopes,
\citet{2020M&PS...55..326H} derived
a neon abundance of $8.06\,\dex$;
\citet{2021ApJ...907...15H} later revised this analysis
to arrive at $8.05\,\dex$.
The second approach is via the EUV spectrum
of the transition region of the quiet Sun,
in which the ratio of neon to oxygen 
has been measured to be $N_{\mathrm{Ne}}/N_{\mathrm{O}}=0.244$
\citep{2018ApJ...855...15Y}.
This suggests a photospheric
neon abundance of $8.09\,\dex$ when using
our recommended photospheric oxygen abundance,
and assuming the same degree of fractionation in neon as in oxygen
in the transition region.
In L25 this second method is adopted to arrive at $8.15\,\dex$
owing to adopting a higher solar photospheric abundance
for oxygen.
Our new recommended result is based on the mean of 
$8.05\,\dex$ and $8.09\,\dex$,
together with the uncertainty estimate from A21:
$\lgeps{Ne}=8.07\pm0.05$.

In A21, the value of $6.38\,\dex$ for argon is based on
Genesis data of the bulk solar wind
\citep{2020M&PS...55..326H,2021ApJ...907...15H},
via a similar analysis as
for neon described above.
The argon abundance in L25 ($6.50\,\dex$) originates from
\citet{2008ApJ...674..607L}, based on various solar and extra solar
sources.  Following a similar procedure but using newer data,
A21 arrive at a value $6.40\,\dex$, in good agreement with the 
Genesis-based analysis.
We thus recommend the result of A21:
$\lgeps{Ar}=6.38\pm0.10$.

For krypton, the value of $3.12\,\dex$ in A21 is again 
based on
Genesis data of the bulk solar wind \citep{2014GeCoA.127..326M}.
A second method is based on the interpolation of
s-process nuclide abundances of neighbouring elements,
from which A21 determine a value of $3.15\,\dex$.
Via a similar approach, L25 infer a value of
$3.31\,\dex$.  
We propose to take the mean of these
three values, and to use the standard deviation to estimate the uncertainty.
This leads to our new recommended result:
$\lgeps{Kr}=3.19\pm0.10$.

In A21, the value of $2.22\,\dex$ for xenon is 
based on interpolation of s-process nuclide abundances of
neighbouring elements.
The xenon abundance in L25 ($2.30\,\dex$) is based
on a similar approach.
Both values are
significantly lower than the value of $2.42\,\dex$
inferred from Genesis data of the bulk solar wind \citep{2020GeCoA.276..289M},
which A21 attributes to
fractionation of xenon relative to hydrogen in the solar wind.
We propose to take the mean of the A21 and L25,
and fold the discrepancy between the two into the uncertainty
from A21.
This leads to our new recommended result:
$\lgeps{Xe}=2.26\pm0.06$.

\subsubsection*{Other elements up to uranium ($Z=92$): CI chondrites.}
For the $15$ long-lived elements not yet discussed,
the solar abundance estimates are based on
the meteoritic abundances from the compilation of L25
following the procedure discussed in
\sect{discussiontcond}.
As discussed there,
there is compelling evidence for a bias in the 
composition of CI chondrites relative to the Sun
that is correlated with the
$50\%$ condensation temperatures of elements in the protosolar nebula,
$T_{\mathrm{c}}$.
The bias can be measured down to $T_{\mathrm{c}}=495\,\mathrm{K}$,
corresponding to the 50\% condensation temperature for lead,
where it amounts to a $-0.06\,\dex$ correction
to the CI chondrite abundance.
We correct for this bias in the recommended results
for the $15$ elements (``CI-Tc'' in \tab{tab:abundances}):
for four of these elements having $240\leq T_{\mathrm{c}}<495\,\mathrm{K}$,
the abundance correction is capped at $-0.06\,\dex$
(see \tab{tab:meteorites}).

\section{Discussion}
\label{discussion}

\subsection{Photosphere versus CI chondrites}
\label{discussiontcond}

The overall composition of 
CI chondrites has long been used as a proxy for the 
protosolar nebula
(see e.g.~the background discussion in \citealt{2024M&PS...59.3193J}).
The mineralogical alterations that they have experienced via
thermal metamorphism and aqueous alteration
are not thought to strongly alter their bulk chemical compositions
\citep{2014pacs.book...15P}.
A key exception to this are the highly volatile elements:
hydrogen, carbon, nitrogen, and oxygen, and the noble gases,
that have low $50\%$ condensation temperatures in the protosolar nebula,
$T_{\mathrm{c}}\lesssim180\,\mathrm{K}$
\citep{2003ApJ...591.1220L,2019AmMin.104..844W}.
They are under-represented in CI chondrites,
by amounts ranging from $0.3\,\dex$ for oxygen,
to $+9.6\,\dex$ for helium.
In particular, hydrogen is depleted by around $3.8\,\dex$.
Consequently, some care must be taken in converting the elemental
abundance ratios measured in CI chondrites onto the scale of the
present-day solar convective envelope,
for which hydrogen serves as the reference element.

\input{{table/meteorites}.tex}

To convert the CI chondrite abundance of a given element
onto the scale of the present-day solar convective envelope,
silicon is the conventional choice for the reference element.
This approach is simple to implement 
and easy to reproduce, but a problem is that
the CI chondrite abundances on the solar scale
are then prone to shifts
when new spectroscopic determinations for silicon
appear in the literature, given that the silicon abundance has
a $1\sigma$ uncertainty of $0.04\,\dex$.
Therefore, in this work we base the conversion on a large set of reference
elements, as described below.

For a given element $\mathrm{X}$ and a given 
reference element $\mathrm{R}$, the CI chondrite elemental abundance
in the scale of the present-day solar convective envelope
was found via
the concentrations by mass, $\conc{X}$ and $\conc{R}$,
provided in Table 4 of
\citet{2025SSRv..221...23L}, together with mean atomic weights
$\awgt{X}$ and $\awgt{R}$:
\begin{equation}
    \lgeps{X;R}_{\mathrm{CI}}=
    \log_{10} \left(\conc{\mathrm{X}}\awgt{X}^{-1}\right)
    + \left[\lgeps{R}_{\mathrm{Phot.}}-
    \log_{10} \left(\conc{\mathrm{R}}\awgt{R}^{-1}\right)\right]\,.
\end{equation}
This was calculated for up to $52$ different
reference elements, with the constraint that
$\mathrm{R}\neq\mathrm{X}$.
These reference elements were chosen to be those with solar abundance
determinations based on spectroscopy of the quiet solar photosphere, excluding
the highly volatile elements carbon, nitrogen, and oxygen, as well as the
fragile elements lithium, beryllium, and boron.
The uncertainty of the square bracket was calculated via:
\begin{equation}
    \sigma_{\mathrm{R}}^2=
    \sigma_{\lgeps{R}_{\mathrm{Phot.}}}^2+
    \left(\frac{1}{\conc{R}\ln{10}}\right)^2
    \sigma_{\conc{R}}^2\,.
\end{equation}
These were used to calculate weights
$w_{\mathrm{R}}=1/\sigma_{\mathrm{R}}^{2}$,
Using normalised weights that sum to unity,
$\bar{w}_{\mathrm{R}}$,
the weighted mean abundance is
\begin{equation}
    \lgeps{X}_{\mathrm{CI}}=
    \sum_{\mathrm{R}}\lgeps{X;R}_{\mathrm{CI}} \times \bar{w}_{\mathrm{R}}\,.
    \label{eq:ciabund}
\end{equation}
Assuming no correlations between different reference
elements,
the weighted uncertainty in the conversion is:
\begin{equation}
    {\sigma}_{\mathrm{conv.}}=\sqrt{\sum_{R}\sigma_{\mathrm{R}}^2
    \times\bar{w}_{\mathrm{R}}^2}\,,
\end{equation}
and amounts to around $0.008\,\dex$,
with some small dependence on the target element $\mathrm{X}$
because it is excluded from the list
of permitted reference elements $\mathrm{R}$.
This is much smaller than the $1\sigma$ uncertainty
on any single reference element
(e.g.~$0.04\,\dex$ for silicon).
The overall uncertainty in the CI chondrite abundance
is then:
\begin{equation}
    \sigma_{\lgeps{X}_{\mathrm{CI}}}^2=
    \left(\frac{1}{\conc{X}\ln{10}}\right)^2
    \sigma_{\conc{X}}^2+\sigma_{\mathrm{conv.}}^2\,.
    \label{eq:cisigma}
\end{equation}
These abundances, \eqn{eq:ciabund}, and 
uncertainties, \eqn{eq:cisigma}, are listed
in \tab{tab:meteorites}.

\begin{figure}[ht]
\centering
    \includegraphics[width=1\textwidth]{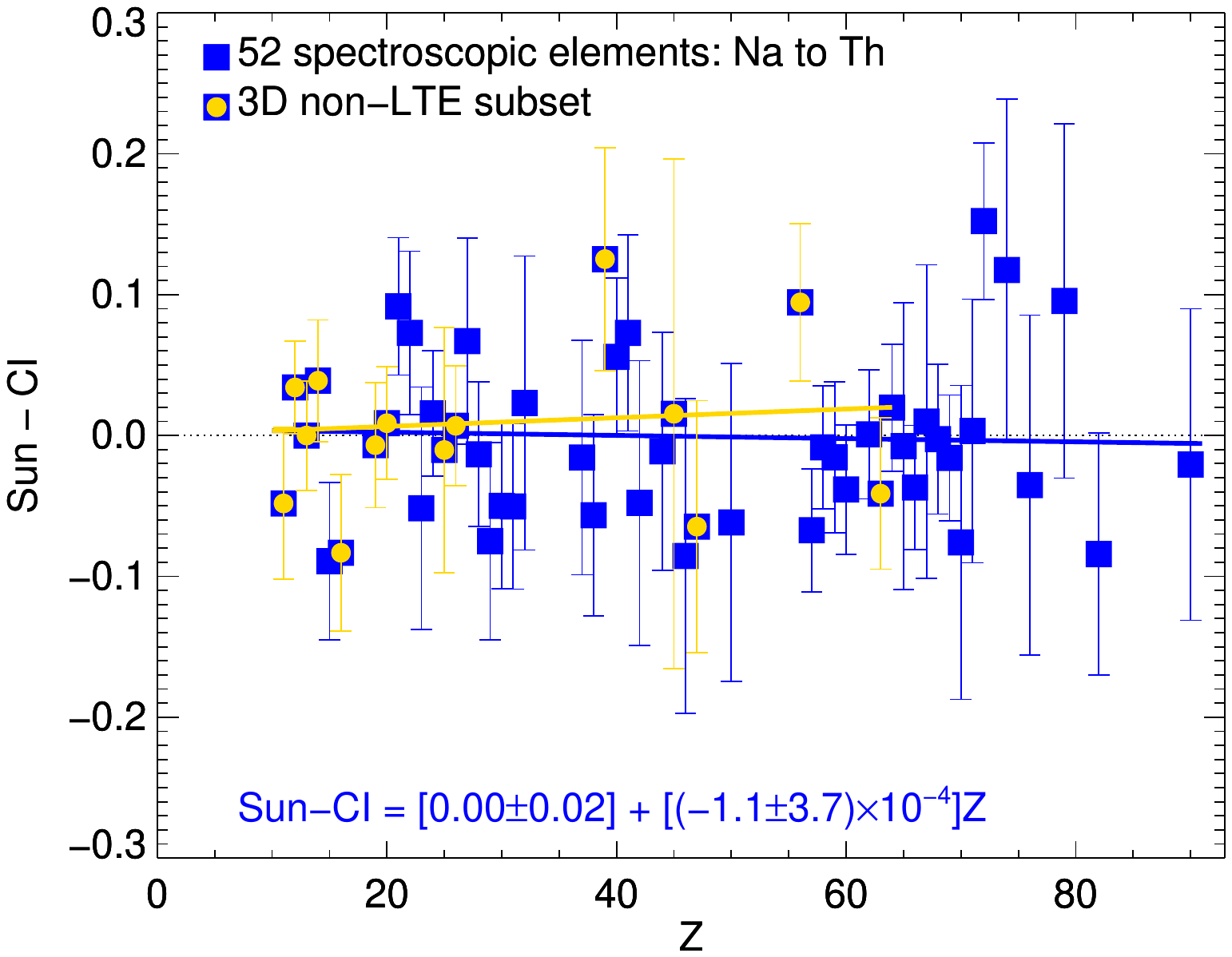}
    \caption{Photospheric abundances from
        \sect{resultsintermediate}
        to \sect{resultsheavy}
        versus those from CI chondrites,
        as a function of atomic number.
        Lines of best fit overplotted for the full set
        of elements and for the subset of elements 
        based on
        consistent 3D non-LTE calculations.
        These fits are weighted by the uncertainties 
        in the photospheric and CI chondrite abundances,
        added in quadrature.}
    \label{fig:atomic}
\end{figure}

In \fig{fig:atomic},
the abundances of the CI chondrites on the 
scale of the present-day solar convective envelope
are compared to
those inferred from spectroscopy of the quiet 
solar photosphere, as a function of atomic number.
Only the $52$ reference elements are shown.
The uncertainties are here given by those of the photospheric
abundances, added in quadrature with
those of the CI chondrite abundances albeit excluding
the systematic contribution of $0.008\,\dex$ from converting
to the solar scale.
Out of these elements, $77\%$ ($40$ out of $52$)
are consistent with $\mathrm{Sun-CI}=0$ to within $1\sigma$.
For the $14$ elements whose abundances are based on
consistent 3D non-LTE modelling, this increases to 
$85\%$.  
Both of these percentages are somewhat larger than
what would be expected from a Normal distribution,
which may indicate that the $1\sigma$ uncertainties
are slightly overestimated overall, or it may reflect
that the
uncertainties in the abundances of different elements are correlated.

\begin{figure}[ht]
\centering
    \includegraphics[width=1\textwidth]{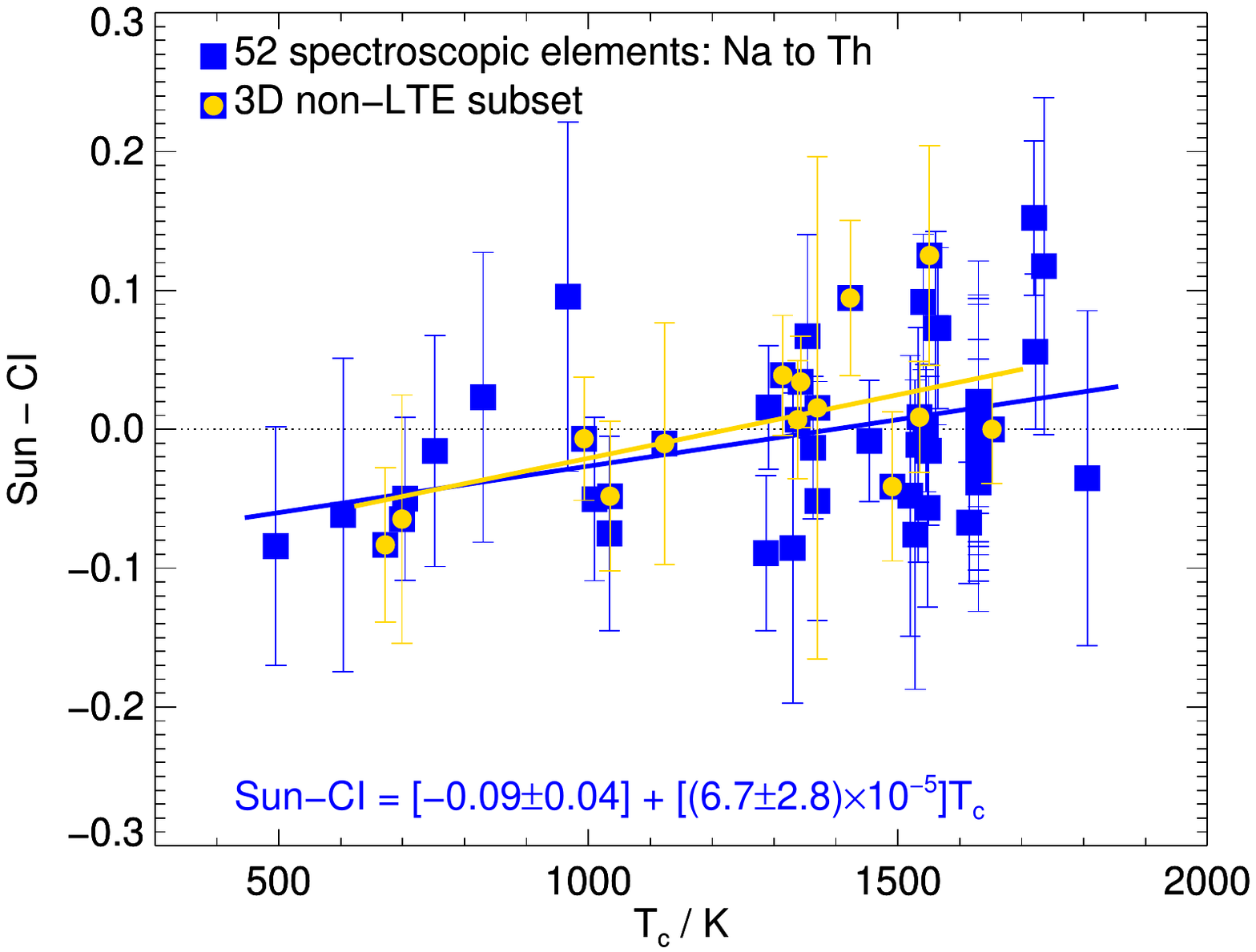}
    \caption{Photospheric abundances from
        \sect{resultsintermediate}
        to \sect{resultsheavy}
        versus those from CI chondrites,
        as a function of condensation temperature 
        \citep{2019AmMin.104..844W}.
        Lines of best fit overplotted for the full set
        of elements and for the subset of elements 
        based on
        consistent 3D non-LTE calculations.
        These fits are weighted by the uncertainties 
        in the photospheric and CI chondrite abundances,
        added in quadrature.}
    \label{fig:tcond}
\end{figure}

The abundances of the CI chondrites and solar photosphere
are compared again in \fig{fig:tcond},
for the same set of $52$ reference elements,
but now as a function of $50\%$ condensation temperature.
As before, the uncertainties are here given by those of the photospheric
abundances, added in quadrature with
those of the CI chondrite abundances, and excluding
the systematic contribution of $0.008\,\dex$ from converting
to the scale of the present-day solar convective envelope.
There is a trend such that the moderately volatile
elements are enriched in the CI chondrites 
by around $0.06\,\dex$
and the refractory elements depleted in the CI chondrites
by $0.03\,\dex$,
relative to the Sun.
It should be stressed that the zero point 
depends on how the CI chondrites are converted to the 
solar scale.
For example, if the moderately volatile element
lead (with $T_{\mathrm{C}}=495\,\mathrm{K}$) 
by itself was used as the sole reference element, then all 
points would shift upwards in \fig{fig:tcond} such that 
the phenomenon would instead appear as a strong
depletion of the refractory elements in the CI chondrites by $0.09\,\dex$.
In any case, the gradient of the trend
is not affected by the conversion method.
Even though our uncertainties may be slightly overestimated or
correlated, the gradient is found to be significant to 
more than $2\,\sigma$;
the true significance may be even greater, given
that our $1\sigma$ uncertainties may be slightly overestimated and that correlations
are neglected.

This phenomenon had first been suggested by
\citet{1997MNRAS.285..403G}, and was later discussed in
detail by \citet{2021A&A...653A.141A}.
\citet{2024M&PS...59.3193J} reported a similar
effect based on solar wind abundances from
the Genesis sample return mission.
In that same study it was shown that the phenomenon
is present in older abundance tables
including \citet{1989GeCoA..53..197A},
\citet{1998SSRv...85..161G}, \citet{2009ARA&A..47..481A}, and
\citet{2011SoPh..268..255C}.
In fact, the phenomenon is also present in
the compilation of \citet{2025SSRv..221...23L}:
when adopting their abundance values together
with our uncertainty estimates, the trend appears 
with a $20\%$ shallower gradient and is
still significant to nearly $2\sigma$.
This historical consistency is remarkable because each of these sets of
solar photospheric abundances has been obtained with rather different
parameters and techniques.
Thus, rather than to confirm or erase the existence of this phenomenon,
future improvements to the spectroscopic methods would likely help 
in constraining the best-fitting profile.  For instance,
the data are not just consistent with a linear trend,
but are equally well-described by a
step function with a break at around
$T_{\mathrm{c}}=1350\,\mathrm{K}$ \citep{2024M&PS...59.3193J,2025A&A...703A..35A}.

As discussed in \citet{2024M&PS...59.3193J},
the observed different behaviour of
volatiles and refractories may be explained by the
evolutionary model of \citet{2018ApJS..238...11D} for
the  protoplanetary disc with the role of the
formation of a planet like Jupiter,
which predicts that CI chondrites are depleted in
refractory elements relative to the photosphere by about $0.05\,\dex$.
We stress again, however, that the zero point
of the Sun versus CI comparison depends on how the CI chondrites are
converted to the scale of the present-day solar convective envelope. 
Nevertheless, it can be said that solar photospheric abundances appear to be
a good test of such models, and may be a good witness
of the formation of the solar system.

\subsection{Mass fractions of the present-day solar convective envelope
and of the protosolar nebula}
\label{discussionproto}

\begin{table}[ht]
    \caption{Recommended mass fractions of the
    present-day solar convective envelope
    and of the protosolar nebula, contrasted with
    A21 \citep{2021A&A...653A.141A}
    and L25 \citep{2025SSRv..221...23L}.}\label{tab:massfractions}
\begin{tabular*}{\textwidth}{@{}lrrrr@{}}
\toprule
    & 
    Hydrogen ($\mathcal{X}$) & 
    Helium ($\mathcal{Y}$) & 
    Metals ($\mathcal{Z}$) &
    Metal-to-hydrogen ($\mathcal{Z}/\mathcal{X}$) \\
\midrule
    \noalign{\smallskip}
    & \multicolumn{4}{c}{Present-day}\\
    \noalign{\smallskip}
    A21 & $0.7438\pm0.0054$ & $0.2423\pm0.0054$ & $0.0139\pm0.0006$ & 
    $0.0187\pm0.0009$ \\
    L25 & $0.7385\pm0.0068$ & $0.2451\pm0.0069$ & $0.0160\pm0.0013$ & 
    $0.0216\pm0.0017$ \\
    Recommended & $0.7354\pm0.0042$ & $0.2506\pm0.0042$ & $0.0140\pm0.0006$ &
    $0.0191\pm0.0009$ \\
    \noalign{\smallskip}
    \hline
    \noalign{\smallskip}
    & \multicolumn{4}{c}{Protosolar}\\
    \noalign{\smallskip}
    A21 & $0.7121\pm0.0058$ & $0.2725\pm0.0059$ & $0.0154\pm0.0007$ &
    $0.0216\pm0.0010$ \\
    L25 & $0.7060\pm0.0065$ & $0.2753\pm0.0077$ & $0.0187\pm0.0015$ &
    $0.0265\pm0.0021$ \\
    Recommended & $0.7170\pm0.0043$ & $0.2681\pm0.0044$ & $0.0149\pm0.0007$ &
    $0.0208\pm0.0010$\\
\botrule
\end{tabular*}
\end{table}

In the usual scale where $\mathcal{X}+\mathcal{Y}+\mathcal{Z}=1$,
the recommended abundances in \tab{tab:abundances}
give a metal-to-hydrogen mass fraction ratio of 
$(\mathcal{Z}/\mathcal{X})_{\mathrm{present}}=0.0191\pm0.0009$
for the present-day solar convective envelope.
Combining with the helium mass fraction
$\mathcal{Y}_{\mathrm{present}}=0.2506\pm0.0042$ from \sect{resultshelium},
the hydrogen mass fraction is $\mathcal{X}_{\mathrm{present}}=0.7354\pm0.0042$ and 
the total metal mass fraction is $\mathcal{Z}_{\mathrm{present}}=0.0140\pm0.0006$
for the present-day solar convective envelope.

Because of the combined effects of thermal diffusion, gravitational settling,
and radiative acceleration (often referred to collectively as atomic
diffusion; e.g. \citealt{2015ads..book.....M})
acting at the base of the
convection zone over the Sun's lifetime of $4.56\,\mathrm{Gyr}$, 
the protosolar abundances, relative to hydrogen, of all the elements are higher than
those of the present-day solar 
convective envelope
\citep[e.g.][]{2002JGRA..107.1442T}.
When using Standard Solar Models
\citep[e.g.][]{1996Sci...272.1286C,1997PhRvL..78..171B},
the predictions of this
difference between protosolar and 
present-day mass fractions are of the order of 
11\% for helium, and 15\% for the metal-to-hydrogen ratio
\citep{1998SSRv...85..161G}. 
However, 
additional mixing processes must be
at work at the base of the convection zone, in
order to explain the large depletion of lithium and 
the small depletion of beryllium compared to what is
found from CI chondrites
(\sect{discussionsolarproblem}).
These additional mixing processes are not
taken into account in the Standard Solar Models.
As a consequence of these
additional mixing processes, the evolutionary corrections 
are decreased from 11\% to about 7\% for the helium mass fraction 
(Section 4 of \citealt{2026A&A...708A.147B}),
and from 15\% down to 9\% for the 
metal-to-hydrogen mass fraction ratio
(Fig.~2 of \citealt{2026A&A...708A.147B}).

In the form $\lgeps{X}$,
the evolutionary corrections amount
to $+0.040\,\dex$
for helium and of $+0.037\,\dex$ for heavier elements
in \tab{tab:abundances}.
For up to nine elements with abundant unstable isotopes
radioactive decay further increases the
inferred abundances of the protosolar nebula and should be taken into
account
(Table B1 of \citealt{2021A&A...653A.141A}).
For the fragile elements,
the abundances of the protosolar nebula 
may be estimated via
the CI chondrite abundances 
converted to the scale of the present-day solar convective envelope
with corrections applied due to
the condensation temperature trend
as listed in \tab{tab:meteorites}
(\sect{discussiontcond}),
together with a correction of $+0.037\,\dex$
for evolutionary effects. The abundances of the protosolar
nebula are thus estimated to be
$3.28\,\dex$, $1.38\,\dex$, and $2.76\,\dex$
for lithium, beryllium, and boron, respectively.

The recommended mass fractions are summarised in \tab{tab:massfractions},
together with those from the complete compilations of
\citet{2021A&A...653A.141A}
and \citet{2025SSRv..221...23L}.
In \citet{2021A&A...653A.141A}, the metal
to hydrogen mass fraction ratio of the present-day solar convective
envelope is 2\% lower, but within the uncertainties, than what
is recommended here, with around half of this difference due to
their $0.01\,\dex$ lower oxygen abundance.
In contrast, their protosolar values are 4\% higher,
because of stronger predicted corrections for evolutionary effects
of $+0.064\,\dex$ (compared to $+0.037\,\dex$) via the 
Standard Solar Models of \citet{2017ApJ...835..202V}.
Already then, \citet{2021A&A...653A.141A}
anticipated that their evolutionary corrections were overestimated,
as discussed in their Section 5.
In \citet{2025SSRv..221...23L}, 
the metal-to-hydrogen mass fraction ratios
are 13\% (present-day) and 27\% (protosolar)
higher than recommended here.
Their $0.06\,\dex$ higher oxygen abundance 
(\sect{resultsoxygen})
is responsible for nearly half of difference in 
the present-day results;
they also adopt much larger evolutionary corrections, amounting
to $+0.09\,\dex$ taken from the models of 
\citet{2019ApJ...873...18Y} 
that are not constrained to reproduce the flat solar rotation profile 
nor the depletion of light elements (lithium and beryllium),
both of which are needed for calibrating
macroscopic mixing at the base of the convective envelope
\citep{2022NatAs...6..788E,2025A&A...694A.285B}.
Turning to helium, the present-day values adopted by
\citet{2025SSRv..221...23L}
via \citet{2004ApJ...606L..85B},
and by \citet{2021A&A...653A.141A} 
via the mean of \citet{2004ApJ...606L..85B} and \citet{2013MNRAS.430.1636V},
are up to 3\% lower than the value recommended here that is drawn from
the mean from seven different analyses (\sect{resultshelium}).
In contrast, their protosolar values are 
up to 3\% higher than what is recommended here,
again due to adopting much stronger evolutionary corrections.

\subsection{The Solar Modelling Problem}
\label{discussionsolarproblem}

At the end of the 1990s, Standard Solar Models
\citep[e.g.][]{1996Sci...272.1286C,1997PhRvL..78..171B},
constructed with the 
Standard Solar Composition with
metal-to-hydrogen mass fraction ratios
ranging from $(\mathcal{Z}/\mathcal{X})_{\mathrm{present}}=0.0274$
\citep{1989GeCoA..53..197A}
to $0.0231$ \citep{1998SSRv...85..161G},
were found to be in
excellent agreement with key observational diagnostics
of helioseismology: 
not least including
the helium abundance in the convective envelope,
the sound speed profile,
and the depth of the convection zone.
In the 2000s, the increasing use of more realistic
3D and non-LTE modelling of the solar photosphere
triggered (in nuanced ways; see \sect{methodeffects})
a downwards revision in the 
metal-to-hydrogen mass fraction ratios:
values as low as $(\mathcal{Z}/\mathcal{X})_{\mathrm{present}}=0.0165$
were reported \citep{2005ASPC..336...25A},
before converging on the range
$0.018<(\mathcal{Z}/\mathcal{X})_{\mathrm{present}}<0.021$
\citep{2009ARA&A..47..481A,2021A&A...653A.141A,
2011SoPh..268..255C,2009LanB...4B..712L,2021SSRv..217...44L}.
Our recommended value of $(\mathcal{Z}/\mathcal{X})_{\mathrm{present}}=0.0191\pm0.0009$
(\sect{discussionproto}) sits within this range, and
is $4\sigma$ below the value presented 
in \citet{1998SSRv...85..161G}.
The same Standard Solar Models built with these
lower metal-to-hydrogen mass fraction ratios
do not satisfactorily reproduce the aforementioned helioseismic constraints.

As we already mentioned in \sect{introduction}, this was
first labelled the Convective Zone Problem
\citep{2004ApJ...614..464B,2004ApJ...615.1042M},
and later the Solar Abundance Problem
\citep[e.g.][]{2008SoPh..251...53C,2009ApJ...704.1174P}.
However, it is
not at all clear that the solar composition is at fault, as we discuss below.

There are a number of other constraints that solar models must satisfy,
in addition to the aforementioned helioseismic contraints.
Actually, even back in the 1990s, the Standard Solar Models
failed to reproduce the lithium and beryllium abundances
in the convective envelope.
With $\lgeps{Li}=0.96$ and $\lgeps{Be}=1.21$ in the 
present-day solar convective envelope
(\sect{resultslight})
and $\lgeps{Li}=3.28$ and $\lgeps{Be}=1.38$ in the 
protosolar nebula (\sect{discussionproto}),
lithium and beryllium are depleted by factors of 
$209$ and $1.48$, respectively.  In order to destroy
these elements by proton captures, they need to reach temperatures of around 
$2.5\times10^{6}\,\mathrm{K}$ and $3.5\times10^{6}\,\mathrm{K}$.
These exceed the
mean temperature at the base of the convective zone,
which was around $2.45\times10^{6}\,\mathrm{K}$ 
on the zero-age main sequence
and is around $2.20\times10^{6}\,\mathrm{K}$ at present-day
\citep{2021LRSP...18....2C}.
Models \citep[e.g.][]{2025A&A...702A.167K}
as well as, for lithium, observations 
(as reviewed in e.g.\ \citealt{2006cams.book..163J}
and \citealt{2021FrASS...8....6R}) indicate that
there is insufficient depletion on the 
pre-main sequence phase to explain the current surface abundances
of these elements.
Thus, there is a need for additional macroscopic transport
at the base of the convection zone that would allow
some material to penetrate deep into the hotter radiative zone,
destroying lithium and beryllium
and gradually depleting them from the solar convective envelope.
But, as already mentioned in \sect{discussionproto},
Standard Solar Models 
do not include the necessary additional mixing processes,
which may include convective overshooting, rotation, internal
gravity waves, and turbulence.

Non-standard solar models with additional mixing processes
calibrated to reproduce the lithium and beryllium abundances
significantly improve the agreement with the helioseismic
helium abundance, and fare just as well as Standard Solar Models
in reproducing the other helioseismic constraints overall
\citep{2025A&A...694A.285B}.
But reproducing the measured solar neutrino fluxes
is a significant challenge, and this is the
case for both low and high metallicity models
\citep{2024A&A...686A.108B}. Including
additional macroscopic mixing at the base of the convection zone,
so as to explain the lithium and beryllium depletions, results in a 
reduction of the efficiency of gravitational settling
which leads to a lower present-day metallicity of the solar core
\citep{2025A&A...702A.167K}.
This significantly reduces the neutrino fluxes
associated with the destruction of
beryllium and boron,
$\phi_{\mathrm{Be}}$
and $\phi_{\mathrm{B}}$ respectively:
\begin{equation}
    \mathrm{^{7}_{4}Be}+e^{-}\rightarrow
    \mathrm{^{7}_{3}Li}+\nu_{e}\,,
\end{equation}
\begin{equation}
    \mathrm{^{8}_{5}B}\rightarrow
    \mathrm{^{8}_{4}Be}+e^{+}+\nu_{e}\,,
\end{equation}
that occur within the proton-proton chain.
The predictions for 
$\phi_{\mathrm{Be}}$ and $\phi_{\mathrm{B}}$
from the low metallicity non-standard solar
model of \citet{2025A&A...694A.285B}
are around 4\% and 8\% lower than the measurements compiled in
\citet{2021ARNPS..71..491O};
the high metallicity 
models presented in \citet{2024A&A...686A.108B}
are also in disagreement with these measurements.
Even more strongly affected are the neutrinos associated
with the CNO cycle.
The first measurement of this neutrino flux was reported by
\citet{2020Natur.587..577B}, with final results published in
\citet{2023PhRvD.108j2005B}:
$\phi_{\mathrm{CNO}}=6.7^{+1.2}_{-0.8}\times10^{-8}\,\mathrm{cm^{-2}s^{-1}}$.
In contrast the low metallicity non-standard solar
model of \citet{2025A&A...694A.285B}
presents a value of $3.84\,\mathrm{cm^{-2}s^{-1}}$, 
that is 43\% lower than the measurement and corresponds to
a $3.5\,\sigma$ discrepancy.
Accretion of matter during the proto-solar phase 
increases the metallicity of the solar core by
only around 5\% \citep{2021A&A...655A..51K},
and is insufficient to resolve the discrepancy
\citep{2025A&A...702A.167K}.
Updated nuclear reaction rates from
\citet{2025RvMP...97c5002A} exacerbate the discrepancy,
regardless of the assumed metallicity of the solar 
model \citep{2026A&A...710A.270S}.

Therefore, further refinements or additions to the microphysics
may be needed in order to reproduce the full suite of observable
constraints. Not least 
among these are improvements to the opacity tables
employed by solar models \citep[e.g.][]{
2025A&A...700A..50B,2025NatCo..16..693B}.
Standard opacity tables, such as 
OPAL \citep{1996ApJ...464..943I},
OP \citep{2005MNRAS.360..458B},
OPAS \citep{2015ApJS..220....2M},
and 
OPLIB \citep{2016ApJ...817..116C}, are constructed by
combining assumed elemental abundances with an
equation of state and cross-sections for photoextinction
by bound-bound and bound-free processes.
At the base of the convection zone the most important
opacity contributors are oxygen and iron: 
they together account for nearly $50\%$ of the total opacity
(Figure 2 of \citealt{2025SoPh..300...97B}).
The opacity of \ion{Fe}{XVII} was measured
in conditions close to that
at the base of the convection zone by \citet{2015Natur.517...56B}.
They reported that standard opacity tables significantly underestimate
the opacity contribution due to iron, by an amount
that resolves around half of the total discrepancy
between Standard Solar Models and helioseismic constraints.
Subsequent studies found 
small overall discrepancies with standard opacity tables
for chromium and nickel 
\citep{2019PhRvL.122w5001N}
and for oxygen \citep{2025PhRvL.135t5101B},
but revealed systematic shortcomings in the different opacity calculations
that may help explain the observed discrepancies for iron
and that may lead to divergences in other temperature regimes.
Any improvements to the microphysics may well
extend to conditions at the solar core, where
iron is the largest single contributor at $30\%$,
and may thus also affect the neutrino fluxes
predicted by solar models.

Shedding light on the problem are the studies of
the metal-to-hydrogen
mass fraction ratio in the present-day solar convective envelope
based directly on
helioseismic inversions of the first adiabatic exponent
\citep{2013MNRAS.430.1636V,2014MNRAS.441.3296V,2017MNRAS.472..751B}.
This approach has the advantage of being 
insensitive to the assumed spectroscopically-determined
abundances and to the treatment of radiative opacities;
the caveat is a sensitivity to
microphysics uncertainties, not least the equation of state
(just like for inferring the helium mass fraction; \sect{resultshelium}).
The most recent study in this direction is that of
\citet{2024A&A...681A..57B}: taking into account
independent observational data sets and theoretical
equation of states, their Table 3 demonstrates that
$0.0175\lesssim(\mathcal{Z}/\mathcal{X})_{\mathrm{present}}\lesssim0.0199$.
Giving equal weight to the two observational data sets
leads to a mean value of $0.0187$ which
agrees with the spectroscopic result recommended here,
$(\mathcal{Z}/\mathcal{X})_{\mathrm{present}}=0.0191\pm0.0009$,
and rules out the higher values of the 1990s.
As already stated above, it thus appears as though
the problem is not with the abundances, and we are rather 
faced with a much more complex Solar Modelling Problem
for which a comprehensive solution still eludes us
\citep{2025SoPh..300...97B}.

\section{Conclusion}
\label{conclusion}

There are various reasons for the departing values
in the solar abundances found by different groups. 
We attempted to shed light on these reasons,
in some cases dissecting the analyses on a 
line-by-line basis
(\sect{results}).
The modelling paradigm can drive some of the discrepancies:
consistent 3D non-LTE modelling is preferred,
and in particular
drawing upon comprehensive and realistic model atoms,
that employ physically-motivated descriptions of the
inelastic hydrogen collisions.
Nevertheless, the traditional challenges of 1D LTE spectroscopy 
cannot be neglected: the most reliable results
are generally based on careful measurements of the disc-centre intensity
spectrum, using diagnostic absorption lines that
are relatively free of blends and
are on the weak regime of the curve of growth,
and for which transition probabilities and broadening parameters
have been accurately measured or calculated by atomic physicists
(\sect{method}).

A broader take-away message from this exercise is on the importance
of continuing to critically question models and analyses, even 
when consistency appears to be reached.
Indeed, careful scrutiny of the Standard Solar Composition
at the turn of the century, aided by 3D and non-LTE models,
is what revealed the Solar Modelling Problem.
Clearly, there is still much exciting
physics to be learned about our Sun,
and ultimately understanding and resolving this Problem
may well have rippling effects on stellar astrophysics
and beyond (\sect{discussion}).

Looking ahead, it is our view that the number one priority in this research
should be the careful and consistent 3D non-LTE modelling
for many more elements. The need for this
is exemplified by the abundances of rhodium and silver,
which here are $+0.28\,\dex$ and $+0.19\,\dex$, respectively,
higher than those in the compilation
of both \citet{2021A&A...653A.141A}
and \citet{2021SSRv..217...44L}.
This is due to severe 3D non-LTE effects,
here accounted for via the detailed modelling of
the neutral silver resonance lines by \citet{2026A&A...711A.155C},
which we have also used to predict the effects on neutral rhodium.
The next largest updates relative to 
\citet{2021A&A...653A.141A} 
are for beryllium ($-0.17\,\dex$), sulphur ($-0.06\,\dex$),
and yttrium ($+0.08\,\dex$);
again, these changes are
largely a result of consistent 3D non-LTE modelling
for the first time
\citep{2024A&A...690A.128A,2025A&A...703A..35A,2024A&A...683A.200S}.

\fig{fig:abundancepattern} and \tab{tab:abundances} illustrate
that up to $19$ elements are now 
based on 3D non-LTE modelling
($17$ elements, if manganese and rhodium are excluded).
This is an increase from around
$12$ elements just five years ago \citep{2021A&A...653A.141A},
and a linear extrapolation would suggest that a complete 
solar abundance pattern based on
3D non-LTE models will take at least fifty more years.
What limits the progress is the lack 
of easily accessible complete and accurate atomic data (energies, and
radiative and collision cross-sections;
\citealt{2016A&ARv..24....9B}), and the person-power
needed to compile and reduce these data into model atoms,
with care and a critical eye.
Efforts to calculate, measure, and compile such data should continue 
to be encouraged.  These efforts  will not only lead to a better
understanding of our Sun, but are also key 
to constraining the parameters and compositions of stars
\citep{2024ARA&A..62..475L}
which in turn provide insight on the physics of stars and the planets
that orbit them and
the formation and evolution of our Galaxy and its neighbours
and thus help us build a better understanding of the nature of the cosmos
\citep{2018A&ARv..26....6N,2025A&ARv..33....3G}.
In the era of increasingly large stellar surveys
\citep{2019ARA&A..57..571J}, we foresee accelerating
progress on 3D non-LTE modelling for stellar astrophysics,
and thus are optimistic that a complete 3D non-LTE 
picture of the solar chemical composition may be obtained
even within the next fifteen years.

\bmhead{Acknowledgements}
We would like to thank 
Elisabetta Caffau and an anonymous referee,
as well as our colleagues listed below, 
for reading the manuscript and providing constructive feedback:
Ga\"{e}l Buldgen, Sema Caliskan,
Bengt Edvardsson, Bengt Gustafsson, Andreas Korn, Masanobu Kunitomo,
Cis Lagae (who also provided the images used in \fig{fig:3dbox}), 
Karin Lind, and Poul Erik Nissen. We would also like to thank
Amy Jurewicz, Martin Laming, and Rainer Wieler
for their insight on data and results from the Genesis sample return mission.
This work was supported by funding 
from the Swedish Research Council (VR 2025-05167).

\section*{Declarations}

\bmhead{Conflict of interest}
The authors declare no conflict of interest.

\phantomsection
\addcontentsline{toc}{section}{References}
\bibliography{bibl.bib}

\end{document}